\documentclass[draft]{agujournal2019}
\usepackage{url} 
\usepackage{lineno}
\usepackage[inline]{trackchanges} 
\usepackage{soul}
\usepackage{amsmath}
\usepackage{amsfonts}
\usepackage{amssymb}
\usepackage{verbatim}
\journalname{JGR: Space Physics}
\renewcommand{\vec}{\mathbf}

\begin{document}

%
%


\title{Kinetic Theory of the Thermal Farley-Buneman Instability in the E-Region Ionosphere}

%
%




\authors{Y.~S.~Dimant and M.~M.~Oppenheim}


\affiliation{}{Center for Space Physics, Boston University, Boston,  Massachusetts, 02215, United States}




\correspondingauthor{Yakov Dimant}{dimant@bu.edu}



\begin{keypoints}
\item A fully kinetic linear theory of the thermal Farley-Buneman instability (TFBI) in the E-region ionosphere has been developed.
\item For the first time in the E-region instability theoretical research, the driving electric field has been included in the kinetic description of ions.
\item The analytical results can enable observers to more accurately interpret observations of electrojet turbulence at
frequency bands for which fluid models are inapplicable.
\end{keypoints}

%
%

%
%


\begin{abstract}
This paper develops a fully kinetic linear theory of the thermal Farley-Buneman instability (TFBI) in the Earth's E-region ionosphere assuming unmagnetized ions. The TFBI combines the Farley-Buneman instability (FBI), ion thermal instability (ITI), and electron thermal instability (ETI). Similar collision-dominated plasma instabilities can also occur in stellar and other planetary ionospheres. This new theory adds a kinetic description of ions that includes a driving electric field, resulting in the automatic inclusion of the ITI. This analytic theory has produced a comprehensive linear wave dispersion relation. It is remarkable that -- similarly to the oversimplified earlier ion-kinetic studies -- this much more general kinetic dispersion relation involves only elementary functions and the standard plasma dispersion function (albeit of several different arguments). This new theory is limited to plasma waves with the frequencies of the order, or larger than, the ion-neutral collision frequency. This kinetic frequency range allows for accurate interpretations of radar signals scattered from relatively high E-region altitudes, but at altitudes where ions are unmagnetized (mostly, below 110 km).
\end{abstract}

%
%

%


%
%
%
%

\section{Introduction\label{Introduction}}

In the lower E-region ionosphere, mostly at the magnetic equator and high latitudes, DC electric fields
$\vec{E}_{0}$ perpendicular to the geomagnetic field $\vec{B}$ create intense
currents called electrojets. When these fields reach sufficient strength, they initiate low-frequency cross-field
plasma instabilities that develop into plasma turbulence. Radars, such as
the Jicamarca Radio Observatory (JRO, Peru) at the magnetic equator \cite{Hysell:Imaging02,Hysell:Combined07}, EISCAT \cite{Schlegel:Auroral90}
and ICEBEAR \cite{Ivarsen:Turbulence24} at high latitudes, and many others frequently detect this plasma turbulence. At high latitudes, during big geomagnetic
storms or under other circumstances, the driving field magnitude, $E_{0}$, may reach hundreds of mV/m, which is many times the typical instability
threshold. The stronger the driving field, the more intense the observed
plasma turbulence and its macroscopic effects [about the latter, see
\citeA{Dimant:Magnetosphere11_2,Wiltberger:Effects17,
St-Maurice:Revisiting21,Liang:Time-dependent21} and references therein]. 
The analysis of this paper mostly applies to these strong electric field conditions.

These electrojet instabilities occur in the lower E region and the top
D region \cite{Blix:Experimental96}, mostly within the altitude range between
70 and 130 km, where electrons are highly magnetized, while ions are fully or partially
unmagnetized due to frequent ion-neutral collisions. Under these conditions,
the strong driving field $\vec{E}_{0}$ causes a fast $\vec{E}_{0}\times\vec
{B}$ drift of electrons, while ions travel mostly with the surrounding neutral gas.
This creates the two-stream condition crucial for driving the instability. 

The most well-known electrojet instabilities are the Farley-Buneman instability (FBI)
\cite{Farley:Two-stream63,Buneman:Excitation63} and the gradient drift
instability (GDI) \cite{Simon:Instability63,Maeda:Theoretical63,Hoh:Instability63}. More
recently, two more instabilities have been predicted theoretically: the
electron thermal instability (ETI)
\cite{Dimant:Kinetic95b,Dimant:Physical97,Robinson:Effects:98,St.-Maurice:Role00}, with possible observational evidences
\cite{Blix:Experimental96,Tsunoda:Radar97}, and the ion thermal instability
(ITI) \cite{Kagan:Thermal00,Dimant:Ion_thermal04,Makarevich:Toward21} (note
that this terminology had not been established until recently). Signatures of
these thermal instabilities have been also detected in fully kinetic
particle-in-cell (PIC) simulations
\cite{Janhunen:Perpendicular94,Janhunen:Recent95,Oppenheim:Hybrid95,
Oppenheim:Saturation96,Oppenheim:Ion_thermal04,Oppenheim:Large08,Oppenheim:Kinetic13}. 
All these instabilities differ by their driving mechanisms: the FBI is
caused by the ion inertia, while the two thermal instabilities are caused by
wave-modulated frictional heating of the corresponding plasma particles
(electrons or ions). Such processes can also occur in the solar chromosphere,
where conditions similar to the E-region ionosphere exist; for the
corresponding literature, \citeA<see>[]{Oppenheim:Newly20,Dimant:Unified23,Evans:Multifluid23} and references
therein. The two last papers, based on a multi-fluid theoretical approach, coined the term TFBI for the combination of
the FBI, ETI, and ITI. This paper develops the linear kinetic theory of this
combined instability.

Starting from the pioneering works by \citeA{Farley:Plasma63,Farley:Two-stream63} and
\citeA{Buneman:Excitation63}, the linear theory of the FBI has been developed
using fluid equations or kinetically (or using a hybrid approach when one 
plasma component was described using a fluid approach, while the other used a kinetic theory); 
for an early review on the theory, \citeA<see>{Fejer:Theory84,Farley:Theory85}. The
fluid approach is simpler, but the kinetic approach is more general and
accurate. The validity of the fluid equations for both electrons and ions is
restricted by the conditions of strong collisionality. Essentially, this means
that the absolute value of the cyclic wave frequency, $\omega=\omega
_{r}+i\gamma$, must be much less than the ion-neutral mean collision frequency,
$\nu_{in}$ (here $\omega_{r}$ is the real frequency, while $\gamma$ is the
instability growth or damping rate, depending on the sign). 
A proper theoretical description of the weakly collisional, medium-to-short
wavelengths, corresponding to frequencies $|\omega|\gtrsim\nu_{in}$, necessarily requires the kinetic theory.

The FBI-related collisional kinetic theory had been developed or
employed by a number of researchers \cite<>[and others]{Farley:Plasma63,Lee:High-Frequency71,Schmidt:Density73,
Ossakow:Parallel75,Schlegel:Short83,Basu:Kinetic95} using for both ions and electrons kinetic equations 
with the simplified BGK collisional model \cite{Bhatnagar:Model54}. None of these models,
however, included the driving electric field, $\vec{E}_{0}$, in the kinetic description of
ions. As a result, the employed ion kinetic theory could only describe the
pure FBI and no thermal processes like the ITI could have been covered.
The kinetic description of electrons included $\vec{E}_{0}$, but only through the $\vec
{E}_{0}\times\vec{B}$ drift velocity (this is usually sufficient for the pure FBI description). 

For electrons, a simplified kinetic equation with the BGK collision term 
is not appropriate, as we discuss in Sec.~\ref{Electron contribution to the wave dispersion relation} below.
\citeA{Dimant:Kinetic95a} developed a consistent kinetic theory for the electron description of the FBI. 
The corresponding kinetic treatment had also automatically predicted the ETI
\cite{Dimant:Kinetic95b,Dimant:Kinetic95c}. That theory, however, was
restricted to the long-wavelength range, where a simpler fluid model is also approximately applicable \cite{Dimant:Physical97}.

There are a number of reasons why a theoretical description of the electrojet
instabilities requires kinetic theory. First, this theory is the most
general approach to describing any wave plasma processes, so that without such
theory no description is complete. Second, for weakly collisional
short-wavelength waves ($|\omega|\gtrsim\nu_{in}$), such purely kinetic factor
as the collisionless Landau damping plays a crucial role for the wave
dissipation. Unfortunately, due to a significant complexity, the analytic kinetic approach can be successfully developed 
mostly for the linear instability theory with limited applications. To
determine full characteristics of developed wave turbulence, one needs the
complete nonlinear theory (still looks unfeasible) or can use advanced fully
kinetic simulations
\cite{Oppenheim:Ion_thermal04,Oppenheim:Large08,Oppenheim:Kinetic13,Oppenheim:Newly20}.

Fully kinetic PIC simulations usually include the entire kinetic physics and,
automatically, all thermal-driven processes. At present time, this approach is
the most powerful tool for modeling the electrojet instabilities, especially
in the highly nonlinear stage of instability-driven plasma turbulence for
which no comprehensive analytical description even exists. PIC simulations,
however, also have limitations: the restricted box sizes, a higher noise
level due to a limited number of particles, often the artificially increased
electron-to-ion mass ratio. Besides, typical multi-processor, highly
parallelized, PIC simulations require significant supercomputer resources and
often a long computational time.

In general, it is unfeasible to fully understand nonlinear processes without proper
understanding of the linear theory. Also, PIC simulations cannot provide
parameter dependencies and proper scaling. To obtain those, one needs
analytical theory. A primary benefit of linear theory is to reveal the ionospheric 
conditions where radars will observe plasma irregularities. However, 
the main goal of the linear kinetic theory developed in
this paper is to help interpret radar signals, which are usually
detected during the nonlinearly saturated stage of instability. 
Applying a simplified linear theory should help understand whether the
detected radar signals were scattered from linearly stable or linearly
unstable plasma waves. A primary benefit of linear theory is to reveal the 
ionospheric conditions where radars will observe plasma irregularities.
Furthermore, for a given wavevector $\vec{k}$
determined by the radar frequency and the line-of-sight (LoS), the accurate
linear theory will also determine the corresponding linear instability
growth/damping rates and the wave phase velocities. The latter should help
explain the observed Doppler frequency shifts. Of course, the linear theory alone
cannot predict such energy-related wave characteristics as the amplitudes of
the turbulent waves and the turbulence spectra, ion line widths, etc.
Without nonlinear theory or simulations, the linear theory remains incomplete.

\begin{figure}
\centering\includegraphics[width=\textwidth]{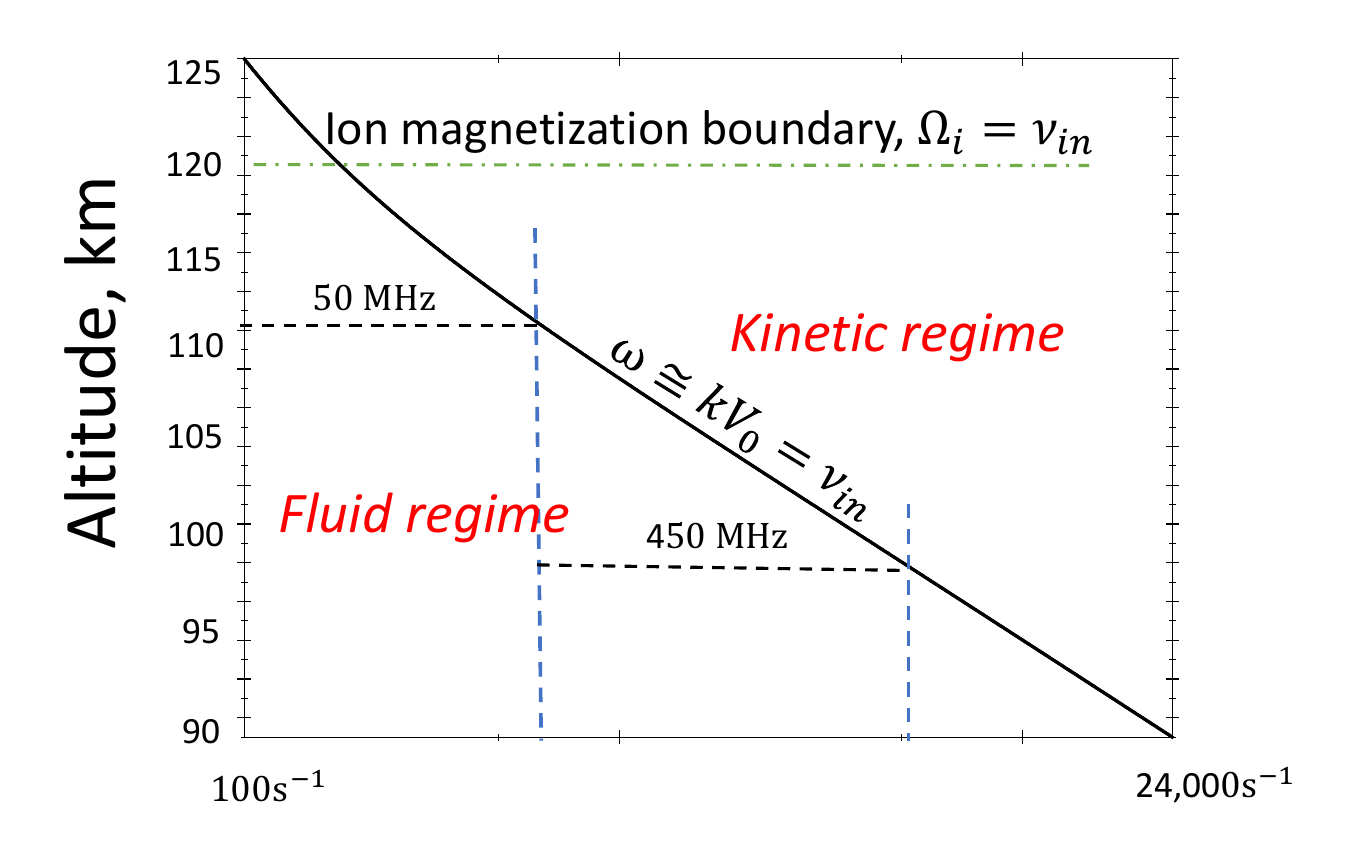}
\caption{Schematic diagram illustrating the areas of applicability for two
different theoretical approaches of the plasma wave description (fluid vs.
kinetic). The solid curve corresponds to $\omega=\nu_{in}$, where the cyclic
wave frequency, $\omega$, is shown on the horizontal axis (in the logarithmic
scale)\ and the \emph{i-n} collision frequency, $\nu_{in}$, is shown
implicitly on the vertical axis using a simplified model of the
altitudinal\ dependence of $\nu_{in}$. The two dashed lines show loosely
defined altitude boundaries for two different radars with the corresponding
frequencies shown above each line. The fluid model is rigorously applicable
only well below the solid curve, while around the curve and above it the more
general kinetic model should be employed. In this paper, we still assume
unmagnetized ions, so that the present kinetic description of ions is
applicable well below the magnetization boundary (shown by the green
dash-dotted horizontal line).}
\label{Figure_1}
\end{figure}

In order to properly interpret the radar signals, one may or may not need the
kinetic theory. This depends upon a specific radar beam frequency and the LoS.
If the detected backscattered wavelength, which equals the half of the
transmitted radar wavelength along the given LoS, falls into the predominantly
collisional long-wavelength range, $|\omega|\ll\nu_{in}$, then one may apply
the much simpler fluid description. However, VHF radars can receive such signals 
only from relatively low altitudes (see below Figure~\ref{Figure_1} and the 
corresponding discussion); from higher altitudes, only HF radars operate at such low 
frequencies, but their waves experience appreciable refraction by the ambient plasma. 
If the detected wavelength is within the weakly collisional, medium-to-short-wavelength range of $|\omega|\gtrsim\nu_{in}$, then
it necessarily requires the more general kinetic theory. 

Furthermore, for typical VHF radars (e.g., 50~MHz, as in the JRO and, approximately, in the
ICEBEAR), which detect $\simeq 3$~m ionospheric irregularities, scattering of the radar
signal from different altitudes may require a different interpretation: one
based on a fluid model if the signal arrives from lower altitudes and another
based on the kinetic theory if the signal arrives from higher altitudes. The
reason is that the collision frequencies, proportional to the neutral
atmosphere density, exponentially decrease with increasing altitude. We may
loosely define an interface between the two regimes as $|\omega|=\nu_{in}(h)$,
where $h$ is the altitude. Given the $\omega(\vec{k})$-dependence [usually,
$\omega(\vec{k})\propto k\equiv|\vec{k}|$, with the proportionality coefficient
varying between the $\vec{E}_{0}\times\vec{B}$ drift speed and the
ion-acoustic speed $C_{s}\simeq[(T_{e}+T_{i})/m_{i}]^{1/2}$, where
$T_{e,i}$ are the electron and ion temperatures and $m_{i}$ is the ion mass,
\citeA<see, e.g.,>{Sahr:Auroral96}], we may relate the wave frequency
$\omega$ to the radar frequency $\Omega_{\mathrm{Radar}}$. Figure~1
illustrates this by plotting the corresponding loosely defined curve
$h(\Omega_{\mathrm{Radar}})$. Given $\Omega_{\mathrm{Radar}}$ and the
corresponding vertical line, we find the corresponding intersection
$h_{0}=h(\Omega_{\mathrm{Radar}})$. For the altitudes well below $h_{0}$, one
may, to a reasonable accuracy, use a fluid model, while for altitudes
$h\gtrsim h_{0}$ one has to employ the more general, and hence much more
accurate results of the kinetic theory, even if they are approximate. Rigorously speaking, 
the fluid model is only applicable at altitudes well below $h_{0}$; 
around this altitude (as well as above it), the accurate description requires kinetic theory.

Earlier works employing the ion kinetics ignored the driving field $\vec{E}_{0}$ and 
hence completely ignored any thermal processes. \citeA{Dimant:Kinetic95a,Dimant:Kinetic95b}
explicitly included $\vec{E}_{0}$ in their kinetic description of electrons
(although for ions they also used an isothermal fluid model). This had allowed those authors to predict the ETI
\cite{Dimant:Kinetic95c,Dimant:Physical97}. However, \citeA{Dimant:Kinetic95a,Dimant:Kinetic95b} kinetic theory has
only been developed for sufficiently long-wavelength waves, for which a fluid
model was also applicable, albeit with less accurate results
\cite{Dimant:Physical97,Robinson:Effects:98,St.-Maurice:Role00,Dimant:Unified23}.

Most probably, the main reason why no previous kinetic studies for ions explicitly included
the driving electric field in their treatment was that including $\vec{E}_{0}$
in the kinetic theory creates significant mathematical difficulties. Indeed, without
$\vec{E}_{0}$, the linear wave perturbations of the particle distribution
function 
relate to the perturbations of the
electrostatic potential, $\Phi_{\omega\vec{k}}$, by a simple algebraic expression.
The resultant relation between the corresponding density perturbations
and $\Phi_{\omega\vec{k}}$ can be expressed in terms of the well-known
plasma dispersion function $Z(x)$ \cite{Fried:Plasma61}. However, if $\vec
{E}_{0}$ is explicitly included in the kinetic equations, then one obtains a much less tractable differential
equation. Under reasonably justified approximations, the ion kinetic equation
leads to a first-order differential equation, while the electron kinetic
equation results in a second-order differential equation. The real
problem, however, is not solving these differential
equations, but rather is the integration of the solution over the
corresponding particle velocities in order to obtain the density
perturbations. For practical applications, it is of paramount importance to
have the final dispersion relation in a form of an explicit functional
expression, rather than in terms of tricky integrals.

In this paper, we have overcome these mathematical difficulties and 
succeeded in including the driving electric field into the kinetic
description of ions in the E-region plasma instabilities. For electrons, the driving 
field was included in \citeA{Dimant:Kinetic95a,Dimant:Kinetic95b},
but in this paper we have shifted that kinetic description into the short-wavelength range. Within the
framework of this fully kinetic approach, we have derived the wave dispersion
relation that describes the combined TFBI. Some derivations (mostly for electrons) have required
approximations, but those are based on the actual physical
conditions in the E-region ionosphere. It is remarkable that the resultant kinetic wave
dispersion relation has a form similar to those in the previous studies: 
this relation includes only elementary functions combined with a few
$Z$-functions of different arguments (for both ions and electrons). 

The main restriction of this paper
(as well as of the earlier ion kinetic studies) is that in this paper we have not included (yet)
the finite ion magnetization. This limits our results to the conditions of $\Omega_{i}\ll\nu_{in}$, 
which are largely valid at altitudes below 110~km  (here $\Omega_{i}$ is
the ion gyrofrequency and $\nu_{in}$ is the ion-neutral collision frequency). 
Usually, this lower altitude range covers most of the
turbulent E region, but some E-region plasma waves exist at higher altitudes as well. 
There are questions that still cannot be answered in the framework of this limited study.
For instance, it is not clear yet how medium-to-short wavelength plasma waves behave near
the ion magnetization boundary, $\Omega_{i}=\nu_{in}$. Above the ion magnetization boundary, where $\nu
_{in}<\Omega_{i}$, in the long-wavelength limit the FBI destabilizing
mechanism, i.e., the ion inertia, becomes, on the contrary, stabilizing, while
the ITI thermal mechanism extends the unstable altitude range up to a few
kilometers over the ion magnetization boundary \cite{Dimant:Ion_thermal04}.
For the VHF and UHF radars, at those altitudes such waves are deeply within
the kinetic range, see Fig. 1, so it is important to understand the
characteristics of the shorter-wavelength waves above the magnetization
boundary. Inclusion of the finite ion magnetization in the collisional kinetic
theory creates additional mathematical difficulties, which we hope will be 
overcome in the future. This would complete the full linear kinetic theory of the E-region plasma
instabilities for the entire E~region.

The theoretical analysis of this paper is quite lengthy and cumbersome, so
that it is restricted only to analytically deriving the kinetic wave
dispersion relation and discussing its limitations. We are planning to perform
a specific analysis of this dispersion relation with application to radar
observations in the near future.

The paper is organized as follows.
Section~\ref{Ion contribution to the wave dispersion relation} derives the
kinetic relation between the linear wave perturbations of the ion particle
density and the electrostatic potential.
Section~\ref{Electron contribution to the wave dispersion relation} derives a
similar kinetic relation between the linear wave perturbations of the electron
particle density and the electrostatic potential (unlike 
\citeA{Dimant:Kinetic95a,Dimant:Kinetic95b}. The treatment of this paper is restricted
to plasma waves with the frequency of the order of, or larger than, the ion-neutral
collision frequency, so that the local collisional cooling can be neglected). 
Section~\ref{TFBI wave dispersion relation} closes the
theory and presents the fully kinetic wave dispersion relation.
Section~\ref{Discussion} discusses the major limitations of our theory.
Section~\ref{Summary and conclusions} concludes the paper. The appendices 
review a simplified analytic model for the background ion distribution function \cite{Koontaweepunya:Non-Maxwellian24} and 
present the most complicated mathematical details of our analytic theory.

\section{Ion Contribution to the Wave Dispersion
Relation\label{Ion contribution to the wave dispersion relation}}

For the ion component of the wave dispersion relation, similarly to the previous ion kinetic 
studies, we will use the kinetic equation with the BGK collision operator:
\begin{linenomath*}
\begin{equation}
\frac{\partial f_{i}}{\partial t}+\frac{q_{i}}{m_{i}}(\vec{E}+\vec
{V}_{i}\times\vec{B})  \cdot\frac{\partial f_{i}}{\partial\vec{V}_{i}
}+\vec{V}_{i}\mathbf{\,\cdot\,}\frac{\partial f_{i}}{\partial\vec{r}}
=-\;\nu_{in}\left(  f_{i}-\frac{n_{i}(\vec{r},t)}{n_{0}}\ \ f_{i0}
^{\mathrm{Coll}}\right).
\label{initial_ion_kinetic_equation}
\end{equation}
\end{linenomath*}
This equation is written in the rest frame of reference for the neutral particles. 
Here $f_{i}(\vec{V}_{i},\vec{r},t)$ is the space ($\vec{r}$) and time ($t$)
dependent ion velocity ($\vec{V}_{i}$) distribution; $\vec{E}$ and $\vec{B}$
are the total electric and magnetic fields, respectively; $m_{i}$ and $q_{i}$
are the mean ion mass and the electric charge, respectively (in the E-region
ionosphere with the vastly predominant singly charged positive ions, $q_{i}=e$, 
where $e$ is the elementary charge); the ion-neutral
collision frequency, $\nu_{in}$, is assumed constant (see below). Further, $n_{i}(\vec{r},t)$
is the time and space-varying ion density, and $f_{i0}^{\mathrm{Coll}}(\vec
{V}_{i})$ is the Maxwellian velocity distribution normalized to the 
spatially uniform and constant background density of electrons and singly
charged ions, $n_{0}=n_{e0}=n_{i0}$,
\begin{linenomath*}
\begin{equation}
f_{i0}^{\mathrm{Coll}}(\vec{V}_{i})  = n_{0}\left(
\frac{m_{i}}{2\pi T_{n}}\right)  ^{3/2}\exp\left(  -\ \frac{m_{i}V_{i}^{2}}{2T_{n}}\right),
\label{ion_cooling _Maxwellian}
\end{equation}
\end{linenomath*}
where $V_{i}=|\vec{V}_{i}|$ and $T_{n}$ is the equilibrium temperature of the
neutral particles (here and across the entire text, all temperatures are given in the energy
units). The two ion distribution functions are normalized according to
\begin{linenomath*}
\begin{equation}
n_{i}(\vec{r},t)=\int\ f_{i}(\vec{V}_{i},\vec{r},t)d^{3}V_{i},\qquad
n_{0}=\int\ f_{i0}^{\mathrm{Coll}}(\vec{V}_{i})d^{3}V_{i}%
.\label{normalization}%
\end{equation}
\end{linenomath*}
The left-hand side (LHS) of equation~(\ref{initial_ion_kinetic_equation})
represents the collisionless part of the ion kinetic equation, while the
right-hand side (RHS) is the collisional term in its simplest BGK form.

The BGK collision term \cite{Bhatnagar:Model54} does not follow from the accurate dual-collision
integro-differential Boltzmann operator, but is rather an artificial construct
convenient for analytical treatment. This model automatically provides thermal
relaxation of arbitrarily unbalanced particles to their thermal equilibrium.
The BGK\ model also conserves the local number of particles and reasonably
well describes the balance of the average momentum and energy. In this paper,
we set the dominant neutral particles (N$_{2}$ and O$_{2}$)\ and the major
ions (O$_{2}^{+}$ and NO$^{+}$) to have approximately the same mass,
$m_{i}\approx m_{n}$. In this approximation, the BGK model even provides
accurate expressions for the mean ion frictional heating and cooling rates
\cite{Dimant:Deriving24}. An important restriction of the BGK\ model is the
assumption of constant, i.e., velocity independent, collision frequency
$\nu_{in}$. For the E-region ions, this approximation called Maxwell 
molecule collisions \cite{Schunk:Ionospheres09} is valid if the ion
energies are well below 1~eV. A more serious problem with the BGK collision 
model, however, is that it includes no angular scattering
of ions through their collisions with the neutral gas particles. In reality,
if the masses of the colliding particles are close to each other, the rates of
the collisional kinetic redistribution of both the particle energies and
velocity angles are comparable. Nevertheless, for the ion-neutral collisions
[unlike the electron-neutral collisions \cite{Dimant:Kinetic95a}, see below],
even the far-from-ideal BGK model is reasonably well applicable and still remains
quite popular among researchers.

In this paper, we restrict our treatment to unmagnetized ions, such that we
disregard the magnetic field $\vec{B}$. In the
Earth's ionosphere, the conditions for such neglect are $\Omega_{i}\ll\nu
_{in}$, where $\Omega_{i}=eB/m_{i}$ is the ion gyrofrequency. These conditions
are reasonably well satisfied at altitudes below $110$~km,
\citeA<see, e.g.,>[ Fig.~2]{Dimant:Ion_thermal04}. Unlike the previous kinetic studies
\cite{Farley:Plasma63,Lee:High-Frequency71,Schmidt:Density73,Ossakow:Parallel75,Schlegel:Short83}, 
we include the background electric field,
$\vec{E}_{0}$, in the ion kinetic equation. This inclusion will automatically add
ion thermal (i.e., heating and cooling) effects for plasma waves, and this is the central
point of our ion-kinetic analysis.

\subsection{The background ion velocity
distribution\label{The background ion velocity distribution}}

A strong driving electric field $\vec{E}_{0}$, perpendicular to the magnetic
field $\vec{B}$, gives rise to plasma instabilities and, simultaneously,
it modifies the background ion distribution function (BIDF), $f_{i0}(\vec{V}_{i})$. If
the driving field is strong, then it heats and shifts the entire ion
velocity distribution $f_{i0}(\vec{V}_{i})$ in such a way that $f_{i0}(\vec
{V}_{i})$ can deviate significantly from $f_{0}^{\mathrm{Coll}}\left(
V_{i}\right)$ \cite{Gurevich:Nonlinear78,Koontaweepunya:Non-Maxwellian24}. The earlier
kinetic studies of the FBI have always assumed $f_{i0}(\vec{V}_{i}%
)=f_{0}^{\mathrm{Coll}}\left(  V_{i}\right)  $. More accurate instability theory requires better
knowledge of the modified background ion distribution function.

Recently, using both an approximate analytic theory and more accurate PIC
simulations, we studied the modification of the BIDF under the action of a strong external electric field
\cite{Koontaweepunya:Non-Maxwellian24}. For the analytic theory, we employed
the above kinetic equation with the BGK collision
term~(\ref{initial_ion_kinetic_equation}), which for the spatially uniform and
stationary background with neglected magnetic field reduces to
\begin{linenomath*}
\begin{equation}
\frac{e}{m_{i}}\ \vec{E}_{0}\cdot\frac{\partial f_{i0}}{\partial\vec{V}_{i}%
}=-\;\nu_{in}\left(  f_{i0}-\ f_{i0}^{\mathrm{Coll}}\right).
\label{zero-order_kinetic_equation}
\end{equation}
\end{linenomath*}
From this equation, it is clear that the imposed electric field $\vec{E}_{0}$
affects the BIDF only in the $\vec{E}_{0}$-direction
(which we will denote the $V_{ix}$ axis), while along the two perpendicular
components, $V_{iy}$ and $V_{iz}$, the velocity distribution contains the same
Maxwellian form as does $f_{i0}^{\mathrm{Coll}}$. This is entirely due to the
fact that the simple BGK collision term includes no angular redistribution of
the colliding ions in the velocity space. 

We briefly review the analytic theory based on
equation~(\ref{zero-order_kinetic_equation}) in
Appendix~\ref{The ion background velocity distribution in the BGK approximation}, 
but for the wave treatment we approximate the actual ion background
distribution function by a simpler bi-Maxwellian distribution
\begin{linenomath*}
\begin{equation}
f_{i0}(\vec{V}_{i})=\frac{1}{\left(  2\pi\right)  ^{3/2}}\frac{n_{0}}{V_{Tx}V_{Ty}^{2}}\exp\left[  -~\frac{(V_{ix}-U_{0})^{2}%
}{2V_{Tx}^{2}}-\frac{V_{iy}^{2}+V_{iz}^{2}}{2V_{Ty}^{2}}\right],
\label{f_i0_Gaussian}
\end{equation}
\end{linenomath*}
where $V_{Tx}=(T_{x}/m_{i})^{1/2}$ and $V_{Ty}=(T_{y}/m_{i})^{1/2}=(T_{z}/m_{i})^{1/2}$, so that
\begin{linenomath*}
\begin{equation}
\int_{-\infty}^{\infty}f_{i0}dV_{iz}=\frac{n_{0}}{2\pi V_{Tx}V_{Ty}}\exp\left[ 
 -~\frac{(V_{ix}-U_{0})^{2}}{2V_{Tx}^{2}}-\frac{V_{iy}%
^{2}}{2V_{Tx}^{2}}\right]  .
\label{Int_f_i0_dV_iz}
\end{equation}
\end{linenomath*}
Here $U_{0}=eE_{0}/(m_{i}\nu_{in})$ is the magnitude of the total Pedersen
drift in the $x$ direction. This approximate distribution function accurately
describes the total background ion Pedersen drift and frictional heating, but
fully ignores any distribution function skewness. The BGK-based analytic model 
predicts this skewness, see \ref{Ion kinetic theory},
Sec.~\ref{The ion background velocity distribution in the BGK approximation}, 
but this model neglects the angular redistribution of ions in the velocity 
space and, as a result, overestimates the BIDF skewness.
We have observed the BIDF skewness in our PIC simulations 
\cite{Koontaweepunya:Non-Maxwellian24}, albeit to a smaller degree than that predicted by the BGK-based model. 
The actual $V_{iz}$-integrated BIDF varies
between those described by equations~(\ref{Int_dz_BGK}) and
(\ref{Int_f_i0_dV_iz}) with no clear preference to the former. Even in spite
of the $f_{i0}$-skewness absence, Eq.~(\ref{Int_f_i0_dV_iz}) is much more
accurate than the conventional assumption of $f_{i0}=f_{i0}^{\mathrm{Coll}}$
(see Sec.~\ref{The background ion velocity distribution}). Note also that for the
highly dynamic harmonic wave perturbations with $\left\vert \omega\right\vert
\gtrsim\nu_{in}$, the effect of the angular scattering should be much less
pronounced than it is for the quasi-stationary background ion distribution.
This means that for the analytic studies of linear wave perturbations, see
Sec.~\ref{Linear wave perturbations: the ion kinetics}, the use of the simple
BGK approximation is justified much better than that for the BIDF.

The quantities $V_{Tx}$ and $V_{Ty}$ are not totally free parameters because 
they are related by the general thermal balance. Indeed, the effective ion temperature for the BIDF given by equation~(\ref{f_i0_Gaussian}) 
is $T_{\mathrm{eff}}=m_i(V_{Tx}^2+2V_{Ty}^2)$. On the other hand, using the general 
ion energy balance equation, see, e.g., equation (1c) in \citeA{Dimant:Deriving24} where 
$s=i$, $m_i=m_n$, $\delta_{in}=1$, $M_{sn}=m_i/2$, and the LHS equals zero, for the same quantity 
we obtain $T_{\mathrm{eff}}=T_{n}+(1/3)m_i U_{0}^{2}$, 
where $U_{0}=eE_0/(m_i\nu_{in})$ is the mean ion drift speed. If we also know the temperature anisotropy,
$T_y/T_x=V_{Ty}^2/V_{Tx}^2$, then we can easily find the explicit value for each parameter $V_{Tx}$ or $V_{Ty}$. 
In this paper, however, we will use approximate equation~(\ref{Int_f_i0_dV_iz}) without
specifying the ratio $V_{Ty}^2/V_{Tx}^2$, because the temperature anisotropy has a still 
unknown dependence on the electric field strength. We leave
such analysis for future studies, but currently we include the temperature anisotropy 
as a free parameter and will investigate in future its possible effect on the linear
instability characteristics, such as the growth/damping rates, wave phase
velocity, and the instability threshold field.

\subsection{Linear wave perturbations: the ion
kinetics\label{Linear wave perturbations: the ion kinetics}}

To find the linear wave dispersion relation, we need to determine the wave
perturbations of the ion and electron densities in terms of a wave perturbation of the corresponding
electrostatic potential and then apply Poisson's equation.
This will allow us to eliminate the electrostatic-potential perturbation and
obtain the unique relation for the complex wave frequency in terms of the real
wavevector. In this section, we deal only with the ion density perturbation. To
find the ion-density wave perturbation, we need to solve the corresponding
linearized kinetic equation for the wave perturbation of the ion distribution
function and integrate this perturbation over velocities. 

Here we derive the kinetic relationship that relates the wave perturbations of
the ion density, $n_{i}$, to the corresponding perturbations of the
electrostatic potential, $\Phi$. In \ref{Ion kinetic theory},
Sec.~\ref{No-field approximation for ions}, we have reproduced the
corresponding no-field ($\vec{E}_{0}=0$) relationship for unmagnetized ions
\cite{Lee:High-Frequency71, Schmidt:Density73} by using two slightly different
approaches, see Sec.~\ref{Useful integrals}. The comparison of the two results have
produced analytic formulas which become crucial for our general
analytic treatment of ions, see equations~(\ref{with_error_functions}) [or
(\ref{Final_integral_w()})] and (\ref{Int(z..)}). The formulas derived provide
exact analytic values for the integral convolutions of Gaussian functions with
the error functions (or the corresponding $Z$-functions), whose complex
arguments are, in general, different and arbitrary.
Popular tables of integrals usually miss such formulas \cite{Dwight:Tables88,Jeffrey:Handbook95,Gradshteyn:Table07} 
or express the results of such integrations
in terms of more general special functions \cite{Prudnikov:Integrals86} with no practical value for our
treatment. 

We start our linear theory from the full kinetic equation for totally unmagnetized
ions with the BGK collision term,
\begin{linenomath*}
\begin{equation}
\frac{\partial f_{i}}{\partial t}+\frac{e}{m_{i}}\ \vec{E}\cdot\frac{\partial
f_{i}}{\partial\vec{V}_{i}}+\vec{V}_{i}\mathbf{\,\cdot\,}
\frac{\partial f_{i}}{\partial\vec{r}}=-\;\nu_{in}\left(  f_{i}-\frac{n_{i}(\vec{r},t)}{n_{0}}
\ f_{i0}^{\mathrm{Coll}}\right).
\label{Initial_kinetic_Review}
\end{equation}
\end{linenomath*}
Using the standard approach, we will derive the relation between small linear
perturbations of the ion density and the electrostatic potential for an
isolated harmonic wave $\propto\exp[i(\vec{k}\cdot\vec{r}-\omega t)]$, where
$\omega=\omega_{r}+i\gamma$ and $\vec{k}$ are the complex wave frequency and
the real wavenumber, respectively. The real wave frequency, $\omega_{r}$, will
always be assumed positive, while the wave growth/damping rate, $\gamma$, can be
either positive (corresponding to the linear instability) or negative (corresponding to a
stable situation when any finite initial wave perturbations will dampen and eventually
disappear). From this point onward, such small harmonic wave perturbations of
any scalar or vector quantity will be denoted by the subscript $\omega\vec{k}$.

For our further linear-wave analysis, it is useful to introduce 
dimensionless expressions for the harmonic wave perturbations of the ion
particle density 
and the electrostatic potential, $\Phi_{\omega\vec{k}}$,
\begin{linenomath*}
\begin{equation}
\phi_{\omega\vec{k}}\equiv\frac{e\Phi_{\omega\vec{k}}}{T_{e}},\qquad
\eta_{\omega\vec{k}}^{i}\equiv\frac{\int f_{\omega\vec{k}}^{i}d^{3}V_{i}%
}{n_{0}},\label{phi,etc_Review}%
\end{equation}
\end{linenomath*}
where the electrostatic wave perturbations are normalized to the electron
temperature, $T_{e}$, for definiteness (the notation $\phi_{\omega\vec{k}}$
will be used for both ions and electrons) 

Now we start deriving the general kinetic relationship between $\phi
_{\omega\vec{k}}$ and $\eta_{\omega\vec{k}}^{i}$ for the arbitrarily large
external DC\ electric field $\vec{E}_{0}$. Linearizing the initial kinetic
equation for ions~(\ref{initial_ion_kinetic_equation}), we obtain
\begin{linenomath*}
\begin{equation}
(\vec{k}\cdot\vec{V}_{i}\mathbf{-}\omega_{i})f_{\omega\vec{k}}^{i}-i\vec{a}%
_{i0}\cdot\frac{\partial f_{\omega\vec{k}}^{i}}{\partial\vec{V}_{i}}%
=\frac{V_{Tn}^{2}}{\beta_{T}}\ \phi_{\omega\vec{k}}\,\left(  \vec{k}\cdot
\frac{\partial f_{i0}}{\partial\vec{V}_{i}}\right)  -i\nu_{in}f_{i0}%
^{\mathrm{Coll}}\eta_{\omega\vec{k}}^{i},\label{linearized_Review}%
\end{equation}
\end{linenomath*}
where $\vec{a}_{i0}=e\vec{E}_{0}/m_{i}$, $\beta_{T}=T_{n}/T_{e}$, 
$V_{Tn}^{2}=T_{n}/m_{n}$ ($m_{n}=m_{i}$), $\omega_{i}\equiv\omega+i\nu_{in}$, and the other
notations were introduced by equations~(\ref{ion_cooling _Maxwellian}) and
(\ref{a_0,u,u_T,V_Ti}). Note that, due to the normalization of $\phi
_{\omega\vec{k}}$, in purely ion equation~(\ref{linearized_Review}) there 
is no actual dependence on any electron parameters. There is also no actual
dependence on the neutral temperature $T_{n}$. Instead of $T_{n}$, we might
have used in $\beta_{T}$ any other constant temperature (with the
corresponding change in the definition for the thermal velocity $V_{Tn}$),
e.g., $T_{x}$ or $T_{y}$. We have picked $T_{n}$ because this
temperature equals the equilibrium ion temperature in the total absence of an
external electric field.

Unlike ``classical'' equation~(\ref{kinetic_no_field}) which represents a simple algebraic
relationship between $f_{\omega\vec{k}}^{i}$, $\phi_{\omega\vec{k}}$, and
$\eta_{\omega\vec{k}}^{i}$, equation~(\ref{linearized_Review}) represents a
linear differential equation. Finding its unique solution for $f_{\omega
\vec{k}}^{i}(\vec{V}_i)$ and the corresponding normalized ion density, $\eta_{\omega
\vec{k}}^{i}\propto \int f_{\omega \vec{k}}^{i}d^3V_i$, is the subject 
of this section with most of the mathematical details given
in \ref{Ion kinetic theory}, see Secs.~\ref{Solving for F_1 and evaluating its contribution to eta_i}
and \ref{Solving for F_2 and evaluating its contribution to eta_i}.

In our treatment, since $\vec{a}_{i0}=e\vec{E}_{0}/m_{i}$ has only one directional component, 
we can reduce the complexity by introducing a coordinate system such that the
wavevector $\vec{k}$ has only two components: $k_{x}$ and $k_{y}$, where the
$x$-axis is directed along $\vec{a}_{i0}$. The absence of
the third vector component, $k_{z}=0$, suggests switching from $f_{\omega
\vec{k}}(\vec{V}_{i})$ to a two-dimensional ion distribution function,
\begin{linenomath*}
\begin{equation}
F_{\omega\vec{k}}(V_{ix},V_{iy})\equiv\int_{-\infty}^{\infty}f_{\omega\vec
{k}}^{i}dV_{iz}.\label{F_omega,k_Review}%
\end{equation}
\end{linenomath*}
For this function, integrating equation~(\ref{linearized_Review}) over $V_{iz}$ we reduce it to
\begin{linenomath*}
\begin{align}
&  \left(  k_{x}V_{x}+k_{y}V_{y}\mathbf{-}\omega_{i}\right)  F_{\omega\vec{k}%
}-ia_{i0}\ \frac{\partial F_{\omega\vec{k}}}{\partial V_{ix}}\nonumber\\
&  =\frac{V_{Ti}^{2}}{\beta_{T}}\ \phi_{\omega\vec{k}}\left(  k_{x}%
\ \frac{\partial}{\partial V_{ix}}+k_{y}\ \frac{\partial}{\partial V_{iy}%
}\right)  \int_{-\infty}^{\infty}f_{i0}dV_{iz}-i\nu_{in}\eta_{\omega\vec{k}%
}^{i}\int_{-\infty}^{\infty}f_{0}^{\mathrm{Coll}}dV_{iz}%
,\label{kinetic_with_field}%
\end{align}
\end{linenomath*}
where, in accord with equations~(\ref{ion_cooling _Maxwellian}) and (\ref{f_i0_Gaussian}), we have
\begin{linenomath*}
\begin{subequations}
\label{Int_f_0,F^Coll_over_V_z}
\begin{align}
\int_{-\infty}^{\infty}f_{i0}dV_{iz} &  =\frac{n_{0}}{2\pi V_{\parallel
}V_{Ty}}\exp\left(  -~\frac{(V_{ix}-U_{0})^{2}}{2V_{Tx}^{2}}%
-\frac{V_{iy}^{2}}{2V_{Ty}^{2}}\right)  ,\label{Int_f_0_over_V_z}\\
\int_{-\infty}^{\infty}f_{0}^{\mathrm{Coll}}dV_{iz} &  =\frac{n_{0}}{2\pi
V_{Tn}^{2}}\exp\left(  -~\frac{V_{ix}^{2}+V_{iy}^{2}}{2V_{Tn}^{2}}\right)
.\label{Int_f_0^Coll_over_V_z}%
\end{align}
\end{subequations}
\end{linenomath*}
Evidently, in the absence of the $\vec{E}_{0}$-field, i.e., when $U_{0}=0$ and
all temperatures are equal, $V_{Ty}=V_{Tx}=V_{Tn}$, these two
integrals coincide.

Since equation~(\ref{kinetic_with_field}) is linear, it is convenient to
separate $F_{\omega\vec{k}}$ into two parts: one proportional to $\eta
_{\omega\vec{k}}^{i}$ and the other proportional to $\phi_{\omega\vec{k}}$,
\begin{linenomath*}
\[
F_{\omega\vec{k}}=F_{\omega\vec{k}}^{(1)}(V_{ix},V_{iy})+F_{\omega\vec{k}}^{(2)}(V_{ix},V_{iy}),
\]
\end{linenomath*}
so that each of them satisfies a separate linear differential equation:
\begin{linenomath*}
\begin{subequations}
\label{Eq_forF^1,2}
\begin{align}
&  (k_{x}V_{ix}+k_{y}V_{iy}\mathbf{-}\omega_{i})F_{\omega\vec{k}}^{(1)}%
-ia_{i0}\ \frac{\partial F_{\omega\vec{k}}^{(1)}}{\partial V_{ix}}%
=-i\ \frac{\nu_{in}n_{0}\eta_{\omega\vec{k}}^{i}}{2\pi V_{Tn}^{2}}%
\ \exp\left(  -\ \frac{V_{ix}^{2}+V_{iy}^{2}}{2V_{T}^{2}}\right)
,
\label{Eq_forF^1}\\
&  (k_{x}V_{ix}+k_{y}V_{iy}\mathbf{-}\omega_{i})F_{\omega\vec{k}}^{(2)}%
-ia_{i0}\ \frac{\partial F_{\omega\vec{k}}^{(2)}}{\partial V_{ix}}\nonumber\\
&  =\frac{V_{Tn}^{2}}{\beta_{T}}\ \phi_{\omega\vec{k}}\left(  k_{x}%
\ \frac{\partial}{\partial V_{ix}}+k_{y}\ \frac{\partial}{\partial V_{iy}%
}\right)  \int_{-\infty}^{\infty}f_{i0}dV_{iz},
\label{Eq_forF^2}
\end{align}
\end{subequations}
\end{linenomath*}
where $\int_{-\infty}^{\infty}f_{i0}dV_{iz}$ was given above by
equation~(\ref{Int_f_0_over_V_z}). After finding $F_{\omega\vec{k}}^{(1)}$ and
$F_{\omega\vec{k}}^{(2)}$, we will calculate $\eta_{\omega\vec{k}}^{i}$ as
\begin{linenomath*}
\begin{equation}
\eta_{\omega\vec{k}}^{i}=\frac{1}{n_{0}}\iint_{-\infty}^{\infty}\left[
F_{\omega\vec{k}}^{(1)}(V_{ix},V_{iy})+F_{\omega\vec{k}}^{(2)}(V_{ix},V_{iy})\right]  dV_{ix}dV_{iy}.
\label{relation_Review}
\end{equation}
\end{linenomath*}

Equations~(\ref{Eq_forF^1}) and (\ref{Eq_forF^2}) are
ordinary differential equations for $F_{\omega\vec{k}}^{(1)}$ and $F_{\omega\vec{k}}^{(2)}$ with respect to the
variable $V_{ix}$. Their formal solutions are rather straightforward, but
evaluating the double integrals of these solutions in order to find $\eta
_{\omega\vec{k}}^{i}$ in accord with equation~(\ref{relation_Review}) is much less trivial. The entire procedure of finding
the $\eta_{\omega\vec{k}}^{i}$ is rather involved mathematically, so that
we present the detailed solutions of equations~(\ref{Eq_forF^1}) and
(\ref{Eq_forF^2}), along with the ensuing integrations, in \ref{Ion kinetic theory}, Secs.
\ref{Solving for F_1 and evaluating its contribution to eta_i} and
\ref{Solving for F_2 and evaluating its contribution to eta_i}. Here, we only
give the resultant linear relation between $\eta_{\omega\vec{k}}^{i}$ and
$\phi_{\omega\vec{k}}$.

Combining equations~(\ref{Int^(1)_Final}) and (\ref{Final_F^(2)}) from Appendices
\ref{Solving for F_1 and evaluating its contribution to eta_i} and
\ref{Solving for F_2 and evaluating its contribution to eta_i}, we obtain the full ion contribution to the wave dispersion
relation:
\begin{linenomath*}
\begin{align}
&  \left[  1-\frac{\sqrt{\pi}\ \nu_{in}}{\sqrt{2(k^{2}V_{Tn}^{2}+i\vec{k}%
\cdot\vec{a}_{i0})}}\ w\left(  \frac{\omega_{i}}{\sqrt{2(k^{2}V_{Ty}%
^{2}+i\vec{k}\cdot\vec{a}_{i0})}}\right)  \right]  \eta_{\omega\vec{k}}%
^{i}\nonumber\\
&  =-\ \frac{k^{2}V_{Tn}^{2}\phi_{\omega k}}{\beta_{T}(k_{x}%
^{2}V_{Tx}^{2}+k_{y}^{2}V_{Ty}^{2}+i\vec{k}\cdot\vec{a}_{i0}%
)}\nonumber\\
&  \times\left[  1+\frac{i\sqrt{\pi}(\omega_{i}-\vec{k}\cdot\vec{U}_{0})}%
{\sqrt{2(k_{x}^{2}V_{Tx}^{2}+k_{y}^{2}V_{Ty}^{2}%
+i\vec{k}\cdot\vec{a}_{i0})}}\ w\left(  \frac{\omega_{i}-\vec{k}\cdot\vec{U}%
_{0}}{\sqrt{2(k_{x}^{2}V_{Tx}^{2}+k_{y}^{2}V_{Ty}%
^{2}+i\vec{k}\cdot\vec{a}_{i0})}}\right)  \right] \phi_{\omega k} .
\label{ion_contribution}
\end{align}
\end{linenomath*}
Equation~(\ref{ion_contribution}) is given in terms of the standard error
function $w(x)$ \cite{Abramowitz:Handbook88}. Using the simple
conversion relation
\begin{linenomath*}
\begin{equation}
w(z)=-\,\frac{i}{\sqrt{\pi}}\,Z(z),
\label{w(x)->Z(x)}
\end{equation}
\end{linenomath*}
we can easily rewrite equation~(\ref{ion_contribution}) in terms of the plasma
dispersion function $Z(x)$ \cite{Fried:Plasma61}. 

Equation~(\ref{ion_contribution}) assumes that $\operatorname{Im}\omega_{i}>0$, i.e., that $\gamma>-\nu_{in}$. 
This condition covers all linearly unstable and moderately damping waves. At the same time, a priori 
we cannot rule out the possibility that heavily damped waves corresponding to $\operatorname{Im}\omega_{i}<0$, 
i.e., $\gamma<-\nu_{in}$, may also exist and be a part of an observed turbulence spectrum. Hypothetically, 
such heavily damped waves might occur, e.g., if the instability-driving mechanisms under certain conditions 
would become, on the contrary, stabilizing. For example, for sufficiently long-wavelength waves, 
above the magnetization boundary of $\nu_{in}=\Omega_i$ (about 120~km altitude) the FBI driving 
mechanism becomes stabilizing \cite{Dimant:Ion_thermal04}. The same pertains to the ETI 
for sufficiently short-wavelength waves \cite{Dimant:Unified23}. If $\operatorname{Im}\omega_{i}<0$, 
then in equation~(\ref{ion_contribution}) the $w$-function should be replaced by $-w(-z) = w(z)-2\exp(-z^2)$.

Equation~(\ref{ion_contribution}) shows that the external electric field
$\vec{E}_{0}$ contributes to the linear $\eta_{\omega\vec{k}}^{i}$%
--$\phi_{\omega k}$ relation only through the additional terms
$i\vec{k}\cdot\vec{a}_{i0}$.
Equation~(\ref{ion_contribution}) also generalizes the corresponding
conventional no-field relation by including the mean ion drift, $\vec{U}_{0}$,
and through the elevated anisotropic ion temperatures, $T_{x}%
>T_{y}>T_{n}$, described by the combination $k_{x}^{2}%
V_{Tx}^{2}+k_{y}^{2}V_{Ty}^{2}>k^{2}V_{Tn}^{2}$ (recall that
$V_{Tx,Ty,n}^{2}=T_{x,y,n}/m_{i}$ and throughout the entire paper we set $m_{n}=m_{i}$).

To conclude this section, we emphasize that equation~(\ref{ion_contribution})
is the exact and unique solution of differential
equation~(\ref{kinetic_with_field}), which satisfies the asymptotic boundary
conditions (the ``good'' behavior of the
solution at $V_{ix}\rightarrow\pm\infty$). Obtaining
this solution has required no other approximations except those which led to
equation~(\ref{Initial_kinetic_Review}) itself.

\section{Electron Contribution to the Wave Dispersion
Relation\label{Electron contribution to the wave dispersion relation}}

Under the E-region conditions, the accurate kinetic description of electrons differs 
significantly from that of ions. For electrons, the simple BGK collision operator
is totally inapplicable for three major reasons: (1) unlike ions that collide with
neutrals of essentially the same mass, the mean rate of the electron energy
losses during one collision of a light electron with a heavy neutral particle
is much less than the corresponding rate of the momentum losses; and (2) the
electron-neutral (\emph{e-n}) collision frequency is strongly
velocity-dependent, $\nu_{en}(V_{e})$, where $V_{e}\equiv|\vec{V}_{e}|$ and
$\vec{V}_{e}$ is the velocity of an individual electron, and (3) e-n collisions 
rapidly isotropize the electron velocity distribution. None of these three factors can be properly described by the BGK
collisional model.

Indeed, in the E-region ionosphere, the effective rate of collisional losses
is given by $\delta_{en}\nu_{en}$, where the fraction of (predominantly
inelastic) collisional energy losses is $\delta_{en}\sim(2$--$4)\times10^{-3}$
(while for the corresponding ion fraction we have $\delta_{in}\simeq1$)
\cite{Gurevich:Nonlinear78}. Furthermore, the \emph{e-n} kinetic frequency in
the lower ionosphere is approximately $\nu_{en}\propto V_{e}^{2\alpha}$, where
$\alpha\simeq1$, while the \emph{i-n} collision frequency, $\nu_{in}$, is
almost constant. These issues have been discussed in detail by
\citeA{Dimant:Kinetic95a,Koontaweepunya:Non-Maxwellian24}, so that we will not
dwell on them here.

In this paper, we employ for electrons the kinetic approach originally
employed by \citeA{Gurevich:Nonlinear78} and later developed further by
\citeA{Dimant:Kinetic95a} and \citeA{Dimant:Effects22}. This approach assumes that due to
the strong magnetic field and intense collisional angular scattering in the
velocity space, the electron distribution function (EDF), $f_{e}(\vec{V}_{e}%
)$, is close to an isotropic one, $f_{e}(\vec{V}_{e})\approx f_{e}(V_{e})$,
along with all its wave perturbations. We will assume that the major isotropic
part of $f_{e}(V_{e})$, at least within the thermal bulk of the electron
energy distribution, is reasonably close to a Maxwellian distribution,
\begin{linenomath*}
\begin{equation}
f_{0}(V_{e})\approx n_{e}\left(  \frac{m_{e}}{2\pi T_{e}}\right)  ^{3/2}%
\exp\left(  -\ \frac{m_{e}V_{e}^{2}}{2T_{e}}\right), 
\label{e_Maxwellian}
\end{equation}
\end{linenomath*}
where, under sufficiently strong electric field $\vec{E}_{0}$, the background
electron temperature $T_{e}$ may differ dramatically from the undisturbed neutral
temperature $T_{n}$. Indeed, the space-averaged electron temperature in the
uniformly dense weakly ionized plasma is determined by the mean-energy balance
equation given by
\begin{linenomath*}
\begin{equation}
\frac{3}{2}\ \frac{dT_{e}}{dt}\approx m_{e}\left\langle \nu_{en}\right\rangle
\vec{V}_{e}^{2} -\frac{3}{2}\left\langle \delta
_{en}\nu_{en}\right\rangle \left(  T_{e}-T_{n}\right)  ,
\label{e_energy_balance}
\end{equation}
\end{linenomath*}
where $\vec{V}_{e}$ is the background electron drift velocity and the angular
brackets denote averaging over the entire electron velocity distribution.
Under stationary conditions, assuming only the regular electron frictional
heating, we have $\vec{V}_{e}\approx$ $\vec{V}_{0}$, where $\vec{V}_{0}%
=\vec{E}_{0}\times\vec{B}/B^{2}$, so that equation~(\ref{e_energy_balance})
yields
\begin{linenomath*}
\begin{equation}
T_{e}(E_{0})\approx T_{n}+\frac{2m_{e}\left\langle \nu_{en}\right\rangle
E_{0}^{2}}{3\left\langle \delta_{en}\nu_{en}\right\rangle B^{2}}.
\label{Elevated_e_temperature}
\end{equation}
\end{linenomath*}
For sufficiently strong $\vec{E}_{0}$, the contribution of the electron
frictional heating, i.e., the second term in the RHS of
Eq.~(\ref{Elevated_e_temperature}) can be comparable to the first one.

Equation~(\ref{Elevated_e_temperature}) represents only the minimal electron
temperature in the presence of the DC electric field $\vec{E}_{0}$. During the
nonlinearly saturated stage of the E-region instability driven by this field
there is additional heating caused by turbulent electric fields, especially
those parallel to $\vec{B}$. This effect of anomalous electron heating (AEH)
has been discussed in many papers, e.g.,
\citeA{Foster:Simultaneous00,Dimant:Model03,
Milikh:Model03,St-Maurice:Revisiting21,Liang:Time-dependent21}, and references
therein. For a sufficiently strong driving field, this anomalous
(i.e., turbulent) heating exceeds significantly the normal (i.e., laminar) heating
produced exclusively by $\vec{E}_{0}$. This paper deals with the linear
dispersion relation which disregards such nonlinear processes as mode
coupling, wave-frequency broadening, etc. We can assume, however, that the
short-wavelength waves under study are not strongly affected by such nonlinear
processes, except the spatially averaged, i.e., macroscopic, effect of AEH. We can
include the latter parametrically by replacing $T_{e}$ given by
equation~(\ref{Elevated_e_temperature}) with the much larger actual electron
temperature $T_{e}$ determined empirically or by using reliable heating models
\cite{Dimant:Model03,Milikh:Model03,Hysell:Implications13,St-Maurice:Revisiting21,Liang:Time-dependent21}.

The main implication of equation~(\ref{Elevated_e_temperature}), especially if
we assume the full electron heating that includes AEH, is as follows. Whatever the
driving field $E_{0}$ (reaching sometimes hundreds mV/m), the
drift speed, $V_{0}=E_{0}/B$, always remains small compared to the modified
thermal speed $V_{Te}=(T_{e}/m_{e})^{1/2}$. This takes place because within
the eV range of the electron energies we always have $\left\langle \nu
_{en}\right\rangle /\left\langle \delta_{en}\nu_{en}\right\rangle \gg1$ (see
above). This means that, unlike the unmagnetized ions whose Pedersen drift,
$|\langle \vec{V}_{i}\rangle| \sim
eE_{0}/(m_{i}\nu_{in}B)$, can be comparable to, or even exceed, the electron
bulk thermal speed $V_{Ti}=(T_{i}/m_{i})^{1/2}$, see
Sec.~\ref{The background ion velocity distribution}, electrons usually allow
one to neglect any velocity-space shifts of their background Maxwellian EDF,
including those connected with the neutral wind (the latter is two-to-three orders
of magnitude smaller than the electron thermal speed, even for the originally
cold ionospheric electrons).

For the linear wave analysis, we will start from \citeA{Dimant:Kinetic95a}
equation~(38),
\begin{linenomath*}
\begin{equation}
\left(  i\Delta_{\omega\vec{k}}+\hat{D}_{\omega\vec{k}}\right)  f_{\omega
\vec{k}}(V_{e})=\hat{B}_{\omega\vec{k}}f_{0}(V_{e})\,\phi_{\omega\vec{k}}\,,
\label{i_Delta_electron}%
\end{equation}
\end{linenomath*}
where $f_{0}(V_{e})$ is the background electron distribution function given by
equation~(\ref{e_Maxwellian}) with the electron temperature $T_{e}$ elevated
by the AEH and often much larger than that given by
equation~(\ref{Elevated_e_temperature}); $f_{\omega\vec{k}}(V_{e})$ is the
single-harmonic linear perturbation of the electron distribution function,
$\Delta_{\omega\vec{k}}\equiv\vec{k}\cdot\vec{V}_{e0}-\omega\approx\vec
{k}\cdot\vec{V}_{0}-\omega$, where $\vec{V}_{0}=(\vec{E}_{0}\times\vec
{B})/B^{2}$; and the two differential operators $\hat{D}_{\omega\vec{k}}$ and
$\hat{B}_{\omega\vec{k}}$ are given by
\begin{linenomath*}
\begin{equation}
\hat{D}_{\omega\vec{k}}=\frac{1}{3V_{e}^{2}}\left[  \left(  \vec{k}_{\perp
}V_{e}+i\vec{a}_{e0}\ \frac{d}{dV_{e}}\right)  \frac{\nu_{en}(V_e)V_{e}^{2}}
{\Omega_{e}^{2}}\cdot\left(  \vec{k}_{\perp}V_{e}+i\vec{a}_{e0}\ 
\frac{d}{dV_{e}}\right)  \right]  +\frac{k_{x}^{2}V_{e}^{2}}{3\nu_{en}(V_e)}, 
\label{D_again}
\end{equation}
\end{linenomath*}
where $\vec{a}_{e0}=e\vec{E}_{0}/m_{e}$ and the speed derivatives $d/dV_e$ are 
assumed to apply (from right to left) to the expressions right of them, including all $V_e$-dependent 
multipliers and $f_{\omega\vec{k}}(V_{e})$,
\begin{linenomath*}
\begin{equation}
\hat{B}_{\omega\vec{k}}=-\ \frac{V_{Te}^{2}}{3}\left[  V_{e}\left(
\frac{k_{y}^{2}\nu_{en}(V_{e})}{\Omega_{e}^{2}}+\frac{k_{x}^{2}%
}{\nu_{en}(V_{e})}\right)  \frac{d}{dV_{e}}+\frac{2i\left(  \vec{k}_{\perp
}\cdot\vec{a}_{e0}\right)  }{\Omega_{e}^{2}V_{e}^{2}}\ \frac{d}{dV_{e}}\left(
V_{e}^{2}\nu_{en}(V_{e})\frac{d}{dV_{e}}\right)  \right] \!\! .
\label{B_omega,k}
\end{equation}
\end{linenomath*}
Compared to equation~(38) from \citeA{Dimant:Kinetic95a}, we have neglected here two
kinetic operators, $\widehat{\delta\nu}_{en}$ and $\hat{\nu}_{ee}$, which
describe the local electron collisional cooling and electron-electron
(\emph{e-e}) collisions, respectively. The former is crucial for the ETI
\cite{Dimant:Kinetic95b}, while the latter may play a role under conditions of
sufficiently dense and cold plasma, as discussed in \citeA{Dimant:Kinetic95a}. The ETI exists
almost exclusively in the ``super-long''-wavelength limit of $|\omega| \lesssim \langle\delta_{en}\nu_{en}\rangle \ll\nu_{in}$
\cite{Dimant:Kinetic95c,Dimant:Physical97,Dimant:Unified23}. In this paper,
however, we are mostly interested in the medium-to-short wavelength range of
$\left\vert \omega\right\vert \gtrsim\nu_{in}\gg\left\langle \delta_{en}\nu_{en}\right\rangle$, 
where the slow local \emph{e-n} collisional cooling plays
no significant role. At the same time, the frictional heating 
may still be of importance, so that we should retain the terms $\propto\vec{a}_{e0}$.
Also, since we are interested in strong driving fields that usually lead to
significant electron heating, we can safely neglect the \emph{e-e} collisions,
described by the quadratically-nonlinear operator of Coulomb collisions
$\hat{\nu}_{ee}$, because the efficiency of Coulomb collisions decreases dramatically with
increasing energies of the colliding particles.

\subsection{Non-thermal approximation for
electrons\label{No-field approximation for electrons}}

Before we present the solution for the full set of electron equations, 
we will develop the linear kinetic theory for electrons in the
limit of $\vec{a}_{e0}\rightarrow0$. Unlike the well-known no-field approximation for
ions (reviewed in \ref{Ion kinetic theory}, Sec.~\ref{No-field approximation for ions}), the
results of this subsection have never been published before. We will use the
$\vec{a}_{e0}\rightarrow0$ results for future references, but these 
results may also be sufficient for a moderately strong electric field $\vec{E}_{0}$.

Unlike in the $\vec{a}_{i0}\rightarrow 0$ limit for ions, we will refer 
to the $\vec{a}_{e0}\rightarrow0$ limit as the non-thermal approximation (NTA) for electrons. 
For magnetized electrons, the driving field $\vec{E}_{0}$
has been implicitly included in $\Delta_{\omega\vec{k}}$ through the
mean electron drift velocity, $\vec{V}_{e0}$ (without this drift, no E-region plasma instability would even be possible). 
With this drift, but no the $\vec{a}_{e0}$-related terms explicitly included,  
all possible electron thermal effects are totally ignored.

For the Maxwellian background electron distribution function $f_{0}(V_{e})$, when neglecting all terms
$\propto\vec{a}_{e0}$, the above differential operators $\hat{D}_{\omega
\vec{k}}$ and $\hat{B}_{\omega\vec{k}}$ reduce to a common algebraic function
of the electron speed $V_{e}$,
\begin{linenomath*}
\begin{equation}
\hat{D}_{\omega\vec{k}}\approx\hat{B}_{\omega\vec{k}}\approx S_{\omega\vec{k}%
}\,\equiv\frac{V_{e}^{2}}{3}\left(  \frac{k_{y}^{2}\nu_{en}(V_{e})}%
{\Omega_{e}^{2}}+\frac{k_{x}^{2}}{\nu_{en}(V_{e})}\right)  .
\label{DBS}
\end{equation}
\end{linenomath*}
This reduction is valid for sufficiently short-wavelength waves satisfying the
conditions (expressed here in terms of the ion variables which are more relevant for the TFBI)
\begin{linenomath*}
\begin{align}
\frac{\left(  kV_{Te}\right)  ^{2}\left\langle \nu_{en}\right\rangle
}{3\Omega_{e}^{2}}  &  =\frac{\left(  kV_{Ti}\right)  ^{2}}{\nu_{in}}%
\frac{\left\langle \psi_{\perp}\right\rangle }{3\beta_{T}}\gg\left\langle
\delta_{en}\nu_{en}\right\rangle ,\;\frac{kV_{0}\left\langle \nu
_{en}\right\rangle }{3\Omega_{e}},\label{init}\\
\frac{\left(  kV_{Ti}\right)  ^{2}}{\beta_{Ti}}  &  \gg kV_{0}\Omega
_{i}=ka_{i0}\nonumber
\end{align}
\end{linenomath*}
where $\beta_{Ti}=T_{i}/T_{e}$ and $\left\langle \psi_{\perp}\right\rangle
=\left\langle \nu_{en}\right\rangle \nu_{in}/(\Omega_{e}\Omega_{i})$.
These conditions mean that the electron diffusion described by the LHS of
inequality~(\ref{init}) dominates over the mean electron energy loss rate,
$\left\langle \delta_{en}\nu_{en}\right\rangle $, and Pedersen conductivity,
$kV_{0}\left\langle \nu_{en}\right\rangle /(3\Omega_{e})$ (these order-of-magnitude conditions are true in general, but more accurate
conditions can be established after analyzing the solution of the general case for
arbitrary $\vec{a}_{e0}$). Under these conditions, equations~(\ref{i_Delta_electron}) and (\ref{DBS}) yield a simple
algebraic relation between $f_{\omega\vec{k}}(V_{e})$ and $\phi_{\omega\vec
{k}}$:
\begin{linenomath*}
\[
f_{\omega\vec{k}}(V_{e})=\frac{S_{\omega\vec{k}}}{S_{\omega\vec{k}}
+i\Delta_{\omega\vec{k}}}\ \,f_{0}(V_{e})\,\phi_{\omega\vec{k}}\,.
\]
\end{linenomath*}
Using $\eta_{\omega\vec{k}}^{e}=\langle f_{\omega\vec{k}}^e(V_{e}) \rangle 
= 4\pi\int_{0}^\infty f_{\omega\vec{k}}^e(V_{e})V_e^2dV_e/n_0$, we obtain
\begin{linenomath*}
\begin{equation}
\eta_{\omega\vec{k}}^{e}=A_{e}\phi_{\omega\vec{k}}\,,
\qquad A_{e}=\left\langle\frac{S_{\omega\vec{k}}}{S_{\omega\vec{k}}+i\Delta_{\omega\vec{k}}}\right\rangle, 
\label{elec}%
\end{equation}
\end{linenomath*}
\begin{linenomath*}
or, equivalently,
\begin{equation}
A_{e}\equiv \frac{\eta_{\omega\vec{k}}^{e}}{\phi_{\omega\vec{k}}}=\int_{0}^{\infty}\frac{S_{\omega\vec{k}}}{S_{\omega\vec{k}}
+i\Delta_{\omega\vec{k}}}\,\exp\left(  -\,\frac{m_{e}V_{e}^{2}}{2T_{e}}\right) 
V_{e}^{2}dV_{e}\,\,\left/  \int_{0}^{\infty}\,\exp\left(  -\,\frac
{m_{e}V_{e}^{2}}{2T_{e}}\right)  V_{e}^{2}dV_{e}\right.  . 
\label{A_3}
\end{equation}
\end{linenomath*} 
In the long-wavelength limit of $|\Delta_{\omega\vec{k}}|\gg S_{\omega\vec{k}}$ (roughly corresponding to
$\psi_{\perp}|\omega|\ll\nu_{in}$), this yields an expression
\begin{linenomath*}
\[
\frac{1}{A_{e}}\approx\frac{\langle S_{\omega\vec{k}}^{2}\rangle
}{\langle S_{\omega\vec{k}}\rangle ^{2}}+i\ \frac{\Delta
_{\omega\vec{k}}}{\langle S_{\omega\vec{k}}\rangle },
\]
\end{linenomath*}
which coincides with the RHS of Eq. (60) from \citeA{Dimant:Kinetic95a}.
In the more general case of larger $|\omega|$ and $k$, we have to explicitly
calculate the RHS of equation~(\ref{A_3}). 

The above relations are valid for the general velocity dependence of $\nu_{en}(V_{e})$, but elsewhere in this 
paper, we will assume the approximate power-law dependence of
$\nu_{en}(V_{e})$ employed by \citeA{Dimant:Kinetic95a}:
\begin{linenomath*}
\begin{equation}
\nu_{en}(V_{e})=\nu_{0}\ \frac{V_{e}^{2}}{V_{Te}^{2}},\qquad V_{Te}^{2}
\equiv\frac{T_{e}}{m_{e}}. 
\label{electron_nu_V^2}
\end{equation}
\end{linenomath*}
Notice that equation~(62) from \citeA{Dimant:Kinetic95a} includes an additional factor 2 in 
the denominator of the expression for $\nu_{en}(V_{e})$, so that their notation $\nu_{0}$ equals twice the parameter
$\nu_{0}$ in this paper.
%
Equation~(\ref{electron_nu_V^2}) yields
\begin{linenomath*}
\begin{equation}
\left\langle \nu_{en}(V_{e})  \right\rangle =3\nu
_{0}
\label{<nu_en>}
\end{equation}
\end{linenomath*}
and
\begin{linenomath*}
\begin{equation}
S_{\omega\vec{k}}=\frac{\nu_{0}}{3}\left(  \frac{k_{y}V_{Te}}%
{\Omega_{e}}\right)  ^{2}\left(\frac{V_{e}^{4}}{V_{Te}^{4}} 
+ \frac{k_{x}^{2}}{k_{y}^{2}}\frac{\Omega_{e}^{2}}{\nu_{0}^{2}}\right)  . 
\label{S_omega,k}
\end{equation}
\end{linenomath*}
Temporarily introducing a dimensionless variable $z=V_e/(\sqrt{2}\,V_{Te})$, we obtain
\begin{linenomath*}
\begin{subequations}\label{A_3,K,tau,beta}
\begin{align}
A_{e}  &  =\frac{4}{\sqrt{\pi}}\int_{0}^{\infty}\frac{z^{4}+\tau}
{z^{4}+\tau+i\beta}\,\ z^{2}e^{-z^{2}}dz\equiv1-i\,\frac{\beta}{\sqrt{\pi}
}\,K(q),
\label{A3}\\
\,K(q)  &  =4\int_{0}^{\infty}\frac{\,z^{2}e^{-z^{2}}}{z^{4}-q^{4}
}\,dz\,,\qquad q^{4}=-(\tau+i\beta)  =Qe^{i\Psi},
\label{Qpsi}\\
\tau &  =\frac{k_{x}^{2}}{4k_{y}^{2}}\frac{\Omega_{e}^{2}}{\nu
_{0}^{2}}\,,\qquad\beta=\frac{3\Delta_{\omega\vec{k}}}{4\nu_{0}}\left(
\frac{\Omega_{e}}{k_{y}V_{Te}}\right)  ^{2},
\label{Tau,beta}
\end{align}
\end{subequations}
\end{linenomath*}
where $Q=|\tau+i\beta| $ and $\Psi=\arg\left[-\,(
\tau+i\beta)\right]$. Usually, the real parameters $\tau$ and $\operatorname{Re}\beta$ 
are positive ($\omega_r < \vec{k}\cdot\vec{V}_0$ and hence $\operatorname{Re}\Delta_{\omega\vec{k}}>0$), so that without loss
of generality we may set the complex phase $\Psi=\arctan(\beta/\tau)$
to vary within the domain
\begin{linenomath*}
\begin{equation}
\pi\leq\Psi\leq\frac{3\pi}{2}, 
\label{quad}
\end{equation}
\end{linenomath*}
and the complex angle $\Psi/4$ belongs to the first quadrant. In this
convention, all four independent roots of the 4-th order equation $q^{4}=Qe^{i\Psi}$, see
(\ref{Qpsi}), are given by $q=q_n$,
\begin{linenomath*}
\begin{align}
q_{n}  &  =Q^{1/4}\exp\left[  i\left(  \frac{\Psi}{4}+\frac{n\pi}{2}\right)
\right]  ,\qquad\frac{\pi}{4}\leq\frac{\Psi}{4}\leq\frac{3\pi}{8},
\label{qn}\\
q_{0}  & = -iq_{1}= -\,q_{2}= iq_{3}\,, \qquad n=0,\,1,\,2,\,3. \nonumber
\end{align}
\end{linenomath*}
Using these representations, we can express in equation~(\ref{Qpsi}) the function $K(q)$ as
\begin{linenomath*}
\begin{equation}
K(q)=\sum_{n=0}^{3}\frac{K_{n}}{q_{n}}\,,\qquad\;K_{n}=\int
_{0}^{\infty}\frac{e^{-z^{2}}}{z-q_{n}}\,dz\,. \label{K(q)}
\end{equation}
\end{linenomath*}
In the summation, we consider first the term with $n=2$. By using the above
relations between the roots and making a change of the integration variable,
we can express this term as
\begin{linenomath*}
\[
\frac{K_{2}}{q_{2}}=\frac{1}{q_{2}}\int_{0}^{\infty}\frac{e^{-z^{2}}}{z-q_{2}%
}\,dz=\frac{1}{q_{0}}\int_{-\infty}^{0}\frac{e^{-z^{2}}}{z-q_{0}}\,dz
\]
\end{linenomath*}
and combine it with the $n=0$ term. Similarly, we proceed with the $n=1$ and $n=3$ terms. As a result, we obtain
\begin{linenomath*}
\begin{align*}
K(q)  &  =\frac{1}{q_{0}}\int_{-\infty}^{\infty}\frac{e^{-z^{2}}}{z-q_{0}%
}\,dz+\frac{1}{q_{1}}\int_{-\infty}^{\infty}\frac{e^{-z^{2}}}{z-q_{1}}\,dz\\
&  =\frac{1}{q_{0}}\left(  \int_{-\infty}^{\infty}\frac{e^{-z^{2}}}{z-q_{0}%
}\,dz-i\int_{-\infty}^{\infty}\frac{e^{-z^{2}}}{z-iq_{0}}\,dz\right)
\end{align*}
\end{linenomath*}
Using equations~(\ref{Iw}) and (\ref{Z_I}) from \ref{Ion kinetic theory}, we obtain
\begin{linenomath*}
\begin{equation}
K(q)=\frac{i\pi}{q_{0}}\left[  w(q_{0})-iw(iq_{0})\right]  = \frac
{\sqrt{\pi}}{q_{0}}\left[  Z(q_{0})-iZ(iq_{0})\right]  . 
\label{Kq}
\end{equation}
\end{linenomath*}

Alternatively, the same result could be obtained using an approach which will 
become helpful when obtaining the general solution (see Sec.~\ref{General case}).
We implement this approach by representing from the very beginning the function $K(q)$ in equation~(\ref{Qpsi}) as
\begin{linenomath*}
\begin{align}
K(q)  & =2\left(  \int_{0}^{\infty}\frac{e^{-z^{2}}}{z^{2}-q^{2}}
\,dz\,+\int_{0}^{\infty}\frac{e^{-z^{2}}}{z^{2}+q^{2}}\,dz\right)
\nonumber\\
& =\int_{-\infty}^{\infty}\frac{e^{-z^{2}}}{z^{2}-q^{2}}\,dz\,+\int_{-\infty
}^{\infty}\frac{e^{-z^{2}}}{z^{2}+q^{2}}\,dz.
\label{K_q_alternative}
\end{align}
\end{linenomath*}
Equation~(\ref{K_q_alternative}) is invariant with respect to the choice of the root of the 
quartic equation $q^{4}=-(\tau+i\beta)$, see equation~(\ref{qn}), but we will choose $q=q_{0}$, so that both the
real and imaginary parts of $q$ are positive.
Temporarily introducing auxiliary functions 
\begin{linenomath*}
\begin{equation}
K^{\pm}(\gamma,q)=\int_{-\infty}^{\infty}\frac{e^{-\gamma z^{2}}}{z^{2}\pm
q^{2}}\,dz,
\label{K^+-}
\end{equation}
\end{linenomath*}
where  $\operatorname{Re}\gamma>0$ and the subscripts $\pm$ correspond to the two different signs in the
denominator of the integrand, we express $K(q)$ in
equation~(\ref{K_q_alternative}) as
\begin{linenomath*}
\begin{equation}
K(q)=K^{+}(1,q)+K^{-}(1,q).\label{via_K+-}
\end{equation}
\end{linenomath*}
We can calculate the integrals $K^{\pm}(\gamma,q)$ by differentiating them
with respect to the parameter $\gamma$. This yields ordinary
differential equations for the auxiliary parameter $\gamma$:
\begin{linenomath*}
\begin{equation}
\frac{dK^{+}}{d\gamma}-q^{2}K^{+}=-\sqrt{\frac{\pi}{\gamma}},\qquad
\frac{dK^{-}}{d\gamma}+q^{2}K^{-}=-\sqrt{\frac{\pi}{\gamma}}.\label{K_diffury}%
\end{equation}
\end{linenomath*}
The general solutions of these differential equations are given by
\begin{linenomath*}
\begin{align*}
K^{+}(\gamma,q)  & =\exp\left(  q^{2}\gamma\right)  \left(  C^{+}-\sqrt{\pi
}\int_{0}^{\gamma}\frac{\exp\left(  -q^{2}t\right)  }{\sqrt{t}}\ dt\right)
\\
& =\exp\left(  q^{2}\gamma\right)  \left(  C^{+}-\frac{\pi}{q}%
\operatorname{erf}\left(  \sqrt{\gamma}q\right)  \right)  ,\\
K^{-}(\gamma,q)  & =\exp\left(  -q^{2}\gamma\right)  \left(  C^{-}-\sqrt{\pi
}\int_{0}^{\gamma}\frac{\exp\left(  q^{2}t\right)  }{\sqrt{t}}\ dt\right)  \\
& =\exp\left(  -q^{2}\gamma\right)  \left(  C^{-}+\frac{i\sqrt{\pi}}%
{q}\operatorname{erf}\left(  i\sqrt{\gamma}q\right)  \right)  ,
\end{align*}
\end{linenomath*}
where the constants of integration, $C^{\pm}$, can be found from the boundary conditions
at $\gamma=0$. 
We will determine these boundary conditions by recalling the fact that
due to our choice of the quartic root $q$ both $\operatorname{Re}q$ and
$\operatorname{Im}q$ are positive. Using the residual theorem, we
obtain
\begin{linenomath*}
\begin{align*}
C^{+}  & =K^{+}(0,q)=\int_{-\infty}^{\infty}\frac{dz}{z^{2}+q^{2}}\,=\frac
{\pi}{q},\\
C^{-}  & =K^{-}(0,q)=\int_{-\infty}^{\infty}\frac{dz}{z^{2}-q^{2}}
\,=\frac{i\pi}{q}.
\end{align*}
\end{linenomath*}
Bearing in mind the fact that the error-function $\operatorname{erf}(x)$ is an odd function of its argument $x$, recalling the definitions of
the functions $w$ and $Z$, and setting $\gamma=1$, we finally obtain relations
\begin{linenomath*}
\begin{subequations}
\label{K^+-(1,q)}
\begin{align}
K^{+}(1,q) &  =\int_{-\infty}^{\infty}\frac{e^{-z^{2}}}{z^{2}+q^{2}}
\,dz=\frac{\pi}{q}\ w(iq)=-\,\frac{i\sqrt{\pi}}{q}\ Z(iq),\label{K^+(1,q)}\\
K^{-}(1,q) &  =\int_{-\infty}^{\infty}\frac{e^{-z^{2}}}{z^{2}-q^{2}}
\,dz=\frac{i\pi}{q}\ w(q)=\frac{\sqrt{\pi}}{q}\ Z(q),
\label{K^-(1,q)}
\end{align}
\end{subequations}
\end{linenomath*}
which also lead to equation~(\ref{Kq}). We will use relations similar to equation~(\ref{K^+-(1,q)})
when obtaining the approximate solution for the general case of arbitrarily large $\vec{a}_{e0}$. 

Using equation~(\ref{Kq}), for $A_{3}$ given by equation~(\ref{A3}), we obtain
\begin{linenomath*}
\begin{equation}
A_{e}\equiv\frac{\eta_{\omega\vec{k}}^{e}}{\phi_{\omega\vec{k}}}=1+\frac{\beta\sqrt{\pi}}{q}\,\left[  w(q)-iw(iq)\right]  =1-i\,\frac
{\beta}{q}\left[  Z(q)-iZ(iq)\right]  . 
\label{A33}
\end{equation}
\end{linenomath*}
where we recall again that $q(\tau,\beta) = q_{0}$ is the only root of equation~$q^{4}+\tau+i\beta=0$ 
with both $\operatorname{Re}(q)>0$ and $\operatorname{Im}(q)\geq0$. 

Note that under conditions of the drift approximation, $k\rho_{e}\ll1$, and
collisional regime, $\omega\ll\nu_{en}$, 
the conventional BGK approach for electrons, e.g., \citeA{Lee:High-Frequency71}, yields a relation that
coincides with the fluid one,
\begin{linenomath*}
\[
\frac{1}{A_{e}}=1-i\beta_{T}\,\frac{\nu_{in}(\omega-\vec{k}\cdot
\vec{V}_{0})}{(kv_{Ti})^{2}\psi_\perp} = 1+i\beta_{T}\,\frac{\nu_{in}\Delta
_{\omega\mathbf{k}}}{(kv_{Ti})^{2}\psi_\perp}.
\]
\end{linenomath*}

The discussed non-thermal approximation for electrons, along with the no-field approximation for unmagnetized ions, 
describe the pure Farley-Buneman instability with the improved kinetic description of electrons. 
Recall that unlike the NFA for ions, see Sec.~\ref{No-field approximation for ions}, the electron 
part of the non-thermal wave dispersion relation for relatively short-wavelength 
waves, $|\omega|\gtrsim \nu_{in}$, and resulting in equation~(\ref{A33}), had never been published before.

\subsection{General case\label{General case}}

In the general case, we should solve full equation~(\ref{i_Delta_electron}), where the operators $\hat{D}_{\omega\vec
{k}}$ and $\hat{B}_{\omega\vec{k}}$ have the differential form of
equations~(\ref{D_again}) and (\ref{B_omega,k}) with the external electric field 
explicitly included via $\vec{a}_{e0}=e\vec{E}_0/m_e$. All terms proportional to
$\vec{a}_{e0}$ describe electron thermal effects responsible
for the ETI \cite{Dimant:Kinetic95b,Dimant:Physical97}. According to the fluid analysis of
\citeA{Dimant:Unified23}, in the short-wavelength kinetic range
defined by equation~(\ref{init}) we might expect no ETI driving. However, the thermal
terms $\propto\vec{a}_{e0}$ could result in the opposite effect of additional
wave damping or may merely modify the wave phase velocity. 
Regardless of whether these effects will be significant,
a quantitatively accurate analysis requires that we include all terms proportional to $\vec{a}_{e0}$.

Assuming for the background electron distribution function the same Maxwellian
function given by equation~(\ref{e_Maxwellian}) with the given electron temperature $T_{e}$,
from equation~(\ref{B_omega,k}) we obtain
\begin{linenomath*}
\begin{align}
& \hat{B}_{\omega\vec{k}}f_{0}=\frac{n_{0}}{3\left(  2\pi\right)  ^{3/2}%
V_{Te}^{3}}\exp\left(  -\ \frac{V_{e}^{2}}{2V_{Te}^{2}}\right)  \nonumber\\
& \times\left[  \frac{\nu_{0}k_{y}^{2}V_{e}^{4}}{\Omega_{e}^{2}V_{Te}^{2}%
}+\frac{k_{x}^{2}V_{Te}^{2}}{\nu_{0}}+\frac{2i\left(  \vec{k}_{\perp
}\cdot\vec{a}_{e0}\right)  \nu_{0}V_{e}^{2}\left(  5V_{Te}^{2}-V_{e}^{2}\right)}{\Omega_{e}^{2}V_{Te}^{4}}\right]  
\label{Bf_0}
\end{align}
\end{linenomath*}
For the further analysis, it is convenient to introduce a dimensionless variable
\begin{linenomath*}
\begin{equation}
\xi\equiv z^2=\frac{V_{e}^{2}}{2V_{Te}^{2}}. 
\label{xi}
\end{equation}
\end{linenomath*}
In terms of this variable, equation~(\ref{i_Delta_electron}) can be recast as
\begin{linenomath*}
\begin{align}
&  \frac{d^{2}f_{\omega\vec{k}}^{e}}{d\xi^{2}}-\left(  \chi-\frac{5}{2\xi
}\right)  \frac{df_{\omega\vec{k}}^{e}}{d\xi}-\left(  C+\frac{5\chi}{4\xi
}+\frac{\mathcal{P}}{\xi^{2}}\right)  f_{\omega\vec{k}}^{e}\nonumber\\
&  =-\left(  N_{0}+\frac{N_{1}}{\xi}+\frac{N_{2}}{\xi^{2}}\right)  e^{-\xi
}\phi_{\omega\vec{k}} 
\label{full_equation_xi_new_1}
\end{align}
\end{linenomath*}
with the parameters given by
\begin{linenomath*}
\begin{align}
\chi &  =\frac{2ik_{y}V_{Te}^{2}\cos\varphi}{a_{e0}},\qquad\mathcal{P}%
=\frac{3\Omega_{e}^{2}V_{Te}^{2}}{4\nu_{0}a_{e0}^{2}}\left(  i\Delta
_{\omega\vec{k}}+\frac{k_{x}^{2}V_{Te}^{2}}{3\nu_{0}}\right)
,\nonumber\\
C  &  =\frac{k_{y}^{2}V_{Te}^{4}}{a_{e0}^{2}},\qquad N_{0}=\frac{\left(
k_{y}^{2}V_{Te}^{2}-2ik_{y}a_{e0}\cos\varphi\right)  n_{0}}{\left(
2\pi\right)  ^{3/2}a_{e0}^{2}V_{Te}},\nonumber\\
N_{1}  &  =\frac{5ik_{y}n_{0}\cos\varphi}{\left(  2\pi\right)
^{3/2}a_{e0}V_{Te}},\qquad N_{2}=\frac{k_{x}^{2}V_{Te}^{2}\Omega
_{e}^{2}n_{0}}{4\left(  2\pi\right)  ^{3/2}a_{e0}^{2}\nu_{0}^{2}V_{Te}}.
\label{pPCxiABD_1}
\end{align}
\end{linenomath*}
Here $\varphi$ is the angle between $\vec{k}_{\perp}$ and 
$\vec{a}_{e0}$: $\vec{k}_{\perp}\cdot\vec{a}_{e0}=k_{y}a_{e0}\cos\varphi$.

Before proceeding with solving equation~(\ref{full_equation_xi_new_1}) in the
general case, we note that the NTA limit (corresponding to $a_{e0}\rightarrow 0$, 
see Sec.~\ref{No-field approximation for electrons})
can be obtained from equation~(\ref{full_equation_xi_new_1}) by
neglecting both $\xi$ derivatives with respect to $\xi$ (compared to the remaining terms) 
and all terms whose $a_{e0}\rightarrow 0$
singularity is lower then $1/a_{e0}^{2}$ (in particular, by neglecting $\chi$ and
$N_{1}$). As a result, all factors $1/a_{e0}^2$ in the LHS and RHS of 
equation~(\ref{full_equation_xi_new_1}) can be mutually canceled out, 
so that the NTA solution for $f_{\omega\vec{k}}^{e}(V_{e})$ becomes
\begin{linenomath*}
\begin{align}
\lim_{a_{e0}\rightarrow0}f_{\omega\vec{k}}^{e}(V_{e}) &  =\frac{\left(
N_{0}+N_{2}/\xi^{2}\right)  e^{-\xi}\phi_{\omega\vec{k}}}{C+\mathcal{P}%
/\xi^{2}}\nonumber\\
&  =\frac{n_{0}\left(  4\nu_{0}k_{y}^{2}V_{Te}^{2}\xi^{2}/\Omega_{e}%
^{2}+k_{x}^{2}V_{Te}^{2}/\nu_{0}\right)  e^{-\xi}}{\left(
2\pi\right)  ^{3/2}V_{Te}^{3}\left(  4\nu_{0}k_{y}^{2}V_{Te}^{2}\xi
^{2}/\Omega_{e}^{2}+k_{x}^{2}V_{Te}^{2}/\nu_{0}+3i\Delta_{\omega
\vec{k}}\right)  }\ \phi_{\omega\vec{k}}.
\label{classical_1}
\end{align}
\end{linenomath*}
This solution agrees with the NTA solution for $f_{\omega\vec{k}}^{e}(V_{e})$ that led to equation~(\ref{A3}).
As mentioned above, equation~(\ref{classical_1}) with no thermal mechanisms explicitly included can be a 
reasonably good approximation for a moderately strong driving electric field.
We will refer to the NTA limit when discussing the general solution for electrons. 

Generally, for electrons (as well as for the ions), instead of a simple
algebraic relation between $f_{\omega\vec{k}}^{e}$ and $\phi_{\omega\vec{k}}$,
we have much less tractable differential equation~(\ref{full_equation_xi_new_1}). This equation is a linear ordinary
differential equation with respect to the variable $\xi$. Unlike the
first-order equation~(\ref{kinetic_with_field}) for ions,
equation~(\ref{full_equation_xi_new_1}) for electrons is a second-order differential
equation. Such equations with non-constant coefficients can be
solved exactly only on rare occasions. Besides, for electrons, similarly to ions, of a special
difficulty becomes the integration of the found solution (with the relevant weight 
function) in order to determine the wave perturbations of the electron density.

We start by obtaining the exact solution of equation~(\ref{full_equation_xi_new_1}) in terms of integrals of special 
functions. This exact solution is inconvenient for practical applications, so that later we will employ
reasonably accurate approximations based on relations between the actual
physical parameters.

The LHS of second-order ODEs like equation~(\ref{full_equation_xi_new_1}), by
using the change of the variable $\xi$ to a new one, $\xi=\lambda x$, and the substitution
\begin{linenomath*}
\begin{equation}
f_{\omega\vec{k}}^{e}(\xi)=e^{\alpha x}x^{\gamma}\bar{w}(x), \qquad x=\frac{\xi}{\lambda},
\label{y(x)_via_w}
\end{equation}
\end{linenomath*}
can be recast in standard Whittaker's form \cite{Abramowitz:Handbook88}:
\begin{linenomath*}
\begin{equation}
\text{LHS}\propto\frac{d^{2}\bar{w}}{dx^{2}}+\left(  -\ \frac{1}{4}%
+\frac{\kappa}{x}+\frac{1/4-\mu^{2}}{x^{2}}\right)  \bar{w}.
\label{Whittaker}
\end{equation}
\end{linenomath*}
Specifically for equation~(\ref{full_equation_xi_new_1}), this procedure 
requires the following values of the parameters $\gamma$, $\lambda$, and $\alpha$:
\begin{linenomath*}
\begin{equation}
\gamma=-\ \frac{5}{4},\qquad\lambda=\frac{a_{e0}}{2k_{y}V_{Te}^{2}
\sin\varphi},\qquad\alpha=\frac{\chi\lambda}{2}=\frac{i\cot\varphi}{2}.
\label{gamma,alpha,kappa}
\end{equation}
\end{linenomath*}
This automatically yields $\kappa=0$, while the parameter $\mu$ is
given by
\begin{linenomath*}
\begin{equation}
\mu=\sqrt{\mathcal{P}+\frac{9}{16}}=\left[  \frac{3\Omega_{e}^{2}V_{Te}^{2}%
}{4\nu_{0}a_{e0}^{2}}\left(  i\Delta_{\omega\vec{k}}+\frac{k_{x}%
^{2}V_{Te}^{2}}{3\nu_{0}}\right)  +\frac{9}{16}\right]  ^{1/2}.
\label{mu==}
\end{equation}
\end{linenomath*}
Here (as well as anywhere else), we choose the branch of the square root with the positive real part.

The fact that in our case Whittaker's parameter $\kappa$ automatically becomes
zero makes the recast equation~(\ref{Whittaker}) simpler than general Whittaker's
equation. In terms of the new variable $x=\xi/\lambda$, the full recast differential equation is given by
\begin{linenomath*}
\begin{subequations}
\label{w,L}%
\begin{align}
&  \frac{d^{2}\bar{w}}{dx^{2}}+\left(  -\ \frac{1}{4}+\frac{1/4-\mu^{2}%
}{x^{2}}\right)  \bar{w}=L\left(  x\right)  ,\label{w_equation_1}\\
&  L\left(  x\right)  =-\lambda^{2}e^{-\left(  \alpha+\lambda\right)
x}x^{\frac{5}{4}}\left(  N_{0}+\frac{N_{1}}{\lambda x}+\frac{N_{2}}%
{\lambda^{2}x^{2}}\right)  \phi_{\omega,\kappa}, 
\label{L(x)}
\end{align}
\end{subequations}
\end{linenomath*}
where the parameters $N_{0,1,2}$ are defined by equation~(\ref{pPCxiABD_1}).

We start solving equation~(\ref{w_equation_1}) by considering 
first the homogeneous equation, i.e., this equation without the RHS:
\begin{linenomath*}
\begin{equation}
\frac{d^{2}\bar{w}}{dx^{2}}+\left(  -\ \frac{1}{4}+\frac{1/4-\mu^{2}}{x^{2}%
}\right)  \bar{w}=0. 
\label{homogeneous}
\end{equation}
\end{linenomath*}
Equation~(\ref{homogeneous}) has exact analytic solutions in terms of special functions. 
As the two linearly independent solutions of this equation, we can choose 
$\sqrt{x}\ \mathrm{I}_{\mu}(x/2)$ and $\sqrt{x}\ \mathrm{K}_{\mu}(x/2)$, where
$\mathrm{I}_{\mu}\left(  z\right)  $ and $\mathrm{K}_{\mu}\left(  z\right)  $
are the modified Bessel functions of the first and second kind, respectively
\cite{Bateman_Erdelyi:Higher3,Abramowitz:Handbook88,Arfken:Mathematical}.

The modified Bessel functions of the integer or half-integer order
\cite{Abramowitz:Handbook88} arise in a significant number of scientific and
engineering problems, but such functions with the general complex order $\mu$ are used less frequently. 
For these functions, we will use relevant relations that can be found in
\citeA{Bateman_Erdelyi:Higher3,Arfken:Mathematical}, or can be easily deduced from some
expressions given there. Since the arguments of the modified Bessel functions in our
case are $x/2$, in order to avoid confusion, we will give all relations for
the functions of this argument, rather than of $x$.

First, we note that the function $\mathrm{K}_{\mu}(x/2)$ can be expressed in
terms of $\mathrm{I}_{\mu}(x/2)$ as
\begin{linenomath*}
\begin{equation}
\mathrm{K}_{\mu}\left(  \frac{x}{2}\right)  =\frac{\pi}{2}\ \frac
{\mathrm{I}_{-\mu}\left(  x/2\right)  -\mathrm{I}_{\mu}\left(  x/2\right)
}{\sin\left(  \mu\pi\right)  }, 
\label{Bessel_K(x/2)}
\end{equation}
\end{linenomath*}
Further, the function $\mathrm{I}_{\mu}(x/2)$ has the following exact series
expansion convergent for arbitrarily large arguments,
\begin{linenomath*}
\begin{equation}
\mathrm{I}_{\mu}\left(  \frac{x}{2}\right)  =\sum_{s=0}^{\infty}\frac
{1}{s!\Gamma(s+\mu+1)}\left(  \frac{x}{4}\right)  ^{\mu+2s},
\label{Bessel_I_mu=Sum}
\end{equation}
\end{linenomath*}
where $\Gamma(x)$ is the standard gamma function. Using
equations~(\ref{Bessel_K(x/2)}) and (\ref{Bessel_I_mu=Sum}), we can write a
similar exact representation for $\mathrm{K}_{\mu}$,
\begin{linenomath*}
\begin{align}
&  \mathrm{K}_{\mu}\left(  \frac{x}{2}\right)  =\frac{\pi}{2\sin\left(  \mu
\pi\right)  }\nonumber\\
&  \times\left[  \sum_{s=0}^{\infty}\frac{1}{s!\Gamma(s-\mu+1)}\left(
\frac{x}{4}\right)  ^{-\mu+2s}-\sum_{s=0}^{\infty}\frac{1}{s!\Gamma(s+\mu
+1)}\left(  \frac{x}{4}\right)  ^{\mu+2s}\right]  . 
\label{Bessel_K_mu=Sum}
\end{align}
\end{linenomath*}
For large $|x|$, according to \citeA{Arfken:Mathematical}, equations~(14.134--14.143), we have
\begin{linenomath*}
\begin{subequations}
\label{IK_asymptotics}
\begin{align}
\sqrt{x}\ \mathrm{I}_{\mu}\left(  \frac{x}{2}\right)   &  =\frac{e^{x/2}%
}{\sqrt{2\pi}}\left[  P_{\mu}\left(  \frac{ix}{2}\right)  -iQ_{\mu}\left(
\frac{ix}{2}\right)  \right]  ,\label{I_asymptotics}\\
\sqrt{x}\ \mathrm{K}_{\mu}(\frac{x}{2})  &  =\sqrt{\frac{\pi}{2}}%
\ e^{-x/2}\left[  P_{\mu}\left(  \frac{ix}{2}\right)  +iQ_{\mu}\left(
\frac{ix}{2}\right)  \right]  , 
\label{K_asymptotics}
\end{align}
\end{subequations}
\end{linenomath*}
where $P_{\mu}$ and $Q_{\mu}$ are non-convergent asymptotic series
\begin{linenomath*}
\begin{subequations}
\label{PQ_mu(z)}
\begin{align}
P_{\mu}(z)  &  \approx1-\frac{\left(  4\mu^{2}-1\right)  \left(  4\mu
^{2}-9\right)  }{2!\left(  8z\right)  ^{2}}\nonumber\\
&  +~\frac{\left(  4\mu^{2}-1\right)  \left(  4\mu^{2}-9\right)  \left(
4\mu^{2}-25\right)  \left(  4\mu^{2}-49\right)  }{4!\left(  8z\right)  ^{4}
}+\ldots,\label{P_mu(z)}\\
Q_{\mu}(z)  &  \approx\frac{4\mu^{2}-1}{1!\left(  8z\right)  }-\frac{\left(
4\mu^{2}-1\right)  \left(  4\mu^{2}-9\right)  \left(  4\mu^{2}-25\right)
}{3!\left(  8z\right)  ^{3}}+\ldots
\label{Q_mu(z)}
\end{align}
\end{subequations}
\end{linenomath*}
(be cautioned that our notation $\mu$ corresponds to
\citeA{Arfken:Mathematical} notation $\eta$, while \citeA{Arfken:Mathematical}
notation $\mu$ corresponds to our $4\mu^{2}$).

Note that $\sqrt{x}\ \mathrm{I}_{\mu}$ and $\sqrt{x}\ \mathrm{I}_{-\mu}$ for
the non-integer order $\mu$ could also form a set of two linearly independent
solutions of equation~(\ref{homogeneous}). However, picking them as the two independent solutions would be
impractical for the following reason. Since $\mu^{2}=(-\mu)^{2}$, equation~(\ref{P_mu(z)}) yields
\begin{linenomath*}
\begin{align*}
\sqrt{x}\ \mathrm{I}_{-\mu}\left(  \frac{x}{2}\right)   &  \approx\frac{e^{\frac
{x}{2}}}{\sqrt{\pi}}\left(  P_{-\mu}\left(  \frac{ix}{2}\right)  -iQ_{-\mu
}\left(  \frac{ix}{2}\right)  \right) \\
&  =\frac{e^{\frac{x}{2}}}{\sqrt{\pi}}\left(  P_{\mu}\left(  \frac{ix}%
{2}\right)  -iQ_{\mu}\left(  \frac{ix}{2}\right)  \right) \approx\sqrt
{x}\ \mathrm{I}_{\mu}\left(  \frac{x}{2}\right)  ,
\end{align*}
\end{linenomath*}
so that for large $x$ the functions $\sqrt{x}\ \mathrm{I}_{\mu}$ and $\sqrt
{x}\ \mathrm{I}_{-\mu}$ merge asymptotically into the same exponentially growing function and hence
are unsuitable to serve as the two linearly independent solutions. On the other hand, the functions $\sqrt{x}\ \mathrm{I}_{\mu}$
and $\sqrt{x}\ \mathrm{K}_{\mu}$ have a very distinct asymptotic behavior at both ends of the domain $0\leq x < \infty$. 
As a result, the choice of $\sqrt{x}\ \mathrm{I}_{\mu}$
and $\sqrt{x}\ \mathrm{K}_{\mu}$ as the two linearly independent solutions
will allow us to obtain the unique solution of full
equation~(\ref{w_equation_1}) more easily than
if we chose the functions $\sqrt{x}\ \mathrm{I}_{\mu}$ and $\sqrt
{x}\ \mathrm{I}_{-\mu}$ as the two independent solutions.

Furthermore, the exact Wronskian of the two linearly independent solutions
$\mathrm{I}_{\mu}$ and $\mathrm{K}_{\mu}$ is given by
\begin{linenomath*}
\begin{equation}
W\equiv\sqrt{x}\ \mathrm{K}_{\mu}\left(  \frac{x}{2}\right)  \frac{d}{dx}\left[
\sqrt{x}\ \mathrm{I}_{\mu}\left(  \frac{x}{2}\right)  \right]  -\sqrt
{x}\ \mathrm{I}_{\mu}\left(  \frac{x}{2}\right)  \frac{d}{dx}\left[  \sqrt
{x}\ \mathrm{K}_{\mu}\left(  \frac{x}{2}\right)  \right]  =1.
\label{Wronskian}
\end{equation}
\end{linenomath*}
Below, we will use this relation for finding the unique solution of full
equation~(\ref{w_equation_1}). Also, combining equation~(\ref{Wronskian}) with
standard properties of the 2nd-order differential equations
\cite{Abramowitz:Handbook88,Arfken:Mathematical}, one can easily find two
exact integral relations that express these linearly independent solutions in terms of each other:
\begin{linenomath*}
\begin{subequations}
\label{Bessel_I,K_via_K,I}
\begin{align}
\sqrt{x}\ \mathrm{K}_{\mu}\left(  \frac{x}{2}\right)   &  =\sqrt
{x}\ \mathrm{I}_{\mu}\left(  \frac{x}{2}\right)  \int_{x}^{\infty}\frac
{dt}{\left[  \sqrt{t}\ \mathrm{I}_{\mu}\left(  t/2\right)  \right]  ^{2}%
},\label{Bessel_K_via_I}\\
\sqrt{x}\ \mathrm{I}_{\mu}\left(  \frac{x}{2}\right)   &  =\sqrt
{x}\ \mathrm{K}_{\mu}\left(  \frac{x}{2}\right)  \int_{0}^{x}\frac{dt}{\left[
\sqrt{t}\ \mathrm{K}_{\mu}\left(  t/2\right)  \right]  ^{2}}\!.
\label{Bessel_I_via_K}
\end{align}
\end{subequations}
\end{linenomath*}
These relations can be used for additional testing the accuracy of 
some approximations implemented in \ref{Simplification of the general solution for electrons}.

Now we proceed to solving equation~(\ref{w_equation_1}) with the parameters
given by equations~(\ref{pPCxiABD_1}) and (\ref{gamma,alpha,kappa}). According
to equations (\ref{Bessel_I_mu=Sum})--(\ref{IK_asymptotics}), the function $\mathrm{I}_{\mu}\left(  x/2\right)  $
behaves regularly at $x\rightarrow0$, but exponentially diverges as
$x\rightarrow\infty$. On the contrary, the function $\mathrm{K}_{\mu}\left(
x/2\right)  $ diverges at $x\rightarrow0$, but exponentially tends to zero as
$x\rightarrow\infty$. Using these properties, along with equation~(\ref{Wronskian}) and standard methods of solving
2nd-order inhomogeneous equations, we arrive at the following exact solution of equation~(\ref{w_equation_1}):
\begin{linenomath*}
\begin{equation}
\bar{w}(x)=-\sqrt{x}\left[  \mathrm{K}_{\mu}\left(  \frac{x}{2}\right)
\int_{0}^{x}\sqrt{z}\ \mathrm{I}_{\mu}\left(  \frac{z}{2}\right)
L(z)dz+\mathrm{I}_{\mu}\left(  \frac{x}{2}\right)  \int_{x}^{\infty}\sqrt
{z}\ \mathrm{K}_{\mu}\left(  \frac{z}{2}\right)  L(z)dz\right]  .
\label{w(x)_L_unique}
\end{equation}
\end{linenomath*}
This exact solution is unique because it is the only solution which behaves regularly within the entire domain
of $0\leq x<\infty$. In addition, the function $\bar{w}(x)\rightarrow0$ as
$x\rightarrow\infty$. Equation~(\ref{w(x)_L_unique}) is our last exact expression. 
From this point on, we will start making approximations.

According to the explicit expression for the function $L$ defined by
equation~(\ref{L(x)}), equation~(\ref{w(x)_L_unique}) involves convolutions of
the modified Bessel functions with power-law and exponential functions.
We cannot evaluate such integrals exactly, so that it is hard to avoid approximations. 
Using the fact that the parameter $|\mu|$ is sufficiently 
large (as explained in the following paragraph), we will construct reasonably accurate
approximations to the special functions $\mathrm{I}_{\mu}$ and $\mathrm{K}_{\mu}$ in
terms of elementary functions. Then, within the framework of the same
accuracy, we will perform all necessary integrations in the RHS of
equation~(\ref{w(x)_L_unique}).

Before proceeding, we will make simple estimates
regarding the crucial parameter $\mu$ given by equation~(\ref{mu==}). Our goal
is to demonstrate that under real physical conditions in the E-region
ionosphere, and for the wave frequency band of interest to us, the absolute value of $\mu$ is always large, regardless of the
driving field magnitude. This fact is crucial for all approximations that will follow.

The main contribution to $\mu$ is given by the quantity
\begin{linenomath*}
\begin{equation}
\Theta\equiv\frac{3\Omega_{e}^{2}V_{Te}^{2}}{4\nu_{0}a_{e0}^{2}}
|\Delta_{\omega\vec{k}}| . 
\label{kappa}
\end{equation}
\end{linenomath*}
By assuming sufficiently short-wavelength waves which are most relevant to the implied kinetic regime,
\begin{linenomath*}
\begin{equation}
|\Delta_{\omega\vec{k}}| =|\omega-\vec{k}\cdot\vec{V}_{e0}| \gtrsim\nu_{in}, 
\label{Delta>nu_in}
\end{equation}
\end{linenomath*}
we consider first the lowest driving field of interest, $E_{0}=m_{e}a_{e0}/e$,
corresponding to the minimum FBI threshold, $E_{0}=BC_{s}$, $C_{s}\sim V_{Ti}$ (for the high-latitude electrojet, typically
$E_{0}\simeq20~$mV/m). This lowest field results in the maximum value of
$\Theta=\Theta_{\max}$. Using equations~(\ref{<nu_en>}) and (\ref{Delta>nu_in}), we
estimate $\Theta_{\max}$ as
\begin{linenomath*}
\begin{equation}
\Theta_{\max}\sim\frac{\Omega_{e}^{2}V_{Te}^{2} 
|\Delta_{\omega\vec{k}}|}{\langle \nu_{en}\rangle \Omega_{e}^{2}V_{Tn}^{2}
}\sim\frac{m_{i}|\Delta_{\omega\vec{k}}|}{m_{e}
\langle \nu_{en}\rangle }\gtrsim\frac{m_{i}\nu_{in}}{m_{e}
\langle \nu_{en}\rangle }\simeq5000. 
\label{K_max}
\end{equation}
\end{linenomath*}
Since the lowest field of interest usually modifies the E-region plasma only weakly, 
here we have used a simple relation $\langle \nu_{en}\rangle
\simeq 10\,\nu_{in}$, which for the undisturbed conditions is approximately valid across the entire lower part
of the E-region ionosphere \cite{Gurevich:Nonlinear78,Kelley:Ionosphere09}.
Thus, we obtain that the maximum value of $|\mu| $ is quite
large, $|\mu|_{\max}\sim\sqrt{\Theta_{\max}}\gtrsim70$.

The parameter $\Theta$ decreases with increasing $a_{e0}$, so that for sufficiently
large $a_{e0}$ it might in principle become so small that $|\mu| $ 
would approach its absolute minimum value of $3/4$, see equation~(\ref{mu==}). However,
this is never the case because as $E_{0}$ increases, $V_{Te}$ also increases, largely in 
proportion with $E_{0}$. Thus, even for the infinitely large electric field 
corresponding to $a_{e0}\rightarrow\infty$, the parameter $|\mu|$ should saturate
at a sufficiently large value. Indeed, for very large $a_{e0}$, according to
(\ref{Elevated_e_temperature}) where we assume $\Delta T_{e}\approx T_{e}\gg
T_{e0}$, we obtain $V_{Te}^{2}\simeq2a_{e0}^{2}/(3\langle \delta
_{en}\rangle \Omega_{e}^{2})$, where $\langle \delta_{en}\rangle \simeq\langle \delta_{en}\nu_{en}\rangle
/\langle \nu_{en}\rangle $. As a result, assuming $|\Delta_{\omega\vec{k}}| \gtrsim\nu_{in}$, we obtain for the
corresponding saturated value of $\Theta$, $\Theta_{\min}\gtrsim\nu_{in}/(2\delta
_{en}\nu_{0})$. Using $\langle \delta_{en}\rangle \sim
3\times10^{-3}$, $\nu_{0}=(1/3)  \langle \nu_{en}
\rangle $, and $\langle \nu_{en}\rangle \simeq10\,\nu_{in}$
(see above), we obtain $\Theta_{\min}\gtrsim50$, so that even the minimum value of
$|\mu|$ should be sufficiently large, $|\mu|_{\min}\sim\sqrt{\Theta_{\min}}\gtrsim7$. In this estimate, however, we have not included
the anomalous electron heating (see above). If we did that, we would obtain
noticeably larger values of $|\mu|_{\min}$. Even for the
extreme, but still possible value of $E_{0}\sim1\,$V/m, by applying currently
available models of the AEH \cite{Dimant:Model03,Liang:Time-dependent21}, we
may assume that the temperature will be an order of magnitude larger that the
$\vec{E}_{0}$-heated value, so that the actual minimum value of $\left\vert
\mu\right\vert $ will be at least $|\mu|_{\min}\gtrsim
20$. For more realistic field values of $E_{0}\lesssim200\ $mV/m, the value of
$|\mu|$ will be somewhat larger, $|\mu|\gtrsim25$. All these estimated values of $|\mu|$ are large enough
to provide the high accuracy of all approximate expressions that will be further used.

We describe the entire procedure of simplifying the solution given by equation~(\ref{full_equation_xi_new_1}) 
in \ref{Simplification of the general solution for electrons}.
Under condition of $|\mu^{2}| \approx \mathcal{P}\gtrsim \Theta\gg 1$, but for an
arbitrary value of $a_{e0}$, we use a series of approximations based on the same 
large parameter $\mathcal{P}$, so that each further approximation keeps the resultant expressions largely within the same level of accuracy.
The final approximate solution of the full differential
equation is given by equation~(\ref{Final_f(xi)}). Structurally, this equation is similar to
equation~(\ref{classical_1}) obtained for $a_{e0}\rightarrow 0$. This might suggest that the finite-$a_{e0}$ solution could also
have been obtained by neglecting in equation~(\ref{full_equation_xi_new_1})
all derivatives of $f_{\omega\vec{k}}^{e}$, i.e., bypassing the direct solution of
the second-order differential equation. However, comparison of
equation~(\ref{Final_f(xi)}) with such an alternative solution shows that the two solutions differ,
although in the $a_{e0}\rightarrow 0$ limit they asymptotically reduce to same 
equation~(\ref{classical_1}). Thus, in the general case of arbitrary $a_{e0}$, 
one cannot avoid solving second-order differential equation~(\ref{full_equation_xi_new_1}).

Finding the solution of equation~(\ref{full_equation_xi_new_1}), however, is not enough. 
To complete the kinetic treatment of electrons, we need to find the electron
density perturbation by integrating the wave perturbation of the electron
velocity distribution (very close to its dominant isotropic part) over the electron speeds,
\begin{linenomath*}
\begin{equation}
n_{\omega\vec{k}}^{e}\approx4\pi\int_{0}^{\infty}f_{\omega\vec{k}}^{e}(V_{e})  V_{e}^{2}dV_{e}. 
\label{n^e_omega,k_via_f^e_omega,k}
\end{equation}
\end{linenomath*}
Using a series of carefully validated approximations, the expression for $f_{\omega\vec{k}}^{e}$ has been 
found in \ref{Simplification of the general solution for electrons} in terms of the 
dimensionless variable $\xi$, see equation~(\ref{Final_f(xi)}). We rewrite equation~(\ref{Final_f(xi)}) in the
following form:
\begin{linenomath*}
\begin{equation}
f_{\omega\vec{k}}^{e}(z)=\frac{z^{4}+\alpha_{1}z^{2}+\alpha_{0}}{z^{4}-q^{4}}
\ \frac{e^{-z^{2}}(k_{y}^{2}V_{Te}^{2}-2i\vec{k}\cdot\vec{a}
_{0})  n_{0}\phi_{\omega\vec{k}}}{(2\pi)  ^{3/2}V_{Te}
^{3}(k_{y}^{2}V_{Te}^{2}-2i\vec{k}\cdot\vec{a}_{0}-a_{e0}^{2}/V_{Te}^{2})},
\label{promka_2}
\end{equation}
\end{linenomath*}
where $z=V_e/(\sqrt{2}\,V_{Te})=\sqrt{\xi}$ 
and
\begin{linenomath*}
\begin{subequations}
\label{alpha_1,0;A}%
\begin{align}
\alpha_{1}  &  =\frac{5i(\vec{k}\cdot\vec{a}_{e0})
}{k_{y}^{2}V_{Te}^{2}-2i\vec{k}\cdot\vec{a}_{0}},\label{alpha_1}\\
\alpha_{0}  &  =\frac{k_{x}^{2}V_{Te}^{2}\Omega_{e}^{2}}{4\nu_{0}%
^{2}(k_{y}^{2}V_{Te}^{2}-2i\vec{k}\cdot\vec{a}_{e0})
},\label{alpha_0}\\
q  &  = \left[ -\ \frac{(3i\nu_{0}\Delta_{\omega,\vec{k}}+k_{x}
^{2}V_{Te}^{2})  \Omega_{e}^{2}}{4\nu_{0}^{2}(k_{y}^{2}
V_{Te}^{2}-2i\vec{k}\cdot\vec{a}_{e0}-a_{e0}^{2}/V_{Te}^{2})}\right]^{1/4}. 
\label{A}
\end{align}
\end{subequations}
\end{linenomath*}
The parameter $q$ generalizes the same parameter introduced in 
Sec.~\ref{No-field approximation for electrons} to the case of arbitrary $\vec{a}_{e0}$.
Bearing in mind that the integrand of equation~(\ref{n^e_omega,k_via_f^e_omega,k}) contains an additional factor
$V_{e0}^2 \propto z^2$, we express the combined pre-exponential $z$-dependent factor as
\begin{linenomath*}
\begin{equation}
\frac{(z^{4}+\alpha_{1}z^{2}+\alpha_{0})z^{2}}{(z^{2}-q^{2})(z^{2}+q^{2})}
=z^{2}+\alpha_{1}+\frac{q^{4}+\alpha_{0}-\alpha_{1}q^{2}}{2(z^{2}
+q^{2})}+\frac{q^{4}+\alpha_{0}+\alpha_{1}q^{2}}{2(z^{2}-q^{2})}.
\label{tozhdestvo}
\end{equation}
\end{linenomath*}
Though equation~(\ref{tozhdestvo}) is invariant with respect to the choice of the quartic-root 
branch for the parameter $q$, we will always imply the only root with $\operatorname{Re}q > 0$ and $\operatorname{Im}q > 0$.

Using equations~(\ref{tozhdestvo}) and ({\ref{K^+-(1,q)}), we implement
the speed integration in equation~(\ref{n^e_omega,k_via_f^e_omega,k}), yielding
\begin{linenomath*}
\begin{align}
&  \frac{\eta_{\omega\vec{k}}^{e}}{\phi_{\omega\vec{k}}}=\frac{4(k_{y}%
^{2}V_{Te}^{2}-2i\vec{k}\cdot\vec{a}_{e0})}{\sqrt{\pi}(k_{y}^{2}V_{Te}%
^{2}-2i\vec{k}\cdot\vec{a}_{e0}-a_{e0}^{2}/V_{Te}^{2})}\int_{0}^{\infty}%
\frac{\left(  z^{4}+\alpha_{1}z^{2}+\alpha_{0}\right)  z^{2}}{z^{4}-q^{4}%
}\ e^{-z^{2}}dz\nonumber\\
&  =\frac{k_{y}^{2}V_{Te}^{2}-2i\vec{k}\cdot\vec{a}_{e0}}{k_{y}^{2}V_{Te}%
^{2}-2i\vec{k}\cdot\vec{a}_{e0}-a_{e0}^{2}/V_{Te}^{2}}\nonumber\\
&  \times\left[  1+2\alpha_{1}+\frac{q^{4}+\alpha_{0}+\alpha_{1}q^{2}}%
{q}\ Z(q)-i\ \frac{(q^{4}+\alpha_{0}-\alpha_{1}q^{2})}{q}\ Z(iq)\right],
\label{A_3_general_corrected}%
\end{align}
\end{linenomath*}
expressed here only in terms of the plasma dispersion function, $Z(x)=i\sqrt{\pi}\,w(x)$ 
(the final dispersion relation will also be given in terms of the $Z$-function). 

In these expressions, we also have the relation
\begin{linenomath*}
\begin{equation}
q^{4}+\alpha_{0}=-\ \frac{\Omega_{e}^{2}\left[  k_{x}^{2}a_{e0}%
^{2}+3i(k_{y}^{2}V_{Te}^{2}-2i\vec{k}\cdot\vec{a}_{e0}%
)\Delta_{\omega,\vec{k}}\nu_{0}\right]  }{4\nu_{0}^{2}(k_{y}^{2}V_{Te}%
^{2}-2i\vec{k}\cdot\vec{a}_{e0})(k_{y}^{2}V_{Te}^{2}-2i\vec
{k}\cdot\vec{a}_{e0}-a_{e0}^{2}/V_{Te}^{2})},
\label{q^4+alpha_0}
\end{equation}
\end{linenomath*}
which simplifies the reduction to the $\vec{a}_{e0}\rightarrow 0$ limit.
As $\vec{a}_{e0}\rightarrow 0$, the quantity $q$ reduces to the same-notation quantity in 
equation~(\ref{Qpsi}); and we also have $\alpha_1\rightarrow 0$, $\alpha_0\rightarrow \tau$, so that, 
according to equation~(\ref{q^4+alpha_0}),  $q^4 + \alpha_0 \rightarrow q^4 + \tau\rightarrow - i\beta$.
As a result, equation~(\ref{A_3_general_corrected}) reduces to NTA equation~(\ref{A33}).

\section{The TFBI Wave Dispersion
Relation\label{TFBI wave dispersion relation}}

To close the derivations of the full TFBI wave dispersion relation, we have to relate
ions and electrons by applying Poisson's equation, $\nabla^{2}\Phi=(n_{e}-n_{i})  /\epsilon_{0}$, where $\epsilon_{0}$ is the
permittivity of the free space. In our notations, for a given
harmonic perturbation this equation reduces to
\begin{linenomath*}
\begin{equation}
\eta_{\omega\vec{k}}^{i}-\eta
_{\omega\vec{k}}^{e}=k^{2}\lambda_{D}^{2}\phi_{\omega\vec{k}}\ , 
\label{Poisson}
\end{equation}
\end{linenomath*}
where $\lambda_{D}=[\epsilon_{0}T_{e}/(e^{2}n_{e})]^{1/2}$ is the
\textquotedblleft electron\textquotedblright\ Debye length [the electron
temperature stems from the definition of $\phi_{\omega\vec{k}}$ given by
equation~(\ref{phi,etc_Review})]. Substituting $\eta_{\omega\vec{k}}^{i}$
from equation~(\ref{ion_contribution}) and $\eta_{\omega\vec{k}}^{e}$ from
equation~(\ref{A_3_general_corrected}), and then canceling $\phi_{\omega\vec{k}}$ from both
sides of equation~(\ref{Poisson}), we obtain the sought-for fully kinetic linear-wave dispersion
relation:
\begin{linenomath*}
\begin{equation}
I_{\mathrm{ion}}+I_{\mathrm{electron}}+k^{2}\lambda_{D}^{2}=0.
\label{Final_dispersion_relation}
\end{equation}
\end{linenomath*}
The first two terms in the LHS represent the ion and electron contributions to the TFBI wave dispersion relation, while the last term 
takes into account the charge separation.

The ion contribution, $I_{\mathrm{ion}}$, is given by
\begin{linenomath*}
\begin{align}
I_{\mathrm{ion}}  & \equiv\frac{k^{2}V_{Tn}^{2}}{\beta_{T}(k^{2}V_{Tn}%
^{2}+i\vec{k}\cdot\vec{a}_{i0})}\left[  1+\frac{i\nu_{in}}{\sqrt{2(k^{2}%
V_{Tn}^{2}+i\vec{k}\cdot\vec{a}_{i0})}}\ \tilde{Z}\left(  \frac{\omega_{i}}{\sqrt
{2(k^{2}V_{Tn}^{2}+i\vec{k}\cdot\vec{a}_{i0})}}\right)  \right]
^{-1}\nonumber\\
& \times\left[  1+\frac{\omega_{i}-\vec{k}\cdot\vec{U}_{0}}{\sqrt{2(k_{\parallel
}^{2}V_{Tx}^{2}+k_{y}^{2}V_{Ty}^{2}+i\vec{k}\cdot\vec{a}_{i0})}%
}\ \tilde{Z}\left(  \frac{\omega_{i}-\vec{k}\cdot\vec{U}_{0}}{\sqrt{2(k_{x}%
^{2}V_{Tx}^{2}+k_{y}^{2}V_{Ty}^{2}+i\vec{k}\cdot\vec{a}_{i0})}%
}\right)  \right],
\label{I_ion}%
\end{align}
\end{linenomath*}
where $\tilde{Z}$ is a piece-wise modification of the regular $Z$-function:
\begin{linenomath*}
\begin{equation}
\tilde{Z}(z)=\left\{
\begin{array}
[c]{ccc}%
Z(z) & \text{if} & \operatorname{Re}(R_i)>0\ \ (\text{i.e.,\ }\gamma > -\nu_{in}),\\
-Z(-z) & \text{if} &  \operatorname{Re}(R_i)<0\ \ (\text{i.e.,\ }\gamma < -\nu_{in}).
\end{array}
\right.
\label{tilde_Z(z)}
\end{equation}
\end{linenomath*}
Here ``$-Z(-z)$'' can be alternatively represented as
\begin{linenomath*}
\begin{equation*}
-Z(-z)=i\sqrt{\pi}\,e^{-z^{2}}(-1+\operatorname{erf}(iz))
=Z(z)-2i\sqrt{\pi}\,e^{-z^{2}}.
\end{equation*}
\end{linenomath*}

The electron contribution, $I_{\mathrm{electron}}$, is given by
\begin{linenomath*}
\begin{align}
I_{\mathrm{electron}}  & =\frac{k_{y}^{2}V_{Te}^{2}-2i\vec{k}\cdot\vec
{a}_{e0}}{k_{y}^{2}V_{Te}^{2}-2i\vec{k}\cdot\vec{a}_{e0}-a_{e0}^{2}%
/V_{Te}^{2}}\nonumber\\
& \times\left(  1+2\alpha_{1}+\frac{q_0^{4}+\alpha_{0}+\alpha_{1}q_0^{2}}%
{q_0}\ Z(q_0)-i\,\frac{q_0^{4}+\alpha_{0}-\alpha_{1}q_0^{2}}{q_0}\ Z(iq_0)\right),
\label{I_electron}
\end{align}
\end{linenomath*}
where $q_0$ is the only quartic root ($q$) of equation~(\ref{A}) with $\operatorname{Re}q > 0$ and $\operatorname{Im}q > 0$.

We are presenting all these relations in terms of the $Z$-functions only; to express them 
in terms of $w(x)$ one can use the simple conversion formula $Z(x) = i\sqrt{\pi}\,w(x)$, see equation (\ref{w(x)->Z(x)}).
Recall that here $\beta_{T}=T_{n}/T_{e}$, $\lambda^2_{D}=\epsilon_{0}T_{e}/(e^{2}n_{e})$, $k^2 = k_{y}^2 + k_{x}^2$,
$\omega_{i}=\omega+i\nu_{in} = \omega_r+i(\gamma + \nu_{in})$, $\vec{a}_{i0} = e\vec{E}_{0}/m_{i}$, $\vec{a}_{e0} = e\vec{E}_{0}/m_{e}$,
$a_{e,i0} = |\vec{a}_{e,i0}|$, and the parameters $V_{\parallel,\perp}$ and $\vec{U}_0=\vec{a}_{i0}/\nu_{in}$ describe the 
parallel/perpendicular thermal velocities and the mean drift velocity in the assumed bi-Maxwellian 
ion velocity distribution given by equation~(\ref{f_i0_Gaussian}), respectively.
The parameters $\alpha_{0}$, $\alpha_{1}$, and $q$ are given by equation~(\ref{alpha_1,0;A}), 
\begin{linenomath*}
\begin{align*}
\alpha_{1} &  =\frac{5i(\vec{k}\cdot\vec{a}_{e0})}{k_{y}^{2}V_{Te}^{2}%
-2i\vec{k}\cdot\vec{a}_{0}},\\
\alpha_{0} &  =\frac{k_{x}^{2}V_{Te}^{2}\Omega_{e}^{2}}{4\nu_{0}^{2}(k_{y}%
^{2}V_{Te}^{2}-2i\vec{k}\cdot\vec{a}_{e0})},\\
q &  =\left[  -\ \frac{(3i\nu_{0}\Delta_{\omega,\vec{k}}+k_{x}^{2}V_{Te}%
^{2})\Omega_{e}^{2}}{4\nu_{0}^{2}(k_{y}^{2}V_{Te}^{2}-2i\vec{k}\cdot\vec
{a}_{e0}-a_{e0}^{2}/V_{Te}^{2})}\right]  ^{1/4},
\end{align*}
\end{linenomath*}
where $\Delta_{\omega\vec{k}}=\vec{k}\cdot\vec{V}_{e0}-\omega\approx\vec
{k}\cdot\vec{V}_{0}-\omega$ and $\vec{V}_{0} = e\vec{E_0}\times \vec{B}/B^2$. 

Without terms proportional to $\vec{a}_{i0}$ and $\vec{a}_{e0}$, equations~(\ref{I_ion}) and (\ref{I_electron})
reduce to those determined by the ``classical'' NFA solution for ions, see equation~(\ref{A_1,2_relation}), and the NTA solution for electrons, see
equations~(\ref{elec}) and (\ref{A33}). These two approximations describe the pure FBI
driving/damping. The terms $\propto \vec{a}_{i0}$ in equation~(\ref{I_ion}) describe the ion thermal effects
leading to the ITI, while the terms $\propto \vec{a}_{e0}$ in equations~(\ref{I_electron}) describe the
corresponding electron thermal effects. Thus, equation~(\ref{Final_dispersion_relation}) 
makes it possible to trace the effects of different physical mechanisms.

Given the wavevector $\vec{k}$ and various plasma parameters, wave dispersion equation~(\ref{Final_dispersion_relation}) 
should be solved for the complex wave frequency, $\omega = \omega_r + i\gamma$. The ion contribution 
to the wave dispersion relation includes the unknown $\omega$ only through the the combination 
$R_i = \omega + i\nu_{in}$, while the electron contribution includes $\omega$ only through the Doppler-shifted wave frequency 
$\Delta_{\omega\vec{k}}=\vec{k}\cdot\vec{V}_{e0}-\omega$, which in turn is involved only through the quantity $q$, 
see equation~(\ref{A}). Equation~(\ref{Final_dispersion_relation}) contains no other $\omega$-dependent quantities.

Unlike the long-wavelength limit of the fluid-model TFBI dispersion relation 
described in Sec.~V of \citeA{Dimant:Unified23}, where the instability
growth/damping rate, $\gamma$, turns out to be automatically small compared to the real wave frequency, $\omega_r$, and in the
former the different instability mechanisms can be linearly separated from
each other, equation~(\ref{Final_dispersion_relation}) is much more involved.
The growth or damping rate can be comparable to the real frequency, especially if
$\omega_r\sim\nu_{in}$. While in the long-wavelength range the wave phase
velocity is largely unaffected by the instability mechanisms, in the kinetic
regime all instability mechanisms can affect the wave phase velocity.
Equation~(\ref{Final_dispersion_relation}) should serve as a basis for such
analyses.

\section{Discussion\label{Discussion}}

In this section, we discuss the limitations of our analytic theory, which has far 
fewer constraints than the previous kinetic studies. The present fully kinetic theory
of the combined TFBI is a serious improvement and a significant step forward in the analytical study of the
electrojet instabilities.

The major limitation of this analysis is that we neglect finite ion magnetization, which
means that it applies to ionospheric altitudes below 110~km. Including the
ion magnetization in the analytic theory is reserved for the future, but the
electron-related part of this dispersion relation remains valid throughout the
entire E region (albeit with a caveat discussed below). If a future kinetic theory includes the ion
magnetization then there should be no need to upgrade the electron-related part.

The next important issue is that for the background velocity distributions of
ions and electrons we assumed modified Maxwellian velocity distributions: a
shifted bi-Maxwellian distribution for ions and an almost isotropic, but strongly heated
Maxwellian distribution for electrons. Note that all previous kinetic studies of the E-region instabilities assumed even
much simpler, fully undisturbed and isothermal, Maxwellian velocity distributions for both
ions and electrons. In our theory, when choosing the parameters for the modified background
velocity distributions, such as the mean particle densities, temperatures, and
the average velocity drift (for ions), we require these parameters to be
consistent with the imposed electric field and plasma parameters. However, some actual
characteristics of the background velocity distributions cannot be fully
mimicked by the modified Maxwellian velocity distributions. Indeed, according
to recent results by \citeA{Koontaweepunya:Non-Maxwellian24} (these results
have been partially reviewed in this paper's Appendix~\ref{The ion background velocity distribution in the BGK approximation}), 
under strong DC electric field, the PIC-determined ion velocity distribution becomes strongly shifted, has an
anisotropic effective temperature, and is skewed in the $\vec{E}_{0}$-direction. The simple BGK-based analytic theory accurately predicts the mean
ion drift and the overall mean effective temperature, but it greatly
exaggerates the temperature anisotropy and skewness. The employed
bi-Maxwellian distribution ansatz can accurately mimic the ion drift and
anisotropic temperatures, provided they are taken from PIC simulations or from a
more accurate analytical model, but the bi-Maxwellian ansatz cannot mimic the
velocity-distribution skewness. How consequential might be this inaccuracy is
not clear yet, but the shifted bi-Maxwellian ion function with the 
consistently chosen parameters seems to be the closest to reality, while remaining still analytically tractable.

Under the action of a strong electric field, the nearly isotropic electron
velocity distribution can deviate significantly from any heated Maxwellian
distribution, but this usually takes place outside of the thermal bulk, i.e.,
closer to the higher-energy tail of the electron velocity distribution, see,
e.g.,~\citeA{Gurevich:Nonlinear78,Milikh:Model03}. The major contributions to
all integrals that resulted in our wave dispersion relation have been made by
the thermal bulk of the electron velocity distribution, $V_{e}\lesssim V_{Te}%
$, so that this inaccuracy does not seem to be crucial. As for the ion
background velocity distribution, there is hardly a better tractable model of the
background electron velocity distribution than the Maxwellian distribution 
with an appropriately chosen temperature.

For electrons, however, there is one more, potentially serious, limitation associated
with the applicability of the entire electron kinetic approach based on
equation~(\ref{i_Delta_electron}). This equation had been originally derived
based on the assumption of strong isotropization of the electron distribution
function \cite{Gurevich:Nonlinear78,Dimant:Kinetic95a}, along with all its
wave perturbations. In the perpendicular to $\vec{B}$ direction, this
isotropization is provided by fast Larmor rotation of electrons, compared to
the wave period, $\Omega_{e}\gg|\omega|$. At the E region, this condition is
easily fulfilled. However, on any plane that includes the
direction parallel to $\vec{B}$, this isotropization usually requires
sufficiently frequent collisions, $\nu_{en}\gg|\omega|$ (or, at least, 
$\nu_{en}\gtrsim|\omega|$). This represents a much stronger
restriction than that in the perpendicular to $\vec{B}$ plane. The electron
kinetic effects are mostly relevant for $|\omega|\gtrsim\nu_{in}$. Under
undisturbed conditions of the cold E-region ionosphere, the electron-to-ion
collision rate ratio, $\nu_{en}/\nu_{in}$, is only about ten (see above), so that
there may not be enough room for the applicability of our kinetic approach for
electrons. However, two favorable circumstances help.

To illustrate the point, we do some estimates for a typical VHF radar,
e.g., for the high-latitude radar ICEBEAR ($f_{\mathrm{Radar}}=49.5~$MHz)
\cite{St-Maurice:Narrow23,Ivarsen:Turbulence24}. The corresponding
scatter from ionospheric irregularities with the wavelength $\lambda
\approx3\ $m and the corresponding wavevector $k=2\pi/\lambda\simeq2\ $%
m$^{-1}$ (along the radar LoS). Assuming first the electric field at the
instability threshold, $E_{0}=20~$mV/m, we have $V_{0}\approx C_{s}%
\approx400\ $m/s. Then the corresponding wave frequency will be $\omega\sim
kV_{0}\approx800\ $s$^{-1}$. Consider first the worst case of the top boundary for the
instability, which is slightly above the ion magnetization boundary, $\nu
_{in}=\Omega_{i}$ \cite{Dimant:Ion_thermal04}. At high latitudes, where
$B_{0}\simeq5\times10^{4}\ $nT, we have $\nu_{in}\simeq\Omega_{i}\approx
160~$s$^{-1}$, so that $\nu_{en}\simeq10\nu_{in}\simeq1,600~$s$^{-1}$. Thus,
at the ion magnetization boundary (typically around 120~km of altitude), the wave frequency $\omega\sim800\ $%
s$^{-1}$ is deeply inside the kinetic regime, $\omega\gg\nu_{in}$, while the same $\omega$ is
only about a half of $\nu_{en}\simeq1,600~$s$^{-1}$. This means that $\omega$
is not much less than $\nu_{en}$, so that the applicability condition is fulfilled only
marginally. However, if the driving field $E_{0}$ is much stronger, $E_{0}\gg
E_{\mathrm{Thr}}$ (the case much more relevant to our theory), then the ion-neutral
collision frequency $\nu_{in}$ stays the same, while $\omega\sim kV_{0}$
increases roughly in proportion with $E_{0}$, and the electron-neutral
collision frequency, $\nu_{en}(T_{e})$, increases even faster, $\nu
_{en}\propto T_{e}\propto E_{0}^{2}$. This improves the
applicability condition dramatically. On the other hand, at altitudes well below the
magnetization boundary, where the ions can be treated as essentially
unmagnetized (in consistency with the ion kinetic treatment of this paper),
the two collision frequencies exponentially grow with decreasing height (in
linear proportion with the neutral density). As a result, while $\nu_{in}$ becomes
closer to $\omega$, $\nu_{en}$ goes farther away from it. This also improves the
applicability of our electron kinetic approach, especially for large $E_0$.

There are also limitations associated with the employed kinetic
approach to the ion-neutral collisions (i.e., the popular BGK collisional model
with constant $\nu_{in}$), the simplified analytical choice of the velocity
dependence of the electron-neutral collision frequency, 
$\nu_{en}(V_{e})\propto V_{e}^{2}$, and some other approximations. We have already
discussed the BGK model for ions, so will not repeat this discussion here. We will only
mention that the biggest BGK model flaw, i.e., the absence of the collisional
scattering that should redistribute the particles over angles in the
ion-velocity space, is worst when describing a stationary ion velocity
distribution \cite{Koontaweepunya:Non-Maxwellian24}. For the time-varying wave
perturbations with $|\omega|\gtrsim\nu_{in}$, the error introduced by this
flaw is less important than using the BGK model for the stationary
distribution (though even for the latter the BGK collisional approach can be
considered as a reasonably satisfactory approximation). As far as the velocity
dependence of $\nu_{en}(V_{e})$ is concerned, it is an average fitting model; the
actual velocity dependence is more complicated and not necessarily well known
for all conditions. \citeA{Gurevich:Nonlinear78} suggested a slightly different
power-law dependence, $\nu_{en}(V_{e})\propto V_{e}^{5/3}$. Direct comparison
of figures in \citeA{Gurevich:Nonlinear78} shows, however, that our approximation of
$\nu_{en}(V_{e})\propto V_{e}^{2}$ works even better. For our specific analytical approach and results, this power-law velocity
dependence is vital.

Finally, the new kinetic theory of this paper does not describe the ETI 
in full measure because, bearing in mind sufficiently short-wavelength waves, 
$|\omega|\gtrsim \nu_{en}\gg \delta_{en}\nu_{en}$, we have neglected local electron collisional 
cooling compared to the wave-related electron frictional heating and thermal transport. For longer-wavelength waves, 
$|\omega|\lesssim \delta_{en}\nu_{en}$, we can use the previously obtained kinetic or 
fluid-model results \cite{Dimant:Kinetic95b,Dimant:Kinetic95c,Dimant:Physical97}. 
Inclusion of the local electron collisional cooling into the present 
kinetic theory would require solving a more general second-order differential 
equation, which is even more complicated than equation~(\ref{full_equation_xi_new_1}). 
If needed, the general electron kinetic theory that covers the entire range of wave frequencies can be developed in the future.

\section{Summary and Conclusions\label{Summary and conclusions}}

In this paper, we have developed a fully kinetic analytical theory of the
thermal Farley-Buneman instability (TFBI). This combined instability includes
all known short-scale E-region instabilities, except the spatially inhomogeneous gradient
drift instability (GDI). The latter usually produces relatively
long-wavelength turbulence, typically tens or hundreds meters, for which the
kinetic theory is not urgently needed, though such theories exist,
e.g., \citeA{Schlegel:Short83,Basu:Kinetic95}.

For the first time in the E-region instability theoretical research, 
the driving electric field $\vec{E}_{0}$ has been
included in the kinetic description of ions. As a result, the ion thermal
instability (ITE) driving has been automatically included in the kinetic
theory. For electrons, in addition to the standard inclusion of $\vec{E}_{0}$ via the $\vec
{E}_{0}\times\vec{B}$-drift velocity, the driving electric field was included
previously in the collisional kinetic theory developed by
\citeA{Dimant:Kinetic95a,Dimant:Kinetic95b}. That theory had allowed the
authors to predict the new effect of the electron thermal instability (ETI).
Compared to that earlier kinetic theory, this paper considers a
shorter-wavelength range, which is especially relevant to the kinetic
treatment. We do not expect this short-wavelength range to be linearly
unstable due to the ETI, but the reverse effect of increased wave stabilization and possible
phase-velocity modifications caused by electron thermal effects associated with
the explicitly included field $\vec{E}_{0}$ can affect dramatically the entire
instability process and may be of importance for the accurate interpretation of radar observations.

The principal outcome of our TFBI-related kinetic theory is the linear wave dispersion relation given by
equation~(\ref{Final_dispersion_relation}). Structurally similar
to the wave dispersion relations in the earlier kinetic theories for ions
\cite{Farley:Plasma63,Lee:High-Frequency71,Schmidt:Density73,
Ossakow:Parallel75,Schlegel:Short83,Basu:Kinetic95}, this much more complicated fully kinetic
dispersion relation also includes only elementary functions combined with the
plasma dispersion functions, $Z(x)$ \cite{Fried:Plasma61}.
Unlike the earlier FBI-related theories, however, our equation~(\ref{Final_dispersion_relation}) 
includes $Z$-functions of different arguments in both the ion and
electron parts of the dispersion equation.

The results of our theory can be of importance for accurate interpretation of radar signals scattered from 
relatively high E-region altitudes where the ionospheric density irregularities with the wavelength 
corresponding to the given radar frequency are within the short-wavelength kinetic range. 
We should bear in mind, however, that this theory is still limited to altitudes where 
ions are essentially unmagnetized (below 110 km). It may also predict the power structure of the 
dissipation region of the spectra measured by rockets, though nonlinear effects may also contribute.

Using the general kinetic wave dispersion relation given by equation~(\ref{Final_dispersion_relation}), one can determine the 
instability threshold for the given short-wavelength waves and establish whether the observed 
radar signals were scattered from linearly unstable or stable waves. For a sufficiently strong 
driving electric field, this equation will also help find the 
fastest growing unstable waves over the entire wavevector $\vec{k}$ space, provided the wave frequency $\omega$ satisfies 
the condition $\delta_{en}\nu_{en}\ll |\omega| \lesssim \nu_{en}$. The right-hand restriction is 
not as severe as it might look because the \emph{e-n} collision frequency $\nu_{en}$ increases with the 
electron temperature, which in turn sharply increases with the strong driving field.
Finally, this linear dispersion equation will provide the wave phase velocities which may approximate the observed radar signal 
Doppler shifts, even if the plasma turbulence is in the highly nonlinear state. To accurately predict the 
full irregularity spectrum and nonlinearly saturated wave amplitudes, fully 
kinetic computer simulations or much more complicated nonlinear models should be employed. 
Nevertheless, this theory enables one to validate these codes and make predictions of where the codes will prove useful.

\appendix

\section{Ion Kinetic Theory: Mathematical Details\label{Ion kinetic theory}}

In this appendix, we review some aspects of the ion kinetic theory and provide mathematical details of the derivations.

\subsection{The ion background velocity distribution in the BGK collision-model
approximation\label{The ion background velocity distribution in the BGK approximation}}

Here we briefly reproduce, and give more details to, the analytical model of
\citeA{Koontaweepunya:Non-Maxwellian24} regarding the background ion distribution function (BIDF) 
modified by a strong electric field $\vec{E}_0$. This simplified analytic model is based on the ion 
kinetic equation with the BGK collision model, see equation~(\ref{zero-order_kinetic_equation}).

Introducing
\begin{linenomath*}
\begin{equation}
F_{0}(u)\equiv\iint_{-\infty}^{\infty}f_{i0}(V_{ix},V_{iy},V_{iz})dV_{iy}dV_{iz} 
\label{F_0_Review}
\end{equation}
\end{linenomath*}
and
\begin{linenomath*}
\begin{equation}
\vec{a}_{i0}=\frac{e\vec{E}_{0}}{m_{i}},\qquad u\equiv\frac{\nu_{in}V_{ix}
}{a_{i0}},\qquad u_{T}\equiv\frac{\nu_{in}V_{Tn}}{a_{i0}},\qquad
V_{Tn}=\left(  \frac{T_{n}}{m_{i}}\right)  ^{1/2} 
\label{a_0,u,u_T,V_Ti}
\end{equation}
\end{linenomath*}
(here $a_{i0}\equiv|\vec{a}_{i0}|$), we have
\begin{linenomath*}
\begin{align}
f_{i0}(\vec{V}_{i})  &  =\frac{F_{0}\left(  u\right)  }{2\pi V_{Tn}^{2}}%
\exp\left(  -\ \frac{V_{iy}^{2}+V_{iz}^{2}}{2V_{Tn}^{2}}\right)
,\label{f_via_F_0}\\
\ f_{i0}^{\mathrm{Coll}}  &  =\frac{F_{0}^{\mathrm{Coll}}\left(  u\right)
}{2\pi V_{Tn}^{2}}\exp\left(  -\ \frac{V_{iy}^{2}+V_{iz}^{2}}{2V_{Tn}^{2}%
}\right)  ,\nonumber\\
F_{0}^{\mathrm{Coll}}\left(  u\right)   &  =\frac{n_{0}}{\sqrt{2\pi}V_{Tn}%
}\exp\left(  -\ \frac{V_{ix}^{2}}{2V_{Tn}^{2}}\right)  =\frac{n_{0}\nu_{in}%
}{a_{i0}\sqrt{2\pi}u_{T}}\exp\left(  -\ \frac{u^{2}}{2u_{T}^{2}}\right)
.\nonumber
\end{align}
\end{linenomath*}
Then, integrating equation~(\ref{zero-order_kinetic_equation}) over the entire
domain of the perpendicular velocities $V_{iy}$ and $V_{iz}$, we reduce it to
\begin{linenomath*}
\begin{equation}
\frac{dF_{0}}{du}=-\left(  F_{0}-F_{0}^{\mathrm{Coll}}\left(  u\right)
\right)  
\label{for_F_0_review}
\end{equation}
\end{linenomath*}
or, equivalently,
\begin{linenomath*}
\begin{equation}
\frac{dG_{0}}{du}=-G_{0}(u)+\frac{1}{\sqrt{2\pi}u_{T}}\ \exp\left(
-\ \frac{u^{2}}{2u_{T}^{2}}\right)  , 
\label{dG_0}
\end{equation}
\end{linenomath*}
where
\begin{linenomath*}
\begin{equation}
G_{0}(u)=\frac{a_{i0}}{n_{0}\nu_{in}}\ F_{0}(V_{x}). 
\label{G_0}
\end{equation}
\end{linenomath*}
The solution of equation (\ref{dG_0}), under the asymptotic boundary
conditions of $\lim_{u\rightarrow\pm\infty}G_{0}(u)=0$, is given by
\begin{linenomath*}
\begin{align}
G_{0}(u)  &  =\frac{1}{2}\exp\left(  -u+\frac{u_{T}^{2}}{2}\right)  \left[
1+\operatorname{erf}\left(  \frac{u-u_{T}^{2}}{\sqrt{2}u_{T}}\right)  \right]
\nonumber\\
&  =\frac{1}{2}\exp\left(  -\ \frac{u^{2}}{2u_{T}^{2}}\right)  w\left(
-i\ \frac{u-u_{T}^{2}}{\sqrt{2}u_{T}}\right)  , \label{G_0_via_erf}%
\end{align}
where $\operatorname{erf}(x)$ and $w(\zeta)$,
\begin{align}
\operatorname{erf}(x)  &  =\frac{2}{\sqrt{\pi}}\int_{0}^{x}e^{-t^{2}%
}dt,\nonumber\\
w(\zeta)  &  =e^{-\zeta^{2}}\left(  1+\frac{2i}{\sqrt{\pi}}\int_{0}^{\zeta
}e^{t^{2}}dt\right)  =e^{-\zeta^{2}}\left[  1+\operatorname{erf}%
(i\zeta)\right]  
\label{erf(x)}
\end{align}
\end{linenomath*}
are standard error functions \cite{Abramowitz:Handbook88}.

Firstly, we discuss the small-field limit of $a_{i0}\rightarrow0$. In this
limit, the LHS of equation~(\ref{dG_0}) should also be small, resulting in
\begin{linenomath*}
\begin{equation}
G_{0}(u)\rightarrow\frac{1}{\sqrt{2\pi}\,u_{T}}\ \exp\left(  -\ \frac{u^{2}%
}{2u_{T}^{2}}\right)  , 
\label{G_0->}
\end{equation}
\end{linenomath*}
corresponding to $f_{i0}\rightarrow\ f_{i0}^{\mathrm{Coll}}$. According to the
definitions given by equation~(\ref{a_0,u,u_T,V_Ti}), the limit of
$a_{i0}\rightarrow0$ corresponds to $u_{T}\gg1$. The asymptotic analysis of
the solution (\ref{G_0_via_erf}) shows that equation~(\ref{G_0->}) is valid
not for the entire ion velocity distribution but only for the bulk of the ion
velocity distribution constrained by the condition
\begin{linenomath*}
\begin{equation}
u\ll u_{T}^{2}. 
\label{u_T<u<u_T^2}
\end{equation}
\end{linenomath*}
Indeed, at large values of $|x|$, the error function has the following
asymptotic behavior \cite{Abramowitz:Handbook88}:
\begin{linenomath*}
\begin{equation}
\operatorname{erf}\left(  x\right)  \approx\left\{
\begin{array}
[c]{ccc}%
1-\frac{\exp\left(  -x^{2}\right)  }{x\sqrt{\pi}} & \text{if} & x\gg1\\
-1+\frac{\exp\left(  -x^{2}\right)  }{\left(  -x\right)  \sqrt{\pi}} &
\text{if} & \left(  -x\right)  \gg1
\end{array}
\right.  . 
\label{erf(x)_asymptotics}
\end{equation}
\end{linenomath*}
According to equation~(\ref{u_T<u<u_T^2}), we have $u-u_{T}^{2}<0$ and
$u_{T}^{2}-u\approx u_{T}^{2}\gg1$, so that for the error function in
equation~(\ref{G_0_via_erf}) we should use the bottom asymptotic expression
(\ref{erf(x)_asymptotics}),
\begin{linenomath*}
\[
1+\operatorname{erf}\left(  \frac{u-u_{T}^{2}}{\sqrt{2}u_{T}}\right)
\approx\frac{1}{u_{T}}\sqrt{\frac{2}{\pi}}\exp\left(  -\ \frac{u^{2}}%
{2u_{T}^{2}}+u-\frac{u_{T}^{2}}{2}\right).
\]
\end{linenomath*}
Then $G_{0}(u)$ does reduce to the asymptotic function given by
equation~(\ref{G_0->}). We should bear in mind, however, that for the ion
distribution outside the range of (\ref{u_T<u<u_T^2}), i.e., for $u\gtrsim
u_{T}^{2}$, the effect of even small electric field cannot be neglected.

Figure~1 from \citeA{Koontaweepunya:Non-Maxwellian24} shows a few examples of the zeroth-order velocity distributions of
$G_{0}(u)$. It reduces to the undisturbed Maxwellian (with the neutral
temperature) under conditions $u_{T}\gg1$ and $u_{T}^{2}\gg u$, or, in the
original variables,
\begin{linenomath*}
\begin{equation}
a_{i0}=\frac{eE_{0}}{m_{i}}\ll\nu_{in}\sqrt{\frac{T_{n}}{m_{i}}},\qquad
V_{ix}\ll\frac{\nu_{in}V_{Tn}^{2}}{a_{i0}}=\frac{\nu_{in}T_{n}}{eE_{0}}
\label{<<<}
\end{equation}
\end{linenomath*}
As the electric field exceeds a critical value,
\begin{linenomath*}
\begin{equation}
E_{\mathrm{cr}}=\frac{\nu_{in}\sqrt{m_{i}T_{n}}}{e}=\frac{B}{\kappa_{i}}%
\sqrt{\frac{T_{n}}{m_{i}}}\approx\frac{14.4}{\kappa_{i}}\left(  \frac{T_{n}%
}{300\mathrm{K}}\right)  ^{\frac{1}{2}}\left(  \frac{B}{0.5\mathrm{G}}\right)
\frac{\mathrm{mV}}{\mathrm{m}}, 
\label{E_cr}
\end{equation}
\end{linenomath*}
the ion distribution function becomes noticeably distorted by the ion heating.
Here we assumed $m_{i}\simeq 30\,m_{p}$, where $m_{p}$ is the proton mass, and
$\kappa_{i}=\Omega_{i}/\nu_{in}$ is the ion magnetization parameter. The
latter increases exponentially with the altitude because the collision
frequency $\nu_{in}$ is proportional to the neutral density obeying mostly the
barometric law. In this paper, we assume unmagnetized ions, $\kappa_{i}\ll1$.
At higher latitudes, this condition is marginally fulfilled near 110~km
($\kappa_{i}\lesssim0.3$) but much better around 100~km ($\kappa_{i}
\simeq0.06$). The critical field, $E_{\mathrm{cr}}$, in this altitude range
varies from $\simeq50~$mV/m at 110~km to $\simeq250$ mV/m at 100~km. During
big geomagnetic storms, strong electric fields $E_{0}$ occur, often $E_{0}
\sim100~$mV/m, but sometimes $E_{0}$ can even reach $250$ mV/m and more. In
such extreme events, one can expect significant distortions of the background
ion distribution function (BIDF). Figure~1 from \citeA{Koontaweepunya:Non-Maxwellian24} shows that the BIDF can be
significantly skewed, growing a prominent tail in the $\vec{E}_{0}$-direction.
In the framework of the BGK approximation, in the two perpendicular to
$\vec{E}_{0}$ directions $f_{i0}(\vec{V}_{i})$ the BIDF remains close to the
undisturbed Maxwellian function $\ f_{i0}^{\mathrm{Coll}}(\vec{V}_{i})$. We
should remind, however, that collisional angular scattering (not included in
the BGK approximation) smears to some degree the extended velocity
distribution towards those directions. This is perfectly clear from 
the PIC simulations also performed in \citeA{Koontaweepunya:Non-Maxwellian24}.

In the original velocity variables, according to the above definitions, the entire BIDF
is given by
\begin{linenomath*}
\begin{align}
&  f_{i0}(\vec{V}_{i})=\frac{n_{0}\nu_{in}}{2\pi a_{i0}V_{Ti}^{2}}\exp\left(
-\ \frac{V_{iy}^{2}+V_{iz}^{2}}{2V_{Ti}^{2}}\right)  G_{0}\left(  u\right)
\nonumber\\
&  =\frac{n_{0}\nu_{in}}{4\pi a_{i0}V_{Ti}^{2}}\exp\left[  -\ \frac{\nu
_{in}V_{ix}}{a_{i0}}+\frac{1}{2}\left(  \frac{\nu_{in}V_{Ti}}{a_{i0}}\right)
^{2}-\frac{V_{iy}^{2}+V_{iz}^{2}}{2V_{Ti}^{2}}\right] \nonumber\\
&  \times\left[  1+\operatorname{erf}\left(  \frac{V_{ix}}{\sqrt{2}V_{Ti}%
}-\frac{\nu_{in}V_{Ti}}{\sqrt{2}a_{i0}}\right)  \right]  .
\label{BIDF_original_variables}
\end{align}
\end{linenomath*}

Directly from equations~(\ref{dG_0}) and (\ref{G_0_via_erf}), we can easily
deduce the following expressions for the three first moments of the BIDF:
\begin{linenomath*}
\begin{equation}
\int_{-\infty}^{\infty}G_{0}(u)du=1,\qquad\int_{-\infty}^{+\infty}%
uG_{0}(u)du=1,\qquad\int_{-\infty}^{\infty}u^{2}G_{0}(u)du=u_{T}^{2}+2.
\label{three_lowest_moments}
\end{equation}
\end{linenomath*}
The first one expresses the particle density conservation, $n_{i}=\int f_{i}d^{3}V_{i}$; 
the second one provides the expression for the
$x$-component of the mean ion drift velocity, $\langle V_{ix}\rangle =a_{i0}/\nu_{in}$, where $\langle \cdots\rangle$
denotes the velocity average, $\langle \cdots\rangle =\left.
\int(\cdots)f_{i}d^{3}V_{i}\right/\int f_{i}d^{3}V_{i}$; and the third
one provides the effective ion temperature, $T_{\mathrm{eff}}=(m_{i}/3)  
(\vec{V}_{i}-\langle \vec{V}_i\rangle)^{2}$, established as the equilibrium balance between the $\vec
{E}_{0}$-field frictional heating and collisional cooling, $T_{\mathrm{eff}%
}=T_{n}+(1/3)m_iU_0^{2}$, $U_0=|\langle\vec{V}_i\rangle| = eE_0/(m_i\nu_{in})$. One can also deduce all
higher-order moments of the BIDF. They characterize the BIDF skewness, but
this is of no importance for our purposes.

The exact BGK solution for the distorted ion background distribution function
$f_{i0}(\vec{V}_{i})$ given by equation~(\ref{Int_dz_BGK}) is hard for
further analytical treatment that involves three velocity integrations. In 
Sec.~\ref{The background ion velocity distribution} we have constructed a
shifted and heated bi-Maxwellian function, see equation~(\ref{f_i0_Gaussian}), that accurately describes the exact
Pedersen drift and total frictional heating by the electric field $\vec{E}_{0}$. 
This analytic construct, however, lacks the skewness of $f_{i0}$ described by
the above BGK analytic solution or, to a much lesser degree, by that observed in the PIC simulations
\cite{Koontaweepunya:Non-Maxwellian24}, as we discuss below.

The exact BGK solution is not exact in the real physical sense, as we already discussed
in Sec.~\ref{The background ion velocity distribution}. The BGK operator has a
major flaw that it does not include collisional angular scattering from the
main direction parallel to the driving $\vec{E}_{0}$-field (i.e., the $x$
direction) to the perpendicular, $y$ and $z$, directions. This leads to the
fact that the BGK-described frictional heating of ions is distributed
extremely uneven between different ion velocity components: in this
approximation, the entire ion heating occurs strictly along the $x$ direction
(purely parallel heating) with zero heating in the two other directions. In
real situations, however, during an ion-neutral collision the effective rates
of the momentum and energy losses are close in value, so that some parallel
heating should inevitably be transferred to the other directions through the
angular scattering. Such scattered heating has been observed in our
PIC simulations \cite{Koontaweepunya:Non-Maxwellian24}. Assuming the Gaussian approximation of $f_{i0}$, we can
mimic a more realistic anisotropic frictional heating by choosing two
different temperatures for the parallel to $\vec{E}_0$ and perpendicular to $\vec{E}_0$
directions ($T_{x}$ and $T_{y}$, respectively). In this paper, these two different temperatures are
just free input parameters that could be specified, e.g., by comparison with
PIC\ simulations. We should bear in mind, however, that these two parameters
are not totally independent because they are constrained by the total balance of
$\vec{E}_{0}$-field frictional heating,
\begin{linenomath*}
\begin{equation}
\frac{T_{x}}{3}+\frac{2T_{y}}{3}=T_{\mathrm{eff}}=T_{n}+\frac{m_i U_{0}^{2}}{3},
\label{Ion_temperature_constrant}
\end{equation}
\end{linenomath*}
see Sec.~\ref{The background ion velocity distribution}), so that only the ratio
$T_{x}/T_{y}$ is really a free parameter (it should vary between 0
and 1, with a possible $E_{0}$-dependence).

\subsection{Linear perturbations: No-field approximation for
ions\label{No-field approximation for ions}}

Here we discuss linear wave perturbations of the ion velocity distribution in the no-field approximation 
(NFA). This approximation produced the well-known ion part of the pure FBI kinetic dispersion relation
\cite{Farley:Plasma63,Lee:High-Frequency71,Schmidt:Density73,Ossakow:Parallel75,Schlegel:Short83,Basu:Kinetic95}.
In the NFA approximation, the total electric field
$\vec{E}$ in the ion kinetic equation includes only the electrostatic wave component, $\vec
{E}=-\nabla\Phi$, but not the driving external DC component, $\vec{E}_{0}$.

In this paper, we use the NFA ion kinetic solution for comparisons and references, but the 
main reason why we review it here is that a small twist in its derivation
produces an unexpected analytic evaluation of two non-conventional integrals. 
The evaluation of these integrals, depicted below in Sec.~\ref{Useful integrals}, 
becomes a crucial step for us when we solve in Sec.~\ref{Ion contribution to the wave dispersion relation} 
the general case of kinetic ions under an arbitrarily strong electric field $\vec{E}_{0}$. 

Now we start the NFA wave kinetic analysis. Linearizing equation~(\ref{initial_ion_kinetic_equation}) with respect to
small wave perturbations, for a separate harmonic wave $\propto\exp[i(\vec{k}\cdot\vec{r}-\omega t)]$, we obtain
\begin{linenomath*}
\begin{equation}
(\vec{k}\cdot\vec{V}_{i}\mathbf{-}\omega_{i})f_{\omega\vec{k}}^{i}=-i\nu
_{in}f_{i0}^{\mathrm{Coll}}\eta_{\omega\vec{k}}+\frac{V_{Ti}^{2}}{\beta_{T}
}\ \phi_{\omega\vec{k}}\left(  \vec{k}\cdot\frac{\partial f_{i0}}{\partial
\vec{V}_{i}}\right). 
\label{kinetic_no_field}
\end{equation}
\end{linenomath*}
Here
\begin{linenomath*}
\begin{equation}
\omega_{i}\equiv\omega+i\nu_{in} = \omega_r+i(\gamma + \nu_{in}),\qquad\beta_{T}\equiv\frac{T_{i}}{T_{e}},\qquad
V_{Ti}^{2}=\frac{T_{i}}{m_{i}}, 
\label{rho,beta_T,T_i}
\end{equation}
\end{linenomath*}
$\ f_{i0}(\vec{V}_{i})$ is the undisturbed uniform and constant BIDF, see
equation~(\ref{f_i0_Gaussian}), and $f_{\omega\vec{k}}^{i}$ is the
corresponding wave perturbation. The parameter $T_{i}$ can be any constant temperature
(e.g., the ion mean temperature $T_{i}=(2T_{\perp}+T_{\parallel})/3$) because,
in accord with the definitions of $\phi_{\omega\vec{k}}$, see 
equation~(\ref{phi,etc_Review}), and parameters $\beta_T$, $V_{Ti}^{2}$,
equation~(\ref{kinetic_no_field}) includes no actual $T_{i,e}$-dependencies (we
have included the inconsequential parameter $T_i$ for convenience).

Without $\vec{E}_{0}$, the LHS\ of equation~(\ref{kinetic_no_field})
includes no $\vec{V}_{i}$-derivatives of $f_{\omega\vec{k}}^{i}$. This
means neglecting the mean ion Pedersen drift and frictional heating.
This simplifies dramatically the linear-wave analysis of the ion
velocity distribution. The total neglect of the driving DC field $\vec{E}_{0}$
also yields $f_{i0}=f_{i0}^{\mathrm{Coll}}$, where $f_{i0}^{\mathrm{Coll}}$ is
given by equation~(\ref{ion_cooling _Maxwellian}) (here we assume $T_{i}=T_{n}$). As a
result, the corresponding purely algebraic equation~(\ref{kinetic_no_field})
yields
\begin{linenomath*}
\begin{equation}
f_{\omega\vec{k}}^{i}=-\ \frac{i\nu_{in}f_{i0}^{\mathrm{Coll}}}{\vec{k}%
\cdot\vec{V}\mathbf{-}\omega_{i}}\ \eta_{\omega\vec{k}}^{i}-\frac{(\vec{k}%
\cdot\vec{V}_{i})f_{i0}^{\mathrm{Coll}}}{\beta_{T}(\vec{k}\cdot\vec{V}%
_{i}\mathbf{-}\omega_{i})}\ R_{\omega\vec{k}}. 
\label{f_omega_classic}
\end{equation}
\end{linenomath*}
Using the expression for $\eta_{\omega\vec{k}}$ in terms of $f_{\omega\vec{k}%
}^{i}$, see equation~(\ref{phi,etc_Review}), after the corresponding
integration of the both sides of equation~(\ref{f_omega_classic}), we obtain
the sought-for linear ion relationship,
\begin{linenomath*}
\begin{equation}
A_{1}\phi_{\omega\vec{k}}=\beta_{T}A_{2}\eta_{\omega\vec{k}}^{i},
\label{A_1,2_relation}
\end{equation}
\end{linenomath*}
where
\begin{linenomath*}
\begin{subequations}
\label{A_1,A_2}%
\begin{align}
A_{1}  &  =-\left\langle \frac{\vec{k}\cdot\vec{V}_{i}}{\vec{k}\cdot\vec
{V}_{i}-\omega_{i}}\right\rangle =-\left(  1+\omega_{i}\left\langle \frac{1}%
{\vec{k}\cdot\vec{V}_{i}-\omega_{i}}\right\rangle \right)  ,\label{A_1}\\
A_{2}  &  =\left\langle \frac{\vec{k}\cdot\vec{V}_{i}-\omega}{\vec{k}\cdot
\vec{V}_{i}-\omega_{i}}\right\rangle =1+i\nu_{in}\left\langle \frac{1}{\vec
{k}\cdot\vec{V}_{i}-\omega_{i}}\right\rangle. 
\label{A_2}
\end{align}
\end{subequations}
\end{linenomath*}
Here the angular brackets denote the velocity average over the Maxwellian
function $f_{i0}^{\mathrm{Coll}}(\vec{V}_{i})$,
\begin{linenomath*}
\begin{align}
&  \left\langle \cdots\right\rangle =\frac{\iiint(\cdots)\exp\left(
-\ \frac{V_{i}^{2}}{2V_{Ti}^{2}}\right)  d^{3}V_{i}}{\iiint\exp\left(
-\ \frac{V_{i}^{2}}{2V_{Ti}^{2}}\right)  d^{3}V_{i}}\nonumber\\
&  =\frac{1}{\left(  2\pi\right)  ^{3/2}V_{Ti}^{3}}\iiint(\cdots)\exp\left(
-\ \frac{V_{i}^{2}}{2V_{Ti}^{2}}\right)  d^{3}V_{i}. 
\label{<...>}
\end{align}
\end{linenomath*}
In accord with absence of the ion frictional heating, we have $T_{i}%
=T_{\parallel}=T_{\perp}=T_{n}$, so that $V_{Ti}^{2}=V_{Tx}^{2}
=V_{n}^{2}$.

Before going any further, we check the long-wavelength (low-frequency) 
limit, loosely defined by the conditions $kV_{Ti}\ll\nu_{in}$,
$|\omega| \ll\nu_{in}$, where $k\equiv|\vec{k}|$. The
linear theory of such waves is usually successfully treated using much simpler fluid-model
equations. It is straightforward to verify that in the long-wavelength limit
the kinetic relationship given by equation~(\ref{A_1,2_relation}) reduces to
the corresponding fluid-model relationship,
\begin{linenomath*}
\begin{equation}
\phi_{\omega\vec{k}}\approx\beta_{T}\left[  \frac{\omega\left(  \omega
+i\nu_{in}\right)  }{k^{2}V_{Ti}^{2}}-1\right]  \eta_{\omega\vec{k}}^{i}.
\label{fluid_model_classical}
\end{equation}
\end{linenomath*}

Now we proceed with the more general NFA kinetic relationship
(\ref{A_1,2_relation}) for arbitrary wavelengths. Temporarily, we set up the
coordinate system in such a way that the wavevector $\vec{k}$ 
is directed along the $x$-axis: $k=k_{x}$ (so that $\vec
{k} \cdot \vec{V}_{i}=kV_{ix}$). From equations (\ref{A_1,A_2}) and
(\ref{<...>}) we see that in order to calculate both multipliers, $A_{1}$ and
$A_{2}$, we need to evaluate just a single integral
\begin{linenomath*}
\begin{equation}
I(p)=\int_{-\infty}^{\infty}\frac{e^{-\xi^{2}}}{\xi-p}\,d\xi,\qquad
p\equiv\frac{\omega_{i}}{\sqrt{2}kV_{Ti}}, 
\label{II(a)}
\end{equation}
\end{linenomath*}
($\omega_{i}\equiv\omega + i\nu_{in}$) so that
\begin{linenomath*}
\begin{equation}
A_{1}=-\left(  1+\frac{p}{\sqrt{\pi}}\,I(p)\right)  ,\qquad A_{2}=1+\frac
{i\nu_{in}}{\omega+i\nu_{in}}\ \frac{p}{\sqrt{\pi}}\,I(p). 
\label{A_1,A_2_Int}
\end{equation}
\end{linenomath*}
Treating $\xi$ as a complex variable and temporarily making a change of the
variable from $\xi$ to $\zeta=\xi-p$, we obtain
\begin{linenomath*}
\begin{equation}
I(p)=e^{-p^{2}}J(p),\qquad J(p)\equiv\int_{C}\frac{e^{-\zeta^{2}-2p\zeta}
}{\zeta}\,d\zeta, 
\label{I_via_J}
\end{equation}
\end{linenomath*}
where the corresponding contour of integration in the complex plane, $C$,
represents a straight line parallel to the real axis and shifted downward or
upward in the complex plane by $i\operatorname{Im}p$, depending on the sign
of $\operatorname{Im}p$. We can deform this contour to another contour
$C^{\prime}$ which is composed of the two semi-infinite segments of the
horizontal $\zeta$ axis with $\operatorname{Im}\zeta=0$ closed from below by
an infinitely small semi-circle of the radius $\epsilon\rightarrow0$ around
$\zeta=0$ if $\operatorname{Im}p>0$ (or from above if $\operatorname{Im}p<0$).
Differentiating $J(p)$ with respect to the parameter $p$, we obtain
\begin{linenomath*}
\[
\frac{dJ}{dp}=-\,2\int_{-i\epsilon-\infty}^{-i\epsilon+\infty}e^{-\zeta
^{2}-2p\zeta}\,d\zeta=-\,2\sqrt{\pi}e^{p^{2}}.
\]
\end{linenomath*}
Integrating this equation, we obtain $J(p)=\mathrm{const}-2\sqrt{\pi}\int%
_{0}^{p}e^{t^{2}}dt$, where `$\mathrm{const}$' is determined by the solution at
$p=0$. Directly from the definition of both $J(p)$ for $p=0$ and the contour
$C^{\prime}$, we see that the odd integrand reduces the entire integration to
a semi-residual $i\pi$ for $\operatorname{Im}p>0$ (anti-clockwise) or $-i\pi$
for $\operatorname{Im}p<0$ (clockwise). As a result, we obtain
\begin{linenomath*}
\begin{subequations}
\begin{align}
I(p)  &  =\int_{-\infty}^{\infty}\frac{e^{-\xi^{2}}}{\xi-p}\,d\xi=i\pi
w(p)\qquad\qquad\text{if}\;\;\operatorname{Im}(p)>0,\label{Im_p>0}\\
I(p)  &  =\int_{-\infty}^{\infty}\frac{e^{-\xi^{2}}}{\xi-p}\,d\xi=-\,i\pi
w(-p)\ \qquad\text{if}\;\;\operatorname{Im}(p)<0,\label{Im_p<0}
\end{align}
\label{Iw}
\end{subequations}
\end{linenomath*}
where the $w$-function, $w(x)=e^{-x^{2}}[1+\operatorname{erf}(ix)]$, 
see equation~(\ref{erf(x)}), can be also written as  \cite{Abramowitz:Handbook88}
\begin{linenomath*}
\begin{equation}
w(\zeta)=e^{-\zeta^{2}}\mathrm{erfc}(-i\zeta)=\frac{2}%
{\sqrt{\pi}}\,e^{-\zeta^{2}}\int_{-i\zeta}^{\infty\times e^{i\varphi}%
}e^{-t^{2}}dt.
\label{ww}
\end{equation}
\end{linenomath*}
Here the phase $\varphi$ in the complex plane lies within the sector
$-\pi/4<\varphi<\pi/4$. 

The $w$-function can be also represented in
terms of the plasma dispersion function, $Z(z)$,
\begin{linenomath*}
\begin{equation}
w(\zeta)\equiv-\,\frac{i}{\sqrt{\pi}}\,Z(\zeta)\,,\qquad Z(\zeta)=i\int%
_{0}^{\infty}\exp\left(  i\zeta t-\frac{t^{2}}{4}\right)  dt=i\sqrt{\pi
}\,w(\zeta). 
\label{w->Z}
\end{equation}
\end{linenomath*}
In terms of the $Z$-functions, equation\,(\ref{Iw}) becomes
\begin{linenomath*}
\begin{subequations}
\begin{align}
I(p)  &  =\int_{-\infty}^{\infty}\frac{e^{-\xi^{2}}}{\xi-p}\,d\xi=\sqrt\pi\,
Z(p)\qquad\qquad\text{if}\;\;\operatorname{Im}(p)>0,\label{Z_Im_p>0}\\
I(p)  &  =\int_{-\infty}^{\infty}\frac{e^{-\xi^{2}}}{\xi-p}\,d\xi = -\sqrt\pi\,Z(-p)
\ \qquad\text{if}\;\;\operatorname{Im}(p)<0.
\label{Z_Im_p<0}
\end{align}
\label{Z_I}
\end{subequations}
\end{linenomath*}
Using an easily provable relation $-Z(-z)=(Z(z^{\ast}))^{\ast}$, the RHS of equation\ (\ref{Z_Im_p<0}) can be also written as $\sqrt\pi\,(Z(p^{\ast}))^{\ast}$.

The earlier kinetic theories of ions were restricted to the case of $\operatorname{Im}p>0$ ($\gamma>-\nu_{in}$), because the opposite case of
$\operatorname{Im}p<0$ corresponds to the linearly stable, highly wave-damping
regime of $\gamma<-\nu_{in}$. In the general case, however, one should bear in mind both cases 
because the major output of radar observations is the wave phase velocity, without definite knowledge of 
whether the observed plasma wave is linearly stable or unstable. 

To conclude this section, we note that for sufficiently large $|\zeta|$ the function
$Z(\zeta)$ has the following asymptotic series \cite{Fried:Plasma61}:
\begin{linenomath*}
\begin{align}
Z(\zeta)  & \approx i\sqrt{\pi}\,\sigma(\zeta)  e^{-\zeta^{2}}-\sum
_{n=0}^{\infty}\frac{(n-1/2)!}{\sqrt{\pi}\,\zeta^{2n+1}}\nonumber\\
& =i\sqrt{\pi}\,\sigma(\zeta)  e^{-\zeta^{2}}-\frac{1}{\zeta}\left(
1+\frac{1}{2\zeta^{2}}+\frac{3}{4\zeta^{4}}+\frac{15}{8\zeta^{6}}+\ldots\right),
\label{Z_asymptotics}
\end{align}
\end{linenomath*}
where
\begin{linenomath*}
\[
\sigma(\zeta)  =\left\{
\begin{array}
[c]{ccc}%
0 & \text{if} & \operatorname{Im}\zeta>0,\\
1 & \text{if} & \operatorname{Im}\zeta=0,\\
2 & \text{if} & \operatorname{Im}\zeta<0,
\end{array}
\right.  
\]
\end{linenomath*}
and the gamma function of a half-integer argument, $(n-1/2)!\equiv\Gamma(n+1/2)$, 
can be expressed in terms of the integer factorials as $\sqrt{\pi}\,(2n)!/(2^{2n}n!)$.
Using equation~(\ref{Z_asymptotics}), one can verify that under conditions of 
$kV_{Ti},|\omega| \ll\nu_{in}$, leading to $|R_i| \approx \nu_{in}$ and $|p|\gg 1$, 
the above expressions agree with fluid-model equation~(\ref{fluid_model_classical}). 

\subsubsection{Useful integrals\label{Useful integrals}}

Now we slightly modify the above derivation by using a more general coordinate
system. Instead of directing the wavevector $\vec{k}$
strictly along the $x$-axis, we will rotate the coordinate axes around the
$z$-axis, so that the wavevector $\vec{k}$ has two non-zero components:
$k_{x}$ and $k_{y}$ (but still $k_{z}=0$). In such coordinate system, one
might use, e.g., the polar coordinates in the velocity space, $V_{i}$, $\theta_{i}$, so
that $\vec{k}\cdot\vec{V}_{i}=kV_{i}\cos\theta_{i}$, $k=(k_{x}^{2}+k_{y}%
^{2})^{1/2}$, and obtain the above result in a more symmetric way with respect
to the $\vec{k}$ components. But such derivation would produce nothing new and useful. If,
however, we keep using the Cartesian coordinate system, then by comparing
the results we will arrive at a non-trivial integral whose exact evaluation,
exclusively in terms of the Gaussian and error functions, will help us obtain
the wave dispersion relation in the general case of arbitrarily large
electric field $\vec{E}_{0}$ (see below).

Now we proceed with the corresponding treatment. We are no longer interested
in obtaining the relationship between the electric potential and density wave
perturbations because we already know the result (see above). What we are
pursuing here is the mutual consistency of two slightly different mathematical
approaches, so that they give the same result. To this end, we restrict
ourselves to the calculation of the following double integral:
\begin{linenomath*}
\begin{equation}
R=\iint_{-\infty}^{\infty}\frac{\exp\left(  -u^{2}\right)  d^{2}u}{\vec
{k}\cdot\vec{u}\mathbf{-}p}. 
\label{R}
\end{equation}
\end{linenomath*}
assuming $\operatorname{Im}p>0$.

For the quantity $R$, the simple first approach with $k_{y}=0$ and positive
$k_{x}=k$, yields
\begin{linenomath*}
\begin{align}
R  &  =\int_{-\infty}^{\infty}\exp\left(  -u_{y}^{2}\right)  du_{y}%
\int_{-\infty}^{\infty}\frac{\exp\left(  -u_{x}^{2}\right)  du_{x}}%
{k_{x}\left(  u_{x}-p\right)  }=\frac{\sqrt{\pi}}{k_{x}}\ I(p)\nonumber\\
&  =\frac{i\pi^{3/2}}{k}\ w\left(  p\right)  =\frac{i\pi^{3/2}}{k}%
\ \exp\left(  -p^{2}\right)  \left[  1+\operatorname{erf}\left(  ip\right)
\right]  . 
\label{R_classical}
\end{align}
\end{linenomath*}

The second approach with two non-zero vector components $k_{x}$ and $k_{y}$
yields
\begin{linenomath*}
\begin{equation}
R=\iint_{-\infty}^{\infty}\frac{\exp\left(  -u^{2}\right)  d^{2}u}{\vec
{k}\cdot\vec{u}\mathbf{-}p}=\int_{-\infty}^{\infty}\exp\left(  -u_{y}%
^{2}\right)  du_{y}\int_{-\infty}^{\infty}\frac{\exp\left(  -u_{x}%
^{2}\right)  du_{x}}{k_{x}u_{x}+k_{y}u_{y}-kp}, 
\label{RR}
\end{equation}
\end{linenomath*}
where the internal integral has the same structure as that in
equation~(\ref{II(a)}) with $p$ replaced by $p-k_{y}u_{y}/k$ (the restriction
$\operatorname{Im}p>0$ easily transfers to $\operatorname{Im}(p-k_{y}%
u_{y}/k)>0$). Applying (\ref{Iw}), the internal integration yields
\begin{linenomath*}
\begin{align}
&  \int_{-\infty}^{\infty}\frac{\exp\left(  -u_{x}^{2}\right)  du_{x}}%
{k_{x}u_{x}+k_{y}u_{y}-kp}=\frac{i\pi}{k_{x}}\ w\left(  \frac{kp-k_{y}u_{y}%
}{k_{x}}\right) \nonumber\\
&  =\frac{i\pi^{3/2}}{k}\ \exp\left[  -\ \left(  \frac{kp-k_{y}u_{y}}{k_{x}%
}\right)  ^{2}\right]  \left[  1+\operatorname{erf}\left(  i\ \frac
{kp-k_{y}u_{y}}{k_{x}}\right)  \right]  . 
\label{internal_integral}
\end{align}
\end{linenomath*}
In the remaining integral of equation~(\ref{RR}), we change the variable
$u_{y}$ to a new complex variable $\tau$,
\begin{linenomath*}
\begin{equation}
\tau=\frac{kp-k_{y}u_{y}}{k_{x}},\qquad u_{y}=\frac{kp-\tau k_{x}}{k_{y}}.
\label{u_y,tau}
\end{equation}
\end{linenomath*}
In terms of the new variable $\tau$, we combine $\exp\left(  -u_{y}%
^{2}\right)  $ with the exponential function in
equation~(\ref{internal_integral}), using $\left(  k_{x}^{2}+k_{y}^{2}\right)
^{1/2}=k$, and then deform the corresponding contour of integration in the
complex $\tau$-plane to make this integration along the new real axis (the
required conditions will be discussed later). This yields
\begin{linenomath*}
\begin{equation}
R=\frac{i\pi^{3/2}}{k}\exp\left(  -p^{2}\right)  \int_{-\infty}^{\infty}%
\exp\left[  -\ \frac{\left(  k\tau-pk_{x}\right)  ^{2}}{k_{y}^{2}}\right]
\left[  1+\operatorname{erf}\left(  i\tau\right)  \right]  d\tau. 
\label{RR=}
\end{equation}
\end{linenomath*}
Equating this $R$ to the same function given by equation~(\ref{R_classical}),
we arrive at an unexpected and fairly simple integral relationship:
\begin{linenomath*}
\begin{equation}
\int_{-\infty}^{\infty}\exp\left[  -\ \frac{\left(  k\tau-pk_{x}\right)
^{2}}{k_{y}^{2}}\right]  \left[  1+\operatorname{erf}\left(  i\tau\right)
\right]  d\tau=1+\operatorname{erf}\left(  ip\right)  . 
\label{relka}
\end{equation}
\end{linenomath*}
In this form, equation~(\ref{relka}) contains a clear asymmetry with respect
to $k_{x}$ and $k_{y}$, which should not be the case. Besides, by its
derivation, it is clear that this relationship may have serious restrictions
associated with the fact that $k_{x}$ and $k_{y}$ have been implied to be real. With these
restrictions and asymmetry, the usefulness of this relationship may be
limited. It turns out, however, that after a simple generalization the 
applicability of a more general relation covers a much broader
domain of complex parameters.

We demonstrate this by introducing complex parameters $A$, $B$, and $C$, which
for equation~(\ref{relka}) reduce to $A=k/k_{y}$, $B=pk_{x}/k_{y}$, $C=i$, but
in general are not limited to these specific values. In terms of these
parameters, equation~(\ref{relka}) acquires a more general form:
\begin{linenomath*}
\begin{equation}
\int_{-\infty}^{\infty}\exp\left[  -\left(  Az-B\right)  ^{2}\right]  \left[
1+\operatorname{erf}\left(  Cz\right)  \right]  dz=\frac{\sqrt{\pi}}{A}\left[
\operatorname{sgn(\operatorname{Re}A)}+\operatorname{erf}\left(  \frac{BC}{\sqrt{A^{2}+C^{2}}}\right)  \right]  ,
\label{Final_integral_ABC}
\end{equation}
\end{linenomath*}
where $\operatorname{sgn}(x)$ is the standard signum function of a real argument $x$: 
$\operatorname{sgn}(x) = x/|x|$ if $x\neq 0$ and $\operatorname{sgn}(x)=0$ if $x=0$.
Below we prove equation (\ref{Final_integral_ABC}) directly, without reference to preceding
equation~(\ref{relka}).

Before performing this proof, we note that equation~(\ref{Final_integral_ABC})
should hold separately for the two separate parts of $\left[
1+\operatorname{erf}\left(  Cz\right)  \right]  $: those proportional to
``1'' on both sides of this equation,
\begin{linenomath*}
\begin{equation}
\int_{-\infty}^{\infty}\exp\left[  -\left(  Az-B\right)  ^{2}\right]
dz=\operatorname{sgn}(\operatorname{Re}A)\,\frac{\sqrt{\pi}}{A}. 
\label{without_error_functions}
\end{equation}
\end{linenomath*}
and those proportional to the error function,
\begin{linenomath*}
\begin{equation}
\int_{-\infty}^{\infty}\exp\left[  -\left(  Az-B\right)  ^{2}\right]
\operatorname{erf}\left(  Cz\right)  dz=\frac{\sqrt{\pi}}{A}\operatorname{erf}%
\left(  \frac{BC}{\sqrt{A^{2}+C^{2}}}\right) 
\label{with_error_functions}
\end{equation}
\end{linenomath*}
(recall that by the square root we always mean the branch with the positive real part). 
Equation~(\ref{without_error_functions}) presents a 
well-known integral whose convergence is restricted by $\left\vert
\arg(A)\right\vert <\pi/4$ or, equivalently,
\begin{linenomath*}
\begin{equation}
\operatorname{Re}(A^{2})>0. 
\label{Re(A^2)>0}
\end{equation}
\end{linenomath*}
Equation~(\ref{with_error_functions}), however, is non-standard and less
trivial. Popular tables of integrals, such as \citeA{Dwight:Tables88,Jeffrey:Handbook95,
Gradshteyn:Table07} do not present the exact analytic result of such integration. 
\citeA{Prudnikov:Integrals86} do this, but they present the result of such integration 
in terms of more complicated special functions, which are of no practical use. 

To proceed with our integral, 
we temporarily denote the LHS of equation~(\ref{with_error_functions}) as
\begin{linenomath*}
\begin{equation}
K(C)=\int_{-\infty}^{\infty}\exp\left[  -\left(  Az-B\right)  ^{2}\right]
\operatorname{erf}\left(  Cz\right)  dz 
\label{K(C)}
\end{equation}
\end{linenomath*}
and differentiate the function $K(C)$ with respect to the complex parameter $C$. This yields
\begin{linenomath*}
\[
\frac{dK}{dC}=\frac{2}{\sqrt{\pi}}\exp\left(  -\ \frac{B^{2}C^{2}}{A^{2}%
+C^{2}}\right)  \int_{-\infty}^{\infty}\exp\left\{  -\ \frac{\left[  \left(
A^{2}+C^{2}\right)  z-AB\right]  ^{2}}{A^{2}+C^{2}}\right\}  zdz.
\]
\end{linenomath*}
Transforming the integral in the RHS\ as follows,
\begin{linenomath*}
\begin{align}
&  \int_{-\infty}^{\infty}\exp\left\{  -\ \frac{\left[  \left(  A^{2}%
+C^{2}\right)  z-AB\right]  ^{2}}{A^{2}+C^{2}}\right\}  zdz\nonumber\\
&  =\int_{-\infty}^{\infty}\exp\left[  -\left(  A^{2}+C^{2}\right)  \left(
z-\frac{AB}{A^{2}+C^{2}}\right)  ^{2}\right]  \left(  z-\frac{AB}{A^{2}+C^{2}%
}\right)  dz\nonumber\\
&  +\frac{AB}{A^{2}+C^{2}}\int_{-\infty}^{\infty}\exp\left[  -\left(
A^{2}+C^{2}\right)  \left(  z-\frac{AB}{A^{2}+C^{2}}\right)  ^{2}\right]  dz,
\label{Transform}
\end{align}
\end{linenomath*}
we see that the contribution of the first integral goes away due to the oddity of the
integrand, while the second integral, under restriction
\begin{linenomath*}
\begin{equation}
\operatorname{Re}\left(  A^{2}+C^{2}\right)  > 0, 
\label{Re(A^2+C^2)>0}
\end{equation}
\end{linenomath*}
can be evaluated in a standard way by deforming the integration path in the
complex plane using, yielding $\sqrt{\pi}\left(  A^{2}+C^{2}\right)  ^{-1/2}$.
As a result, we obtain
\begin{linenomath*}
\begin{equation}
\frac{dK}{dC}=\frac{2AB}{\left(  A^{2}+C^{2}\right)  ^{3/2}}\exp\left(
-\ \frac{B^{2}C^{2}}{A^{2}+C^{2}}\right)  . 
\label{dK/dC}
\end{equation}
\end{linenomath*}
The definition of the error function is such that $\operatorname{erf}(0)=0$, so that,
according to equation~(\ref{K(C)}), we have $K(0)=0$. Using this as the
boundary condition, we integrate equation~(\ref{dK/dC}) and obtain
\begin{linenomath*}
\begin{align}
K(C)  &  =\int_{0}^{C}\frac{2AB}{\left(  A^{2}+C^{2}\right)  ^{3/2}}%
\exp\left(  -\ \frac{B^{2}C^{2}}{A^{2}+C^{2}}\right)  dC\nonumber\\
&  =\frac{\sqrt{\pi}}{A}\operatorname{erf}\left(  \frac{BC}{\sqrt{A^{2}+C^{2}%
}}\right)  . 
\label{K(C)=}
\end{align}
\end{linenomath*}
This result can be easily verified by differentiating the RHS of
equation~(\ref{K(C)=}) with respect to $C$. 

Equation~(\ref{K(C)=}) closes the
proof of equation~(\ref{with_error_functions}) and hence of fuller
equation~(\ref{Final_integral_ABC}) for complex parameters $A$, $B$, and $C$,
under constraints imposed by equations~(\ref{Re(A^2)>0}) and
(\ref{Re(A^2+C^2)>0}). In the general case, reduction of the original integral
to this form may involve a linear transform of the original variable in the
complex plane with the corresponding deformation of the integration path to
the real axis of $z$. This makes the total number of parameters (three)
somewhat redundant. To avoid this redundancy and simplify the result, we assign for the parameter
$C$ a specific value $C=i$, so that the two remaining parameters ($A$ and $B$)
could be determined from the original integral. Then
equation~(\ref{Final_integral_ABC}) reduces to a more compact relationship:
\begin{linenomath*}
\begin{equation}
\int_{-\infty}^{\infty}\exp\left[  -\left(  Az-B\right)  ^{2}\right]  \left(
1+\operatorname{erf}\left(  iz\right)  \right)  dz=\frac{\sqrt{\pi}}{A}\left[
\operatorname{sgn}(\operatorname{Re}A) + \operatorname{erf}\left(  \frac{iB}{\sqrt{A^{2}-1}}\right)  \right]  .
\label{Final_integral_AB}
\end{equation}
\end{linenomath*}
The corresponding integral convergence conditions given by separate
equations~(\ref{Re(A^2)>0}) and (\ref{Re(A^2+C^2)>0}) reduce now to a single
condition:
\begin{linenomath*}
\begin{equation}
\operatorname{Re}(A^{2}-1)>0. 
\label{Re(A^2-1)>0}
\end{equation}
\end{linenomath*}

Recalling the definition of the standard function $w(x)$ given by
equation~(\ref{ww}) and using transformations similar to those in
equation~(\ref{Transform}), we rewrite equation~(\ref{Final_integral_AB}) as
\begin{linenomath*}
\begin{align}
&  \int_{-\infty}^{\infty}\exp\left[  -\left(  A^{2}-1\right)  
\left(z-\frac{AB}{A^{2}-1}\right)  ^{2}\right]  w\left(  z\right)  dz\nonumber \\
&  =\frac{\sqrt{\pi}}{A}\ \tilde{w}\left(  \frac{B}{\sqrt{A^{2}-1}}\right),
\label{Final_integral_w()}
\end{align}
\end{linenomath*}
where the extended $w$-function, $\tilde{w}(z)$, is given by
\begin{linenomath*}
\begin{equation}
\tilde{w}\left(  \frac{B}{\sqrt{A^{2}-1}}\right)  =\left\{
\begin{array}
[c]{ccc}%
w\left(  {B}/{\sqrt{A^{2}-1}}\,\right)   & \text{if} & \operatorname{Re}%
A>0,\\
-w\left(  - {B}/{\sqrt{A^{2}-1}}\,\right)   & \text{if} & \operatorname{Re}A<0
\label{tilde_w}
\end{array}
\right.
\end{equation}
\end{linenomath*}
and we ignore the degenerate case of $\operatorname{Re}A=0$. Here ``$-w(-z)$'' can be identically represented as
\begin{linenomath*}
\begin{equation*}
-w(-z)=e^{-z^{2}}(-1+\operatorname{erf}(iz))
=w(z)-2e^{-z^{2}},
\end{equation*}
\end{linenomath*}
which means that $-w(-z)$ is shifted from $w(z)$ by a simple Gaussian function.

Due to the proportionality of the $w$ and $Z$-functions, 
the two $w$-functions on both sides of equation~(\ref{Final_integral_w()}) can be
simultaneously replaced by the corresponding $Z$-functions:
\begin{linenomath*}
\begin{align}
& \int_{-\infty}^{\infty}\exp\left(  -(A^{2}-1)\left(  z-\frac{AB}{A^{2}%
-1}\right)  ^{2}\right)  Z(z)dz\nonumber\\
& =\frac{\sqrt{\pi}}{A}\ \tilde{Z}\left(  \frac{B}{\sqrt{A^{2}-1}}\right),
\label{Final_integral_Z()}%
\end{align}
where
\begin{linenomath*}
\begin{equation}
\tilde{Z}\left(  \frac{B}{\sqrt{A^{2}-1}}\right)  =\left\{
\begin{array}
[c]{ccc}%
Z\left(  {B}/{\sqrt{A^{2}-1}}\,\right)   & \text{if} & \operatorname{Re}%
A>0,\\
-Z\left(  - {B}/{\sqrt{A^{2}-1}}\,\right)   & \text{if} & \operatorname{Re}A<0
\label{tilde_Z}
\end{array}
\right.
\end{equation}
\end{linenomath*}
and ``$-Z(-z)$'' can be alternatively represented as
\begin{equation*}
-Z(-z)=i\sqrt{\pi}\,e^{-z^{2}}(-1+\operatorname{erf}(iz))
=Z(z)-2i\sqrt{\pi}\,e^{-z^{2}}.
\end{equation*}
\end{linenomath*}

We emphasize that using proper linear transformation of the integration
variable and contour deformation, any integrable convolution of the Gaussian and $w$ or $Z$-functions
with two different complex arguments can always be reduced to the form given by
equations~(\ref{Final_integral_AB}), (\ref{Final_integral_w()}), or (\ref{Final_integral_Z()}) with only two adjustable
parameters $A$ and $B$.

Now we evaluate another useful integral,
\begin{linenomath*}
\begin{equation}
I_{1}\equiv\int_{-\infty}^{\infty}z\exp\left[  -(A^{2}-1)\left(  z-\frac
{AB}{A^{2}-1}\right)  ^{2}\right]w(z)dz, 
\label{I_1}
\end{equation}
\end{linenomath*}
which differs from that in equation~(\ref{Final_integral_w()}) by the additional multiplier $z$ in the integrand.
To proceed with this integral evaluation, we temporarily introduce an auxiliary
integral,
\begin{linenomath*}
\begin{equation*}
I_{2}\equiv(A^{2}-1)\int_{-\infty}^{\infty}\left(  z-\frac{AB}{A^{2}-1}\right)\exp\left[  -(A^{2}-1)\left(  z-\frac
{AB}{A^{2}-1}\right)  ^{2}\right]w(z)dz
\end{equation*}
\end{linenomath*}
On the one hand, this definition combined with
equation~(\ref{Final_integral_w()}) gives
\begin{linenomath*}
\begin{equation}
I_{2}=(A^{2}-1)I_{1}-B\sqrt{\pi}\,w\left(  \frac{B}{\sqrt{A^{2}-1}}\right)  .
\label{I_2_2}
\end{equation}
\end{linenomath*}
On the other hand, we can evaluate $I_{2}$ directly, using integration by
parts,
\begin{linenomath*}
\begin{align*}
I_{2}  &  =-\ \frac{1}{2}\int_{-\infty}^{\infty}\frac{d}{dz}\exp\left[
-(A^{2}-1)\left(  z-\frac{AB}{A^{2}-1}\right)  ^{2}\right]  w\left(  z\right)
dz\\
&  =\frac{1}{2}\int_{-\infty}^{\infty}\exp\left[  -(A^{2}-1)\left(
z-\frac{AB}{A^{2}-1}\right)  ^{2}\right]  \frac{dw}{dz}\ dz,
\end{align*}
\end{linenomath*}
where, as usual, we imply a good behavior of the integrand at $\pm\infty$.
Then, using the easily verifiable differential equation for the $w$-function,
\begin{linenomath*}
\[
\frac{dw}{dz}=2\left(  \frac{i}{\sqrt{\pi}}-zw\left(  z\right)  \right)  ,
\]
\end{linenomath*}
we obtain
\begin{linenomath*}
\begin{align}
&  I_{2}=\int_{-\infty}^{\infty}\exp\left[  -(A^{2}-1)\left(  z-\frac
{AB}{A^{2}-1}\right)  ^{2}\right] \nonumber\\
&  \times\left(  \frac{i}{\sqrt{\pi}}-zw\left(  z\right)  \right)  dz=\frac
{i}{\sqrt{A^{2}-1}}-I_{1}. 
\label{I_2_1}
\end{align}
\end{linenomath*}
Comparison of equations~ (\ref{I_2_2}) and (\ref{I_2_1}) yields the explicit
expression for $I_{1}$, i.e., another useful integral relation:
\begin{linenomath*}
\begin{align}
&  \int_{-\infty}^{\infty}z\exp\left[  -(A^{2}-1)\left(  z-\frac{AB}{A^{2}%
-1}\right)  ^{2}\right]w(z)  dz\nonumber\\
&  =\frac{B\sqrt{\pi}}{A^{2}}\ \tilde{w}\left(  \frac{B}{\sqrt{A^{2}-1}}\right)
+\frac{i}{A^{2}\sqrt{A^{2}-1}}. 
\label{Int(z..)}
\end{align}
\end{linenomath*}
In terms of the $Z$-functions, the equivalent integral relation is given by
\begin{linenomath*}
\begin{align}
&  \int_{-\infty}^{\infty}z\exp\left(  -(A^{2}-1)\left(  z-\frac{AB}{A^{2}%
-1}\right)  ^{2}\right)  Z(z)dz\nonumber\\
&  =\frac{B\sqrt{\pi}}{A^{2}}\ \tilde{Z}\left(  \frac{B}{\sqrt{A^{2}-1}%
}\right)  -\frac{\sqrt{\pi}}{A^{2}\sqrt{A^{2}-1}}.
\label{Int(Z...)}
\end{align}
\end{linenomath*}
One can easily verify that these integrals remain invariant with respect to the simultaneous change of 
$A$, $B$ to $-A$, $-B$, respectively. In doing so, we should bear in mind that this change toggles the piece-wise conditions 
of the extended function $\tilde{w}$, as defined in equation~(\ref{tilde_w}), because
$\operatorname{Re}(-A)>0$ becomes $\operatorname{Re}A<0$ and vice versa.
 
Thus, any integral convolution of a Gaussian function with an error function (or the related $w$ and $Z$-functions) of  
linearly independent complex arguments, either with or without an additional weight function $z$,  can 
always be reduced to the two non-conventional integrals given by equations~(\ref{without_error_functions}), 
(\ref{with_error_functions}), (\ref{Final_integral_w()}) 
and (\ref{Int(z..)}) (or equivalent equations (\ref{Final_integral_Z()}) and (\ref{Int(Z...)}) 
in terms the $Z$-function). This reduction may require a simple linear transformation of the integration 
variable $z$, followed by the standard deformation of 
the integration contour in the new complex plane to make it the real axis. 
Since the integrand functions are analytical, the main restriction 
is the good behavior of the integrand at $|z|\rightarrow\infty$ (the integrand there should tend to zero faster 
than $1/|z|$, providing the convergence of the integral with no contribution from the infinitely large contour). 
It is important that the results of these integrations are also represented in terms 
of the $w$ or $Z$ functions, albeit with simple piece-wise conditions.
Equations~(\ref{Final_integral_w()}) and (\ref{Int(z..)}), or equivalent equations 
(\ref{Final_integral_Z()}) and (\ref{Int(Z...)}), play a crucial part in our calculations 
of the ion contribution to the general kinetic wave dispersion relation 
(see below). 

\subsection{Solving for $F_{\omega\vec{k}}^{(1)}$ and evaluating its
contribution to $\eta_{\omega\vec{k}}^{i}$%
\label{Solving for F_1 and evaluating its contribution to eta_i}}

Here we find the solution for $F_{\omega\vec{k}}^{(1)}$ and then evaluate the
corresponding double integral. Equation~(\ref{Eq_forF^1}) can be rewritten as
\begin{linenomath*}
\begin{equation}
\frac{\partial F_{\omega\vec{k}}^{(1)}}{\partial V_{ix}}+\frac{i}{a_{i0}%
}(k_{x}V_{ix}+k_{y}V_{iy}\mathbf{-}\omega_{i})F_{\omega\vec{k}}^{(1)}=\frac
{P}{a_{i0}}\exp\left(  -\ \frac{V_{ix}^{2}+V_{iy}^{2}}{2V_{Tn}^{2}}\right)  ,
\label{F(1)_viua_rho,P_Review}
\end{equation}
\end{linenomath*}
where
\begin{linenomath*}
\begin{equation}
P\equiv\frac{\nu_{in}n_{0}\eta_{\omega\vec{k}}^{i}}{2\pi V_{Tn}^{2}}.
\label{P}
\end{equation}
\end{linenomath*}

Before proceeding with solving general differential equation~(\ref{F(1)_viua_rho,P_Review}), we
consider its no-field\ limit of $a_{i0}\rightarrow0$. In this limit, the term
$\partial F_{\omega\vec{k}}^{(1)}/\partial V_{ix}$, can be neglected compared
to the other, infinitely large terms, so that $F_{\omega\vec{k}}^{(1)}$
reduces to
\begin{linenomath*}
\[
F_{\omega\vec{k}}^{(1)}=-\ \frac{iP}{k_{x}V_{ix}+k_{y}V_{iy}\mathbf{-}\omega_{i}}\exp
\left(  -\ \frac{V_{ix}^{2}+V_{iy}^{2}}{2V_{Tn}^{2}}\right)  .
\]
\end{linenomath*}
Assuming $k_{y}=0$ and using equations~(\ref{II(a)})--(\ref{ww}) leading to
\begin{linenomath*}
\[
\int_{-\infty}^{\infty}\frac{e^{-\xi^{2}}}{\xi-p}\,d\xi=i\pi \tilde{w}(p),
\]
\end{linenomath*}
we easily obtain
\begin{linenomath*}
\begin{align}
\iint_{-\infty}^{\infty}F_{\omega\vec{k}}^{(1)}dV_{ix}dV_{iy}  &
=\frac{\sqrt{2}\,\pi^{3/2}PV_{Tn}}{k}\ \tilde{w}\left(  \frac{\omega_{i}}{\sqrt{2}\,kV_{Tn}%
}\right) \nonumber\\
&  =\sqrt{\frac{\pi}{2}}\,\frac{\nu_{in}n_{0}\eta_{\omega\vec{k}}^{i}}{kV_{Tn}%
}\ \tilde{w}\left(  \frac{\omega_{i}}{\sqrt{2}\,kV_{Tn}}\right)  
\label{Int_F^1_classical}
\end{align}
\end{linenomath*}
We will use this equation for comparison with the corresponding limit
following from the solution of general equation\ (\ref{F(1)_viua_rho,P_Review}).

Now we proceed with the general case of arbitrary $a_{i0}$. We start from the most important case of $\operatorname{Re}(R_i)>0$. 
After a long and detailed description of the corresponding solution, in the end of this section, we will outline the opposite case of $\operatorname{Re}(R_i)<0$.

The case of $\operatorname{Re}(R_i)>0$ corresponds to $\gamma>-\nu_{in}$. This case covers all linearly 
unstable waves and probably the most of (if not all) other linear waves. 
In this case, a simple analysis of linear differential equation~(\ref{F(1)_viua_rho,P_Review}) suggests choice of the asymptotic
boundary condition of $F_{\omega\vec{k}}^{(1)}\rightarrow0$ as $V_{x}\rightarrow-\infty$, resulting in
\begin{linenomath*}
\begin{align}
&  F_{\omega\vec{k}}^{(1)}(V_{ix},V_{iy})\nonumber\\
&  =\frac{P}{a_{i0}}\int_{-\infty}^{V_{ix}}\exp\left[  -\ \frac{U_{x}%
^{2}+V_{iy}^{2}}{2V_{Tn}^{2}}+\frac{ik_{x}\left(  U_{x}^{2}-V_{ix}^{2}\right)
}{2a_{i0}}+\frac{i(k_{y}V_{iy}\mathbf{-}\omega_{i})\left(  U_{x}\mathbf{-}%
V_{ix}\right)  }{a_{i0}}\right]  dU_{x}. 
\label{F^(1)=_Review}
\end{align}
\end{linenomath*}
To obtain the corresponding contribution to $\eta_{\omega\vec{k}}^{i}$ in
equation~(\ref{relation_Review}), we should integrate $F_{\omega\vec{k}}%
^{(1)}$ with respect to both $V_{ix}$ and $V_{iy}$ over the entire 2-D
velocity domain. Integrating first with respect to $V_{iy}$, for the
corresponding $V_{iy}$-dependent multiplier we obtain
\begin{linenomath*}
\begin{align*}
&  \int_{-\infty}^{\infty}\exp\left[  -\ \frac{V_{iy}^{2}}{2V_{Tn}^{2}}%
+\frac{ik_{y}V_{iy}\left(  U_{x}-V_{ix}\right)  }{a_{i0}}\right]  dV_{iy}\\
&  =\sqrt{2\pi}\,V_{Tn}\exp\left[  -\ \frac{k_{y}^{2}V_{Tn}^{2}\left(
U_{x}-V_{ix}\right)  ^{2}}{2a_{i0}^{2}}\right]  .
\end{align*}
\end{linenomath*}
To evaluate this integral, we have changed the integration variable to a new
complex variable in order to reduce the integrand to an easily integrable
Gaussian function. Then, in the new complex plane we have deformed the
corresponding integration path using a standard approach by making the final
integration path to lie fully along the new real axis. As usual, this has been
done by using the good analytical properties of the integrand, combined with
the proper behavior of the Gaussian function at an infinite contour within the
allowed domains of the complex plane that provide the integral convergence. We
will do similar manipulations in a few other places without dwelling on
such standard and well-known details.

Further, we have to integrate over $V_{ix}$. When doing this, it is more
beneficial to swap the order of integrations between $V_{ix}$ and
$U_{x}$. Adjusting the internal integration limits accordingly, we obtain
\begin{linenomath*}
\begin{align}
&  \iint_{-\infty}^{\infty}F_{\omega\vec{k}}^{(1)}(V_{ix},V_{iy}%
)dV_{iy}dV_{ix}=\frac{PV_{Tn}\sqrt{2\pi}}{a_{i0}}\int_{-\infty}^{\infty
}dU_{x}\exp\left(  -\ \frac{U_{x}^{2}}{2V_{Tn}^{2}}\right) \nonumber\\
&  \times\int_{U_{x}}^{\infty}\exp\left[  -\ \frac{k_{y}^{2}V_{Tn}^{2}\left(
V_{ix}-U_{x}\right)  ^{2}}{2a_{i0}^{2}}-\frac{ik_{x}\left(  V_{ix}^{2}%
-U_{x}^{2}\right)  }{2a_{i0}}+\frac{i\left(  V_{ix}-U_{x}\right)  \omega_{i}%
}{a_{i0}}\right]  dV_{ix}. 
\label{begin}
\end{align}
\end{linenomath*}
Redistributing terms within the exponents as follows,
\begin{linenomath*}
\begin{align*}
&  -\ \frac{k_{y}^{2}V_{Tn}^{2}\left(  V_{ix}-U_{x}\right)  ^{2}}{2a_{i0}^{2}%
}-\frac{ik_{x}\left(  V_{ix}^{2}-U_{x}^{2}\right)  }{2a_{i0}}+\frac{i\left(
V_{ix}-U_{x}\right)  \omega_{i}}{a_{i0}}\\
&  =-\ \frac{k_{y}^{2}V_{Tn}^{2}+ik_{x}a_{i0}}{2a_{i0}^{2}}\left(
V_{ix}-\frac{k_{y}^{2}V_{Tn}^{2}U_{x}+i\omega_{i}a_{i0}}{k_{y}^{2}V_{Tn}%
^{2}+ik_{x}a_{i0}}\right)  ^{2}-\frac{\left(  \omega_{i}-k_{x}U_{x}\right)
^{2}}{2\left(  k_{y}^{2}V_{Tn}^{2}+ik_{x}a_{i0}\right)  }
\end{align*}
\end{linenomath*}
we reduce equation~(\ref{begin}) to
\begin{linenomath*}
\begin{align}
&  \iint_{-\infty}^{\infty}F_{\omega\vec{k}}^{(1)}(V_{ix},V_{iy}%
)dV_{iy}dV_{ix}=\frac{PV_{Tn}\sqrt{2\pi}}{a_{i0}}\nonumber\\
&  \times\int_{-\infty}^{\infty}\exp\left[  -\ \frac{U_{x}^{2}}{2V_{Tn}^{2}%
}-\frac{\left( \omega_{i}-k_{x}U_{x}\right)  ^{2}}{2\left(  k_{y}^{2}V_{Tn}%
^{2}+ik_{x}a_{i0}\right)  }\ \right]  dU_{x}\nonumber\\
&  \times\int_{U_{x}}^{\infty}\exp\left[  -\ \frac{k_{y}^{2}V_{Tn}^{2}%
+ik_{x}a_{i0}}{2a_{i0}^{2}}\left(  V_{ix}-\frac{k_{y}^{2}V_{Tn}^{2}U_{x}%
+i\omega_{i}a_{i0}}{k_{y}^{2}V_{Tn}^{2}+ik_{x}a_{i0}}\right)  ^{2}\right]
dV_{ix}, 
\label{promezh}
\end{align}
\end{linenomath*}
Then, in accord with the definition of the $w$-function given by
equation~(\ref{erf(x)}), the internal integral gives
\begin{linenomath*}
\begin{align*}
&  \int_{U_{x}}^{+\infty}\exp\left[  -\ \frac{k_{y}^{2}V_{Tn}^{2}+ik_{x}%
a_{i0}}{2a_{i0}^{2}}\left(  V_{ix}-\frac{k_{y}^{2}V_{T}^{2}U_{x}+iR
_{i}a_{i0}}{k_{y}^{2}V_{Tn}^{2}+ik_{x}a_{i0}}\right)  ^{2}\right]  dV_{ix}\\
&  =a_{i0}\sqrt{\frac{\pi}{2(k_{y}^{2}V_{Tn}^{2}+ik_{x}a_{i0})}}\exp\left[
\frac{\left(\omega_{i}-k_{x}U_{x}\right)  ^{2}}{2(k_{y}^{2}V_{Tn}^{2}%
+ik_{x}a_{i0})}\right]  w\left[  \frac{\omega_{i}-k_{x}U_{x}}{\sqrt{2(k_{y}%
^{2}V_{Tn}^{2}+ik_{x}a_{i0})}}\right].
\end{align*}
\end{linenomath*}
Here, and elsewhere across the text, we choose the branch of the square
root with the positive real part. As a result, equation~(\ref{promezh}) yields
\begin{linenomath*}
\begin{align}
&  \iint_{-\infty}^{\infty}F_{\omega\vec{k}}^{(1)}(V_{ix},V_{iy}%
)dV_{iy}dV_{ix}=\frac{\pi PV_{Tn}}{\sqrt{k_{y}^{2}V_{Tn}^{2}+ik_{x}a_{i0}}%
}\nonumber\\
&  \times\int_{-\infty}^{\infty}\exp\left(  -\ \frac{U_{x}^{2}}{2V_{Tn}^{2}%
}\right)  w\left[  \frac{\omega_{i}-k_{x}U_{x}}{\sqrt{2(k_{y}^{2}V_{Tn}%
^{2}+ik_{x}a_{i0})}}\right]  dU_{x} 
\label{promka}
\end{align}
\end{linenomath*}
To evaluate the remaining integral, we make the argument of the error function
a new integration variable
\begin{linenomath*}
\begin{equation}
z=\frac{\omega_{i}-k_{x}U_{x}}{\sqrt{2(k_{y}^{2}V_{Tn}^{2}+ik_{x}a_{i0})}}
\label{z}
\end{equation}
\end{linenomath*}
and then deform the integration path is such a way that it becomes the real
axis in the complex $z$-plane. Then equation~(\ref{promka}) becomes
\begin{linenomath*}
\begin{align}
&  \iint_{-\infty}^{\infty}F_{\omega\vec{k}}^{(1)}(V_{ix},V_{iy}%
)dV_{iy}dV_{ix}\nonumber\\
&  =\frac{\sqrt{2}\,\pi PV_{Tn}}{k_{x}}\int_{-\infty}^{\infty}\exp\left[
-\ \frac{k_{y}^{2}V_{Tn}^{2}+ik_{x}a_{i0}}{k_{x}^{2}V_{Tn}^{2}}\left(
z-\frac{\omega_{i}}{\sqrt{2(k_{y}^{2}V_{Tn}^{2}+ik_{x}a_{i0})}}\right)
^{2}\right]  w(z)dz. 
\label{promka_1}
\end{align}
\end{linenomath*}
Designating
\begin{linenomath*}
\[
\frac{k_{y}^{2}V_{Tn}^{2}+ik_{x}a_{i0}}{k_{x}^{2}V_{Tn}^{2}}=A^{2}%
-1,\qquad\frac{\omega_{i}}{\sqrt{2(k_{y}^{2}V_{Tn}^{2}+ik_{x}a_{i0})}}=\frac
{AB}{A^{2}-1},
\]
\end{linenomath*}
we obtain
\begin{linenomath*}
\begin{subequations}
\label{AABB}
\begin{align}
A  &  =\frac{\sqrt{k^{2}V_{Tn}^{2}+ik_{x}a_{i0}}}{|k_{x}|V_{Tn}},\label{AA}\\
B  &  =\frac{\omega_{i}}{|k_{x}|V_{Tn}}\sqrt{\frac{k_{y}^{2}V_{Tn}^{2}%
+ik_{x}a_{i0}}{2(k^{2}V_{Tn}^{2}+ik_{x}a_{i0})}}, \label{BB}%
\end{align}
\end{subequations}
\end{linenomath*}
where $k^{2}=k_{x}^{2}+k_{y}^{2}$ (recall that in the assumed coordinate
system $k_{z}=0$). Now we can apply to equation~(\ref{promka_1}) the integral
relation given by equation~(\ref{Final_integral_w()}). Then, recalling the definition of $P$ given by equation~(\ref{P}), we
obtain
\begin{linenomath*}
\begin{equation}
\iint_{-\infty}^{\infty}F_{\omega\vec{k}}^{(1)}dV_{ix}dV_{iy}=\frac{\sqrt
{\pi}\,\nu_{in}n_{0}\eta_{\omega\vec{k}}^{i}}{\sqrt{2(k^{2}V_{Tn}^{2}+i\vec
{k}\cdot\vec{a}_{i0})}}\ w\left(  \frac{\omega_{i}}{\sqrt{2(k^{2}V_{Tn}^{2}+i\vec{k}\cdot\vec{a}_{i0})}}\right)  , 
\label{Int^(1)_Final}
\end{equation}
\end{linenomath*}
where we have returned to the coordinate-invariant format, $k_{x}a_{i0}%
=\vec{k}\cdot\vec{a}_{i0}$. In the limit of $\vec{a}_{i0}\rightarrow0$,
equation~(\ref{Int^(1)_Final}) reduces to
equation~(\ref{Int_F^1_classical}).

Now we outline the solution for $F_{\omega\vec{k}}^{(1)}$ in the opposite case of $\operatorname{Re}(R_i)<0$, 
corresponding to $\gamma < -\nu_{in}$. Although this case may never be realized for any linear waves, 
we need to include it for completeness of the wave dispersion relation. 

The major difference from the $\operatorname{Re}(R_i)>0$ case is that if $\operatorname{Re}(R_i)<0$
then equation~(\ref{F(1)_viua_rho,P_Review}) should be solved using a different 
asymptotic boundary condition, namely $F_{\omega\vec{k}}^{(1)}\rightarrow0$ as $V_{x}\rightarrow\infty$. 
After making the same steps that lead to equation~(\ref{begin}), we arrive now at a slightly modified solution,
\begin{linenomath*}
\begin{align}
&  \iint_{-\infty}^{\infty}F_{\omega\vec{k}}^{(1)}(V_{ix},V_{iy}%
)dV_{iy}dV_{ix}=-\ \frac{PV_{Tn}\sqrt{2\pi}}{a_{i0}}\int_{-\infty}^{\infty
}dU_{x}\exp\left(  -\ \frac{U_{x}^{2}}{2V_{Tn}^{2}}\right) \nonumber\\
&  \times\int_{-\infty}^{U_{x}}\exp\left[  -\ \frac{k_{y}^{2}V_{Tn}^{2}\left(
V_{ix}-U_{x}\right)  ^{2}}{2a_{i0}^{2}}-\frac{ik_{x}\left(  V_{ix}^{2}%
-U_{x}^{2}\right)  }{2a_{i0}}+\frac{i\left(  V_{ix}-U_{x}\right)  \omega_{i}%
}{a_{i0}}\right]  dV_{ix}.
\label{begin-}
\end{align}
\end{linenomath*}
In this equation, the integral $\int_{U_x}^{\infty}(\cdots)dV_{ix}$ from equation~(\ref{begin}) 
is replaced by the same integral, but with different integration limits, $-\,\int_{-\infty}^{U_x}(\cdots)dV_{ix}$. Making the 
same steps as described above for the $\operatorname{Re}(R_i)>0$ case, we arrive eventually at an expression 
\begin{linenomath*}
\begin{equation}
\iint_{-\infty}^{\infty}F_{\omega\vec{k}}^{(1)}dV_{ix}dV_{iy}=-\ \frac{\sqrt
{\pi}\,\nu_{in}n_{0}\eta_{\omega\vec{k}}^{i}}{\sqrt{2(k^{2}V_{Tn}^{2}+i\vec
{k}\cdot\vec{a}_{i0})}}\ w\left(-\ \frac{\omega_{i}}{\sqrt{2(k^{2}V_{Tn}^{2}+i\vec{k}\cdot\vec{a}_{i0})}}\right) 
\label{Int^(1)_Final_-}
\end{equation}
\end{linenomath*}
where the function $w(z)$ in the RHS of equation~(\ref{Int^(1)_Final}) is replaced by $-w(-z)$; 
recall that $-w(-z)=w(z)-2\exp(-z^{2})$. The existence of two different functional expressions, 
depending on the sign of $\operatorname{Re}(R_i)$, is the same as in equations~(\ref{Iw}) 
and (\ref{tilde_w}), although the latter has a different parameter condition that separates them.

\subsection{Solving for $F_{\omega\vec{k}}^{(2)}$ and evaluating its
contribution to $\eta_{\omega\vec{k}}^{i}$%
\label{Solving for F_2 and evaluating its contribution to eta_i}}

Here we find the solution for $F_{\omega\vec{k}}^{(2)}$ and evaluate the
corresponding contribution to $\eta_{\omega\vec{k}}^{i}$. Similarly to the 
above treatment for $F_{\omega\vec{k}}^{(1)}$, we provide detailed calculations 
for the major case of $\operatorname{Re}(R_i)>0$ (i.e., $\gamma>-\nu_{in}$), and then present the 
result for the heavily damped waves of $\operatorname{Re}(R_i)<0$ ($\gamma<-\nu_{in}$).

Equation~(\ref{Eq_forF^2}) can be rewritten as
\begin{linenomath*}
\begin{align}
&  \frac{\partial F_{\omega\vec{k}}^{(2)}}{\partial V_{ix}}+\frac{i}{a_{i0}%
}(k_{x}V_{ix}+k_{y}V_{iy}\mathbf{-}\omega_{i})F_{\omega\vec{k}}^{(2)}\nonumber\\
&  =\frac{iV_{Tn}^{2}}{\beta_{T}a_{i0}}\ \phi_{\omega\vec{k}}\left(
k_{x}\ \frac{\partial}{\partial V_{ix}}+k_{y}\ \frac{\partial}{\partial
V_{iy}}\right)  \int_{-\infty}^{\infty}f_{i0}dV_{iz},
\label{F(2)_via_rho,P_Review}
\end{align}
\end{linenomath*}
where, according to Eq.~(\ref{f_i0_Gaussian}),
\begin{linenomath*}
\begin{equation}
\int_{-\infty}^{\infty}f_{i0}dV_{iz}=\frac{n_{0}}{2\pi V_{Tx}V_{\perp
}}\exp\left(  -~\frac{(V_{ix}-U_{0})^{2}}{2V_{Tx}^{2}}-\frac{V_{iy}%
^{2}}{2V_{Ty}^{2}}\right)  . 
\label{Int_dz_BGK}
\end{equation}
\end{linenomath*}
Temporarily denoting the RHS of Eq.~(\ref{F(2)_via_rho,P_Review}) by
$R(V_{ix})$,
\begin{linenomath*}
\[
R(V_{ix})\equiv\frac{iV_{Tn}^{2}}{\beta_{T}a_{i0}}\ \phi_{\omega\vec{k}%
}\left(  k_{x}\ \frac{\partial}{\partial V_{ix}}+k_{y}\ \frac{\partial
}{\partial V_{iy}}\right)  \int_{-\infty}^{\infty}f_{i0}dV_{iz},
\]
\end{linenomath*}
we obtain the solution for $F_{\omega\vec{k}}^{(2)}$,
\begin{linenomath*}
\begin{align*}
&  F_{\omega\vec{k}}^{(2)}(V_{ix})=\exp\left\{  -\ \frac{i}{a_{i0}}\left[
\frac{k_{x}V_{ix}^{2}}{2}+(k_{y}V_{iy}\mathbf{-}\omega_{i})V_{ix}\right]
\right\} \\
&  \times\int_{-\infty}^{V_{ix}}R(U_{x})\exp\left\{  \frac{i}{a_{i0}}\left[
\frac{k_{x}U_{x}^{2}}{2}+(k_{y}V_{iy}\mathbf{-}\omega_{i})U_{x}\right]
\right\}  dU_{x},
\end{align*}
\end{linenomath*}
that satisfies the boundary condition of $\lim_{V_{ix}\rightarrow-\infty}F_{\omega\vec{k}}^{(2)}(V_{ix})=0$.

In order to find $F_{\omega\vec{k}}^{(2)}$, we need to solve differential
equation~(\ref{Eq_forF^2}), where $\int_{-\infty}^{\infty}f_{i0}dV_{iz}$ is
given by equation~(\ref{Int_f_i0_dV_iz}). Under assumption of $\lim
_{V_{ix}\rightarrow-\infty}F_{\omega\vec{k}}^{(2)}=0$, the formal solution can
be written as
\begin{linenomath*}
\begin{align}
&  F_{\omega\vec{k}}^{(2)}=Q\int_{-\infty}^{V_{ix}}dU_{x}\left(  \frac
{k_{x}(U_{x}-U_{0})}{V_{Tx}^{2}}+\frac{k_{y}V_{iy}}{V_{Ty}^{2}%
}\right) \nonumber\\
&  \times\exp\left\{  -~\frac{(U_{x}-U_{0})^{2}}{2V_{Tx}^{2}}%
-\frac{V_{iy}^{2}}{2V_{Ty}^{2}}+\frac{i}{a_{i0}}\left[  \frac{k_{x}\left(
U_{x}^{2}-V_{ix}^{2}\right)  }{2}+(k_{y}V_{iy}-\omega_{i})\left(  U_{x}%
-V_{ix}\right)  \right]  \right\}  , 
\label{F^2_solution}%
\end{align}
\end{linenomath*}
where
\begin{linenomath*}
\begin{equation}
Q\equiv-\ \frac{iV_{Tn}^{2}n_{0}\phi_{\omega\vec{k}}}{2\pi\beta_{T}%
a_{i0}V_{Tx}V_{Ty}}. 
\label{Q}
\end{equation}
\end{linenomath*}
To obtain the ion contribution to the linear dispersion relations, we need to
integrate $F_{\omega\vec{k}}^{(2)}$ over the two remaining components in the
ion velocity space ($V_{ix}$ and $V_{iy}$). First, we integrate $F_{\omega
\vec{k}}^{(2)}$ over $V_{iy}$. This straightforward integration yields
\begin{linenomath*}
\begin{align}
&  \int_{-\infty}^{\infty}F_{\omega\vec{k}}^{(2)}dV_{iy}\nonumber\\
&  =\sqrt{2\pi}V_{T}Q\int_{-\infty}^{V_{ix}}dU_{x}\left[  \frac{k_{x}%
(U_{x}-U_{0})}{V_{Tx}^{2}}+\frac{ik_{y}^{2}\left(  U_{x}-V_{ix}\right)
}{a_{i0}}\right] \label{Integration_over_V_y}\\
&  \times\exp\left\{  -~\frac{(U_{x}-U_{0})^{2}}{2V_{Tx}^{2}}+\frac
{i}{a_{i0}}\left[  \frac{k_{x}\left(  U_{x}^{2}-V_{ix}^{2}\right)  }{2}%
-\omega_{i}\left(  U_{x}-V_{ix}\right)  \right]  -\frac{k_{y}^{2}V_{Ty}%
^{2}\left(  U_{x}-V_{ix}\right)  ^{2}}{2a_{i0}^{2}}\right\}  .\nonumber
\end{align}
\end{linenomath*}
Second, we need to integrate this expression over $V_{ix}$. This will be a
multi-step procedure.

We start by introducing an intermediate quantity
\begin{linenomath*}
\begin{align*}
&  M\equiv\int_{-\infty}^{\infty}dV_{ix}\int_{-\infty}^{V_{ix}}dU_{x}\left[
\frac{k_{x}(U_{x}-U_{0})}{V_{Tx}^{2}}+\frac{ik_{y}^{2}\left(
U_{x}-V_{ix}\right)  }{a_{i0}}\right] \\
&  \times\exp\left\{  -~\frac{(U_{x}-U_{0})^{2}}{2V_{Tx}^{2}}+\frac
{i}{a_{i0}}\left[  \frac{k_{x}\left(  U_{x}^{2}-V_{ix}^{2}\right)  }{2}%
-\omega_{i}\left(  U_{x}-V_{ix}\right)  \right]  -\frac{k_{y}^{2}V_{Ty}%
^{2}\left(  U_{x}-V_{ix}\right)  ^{2}}{2a_{i0}^{2}}\right\}  ,
\end{align*}
\end{linenomath*}
so that
\begin{linenomath*}
\begin{equation}
\int_{-\infty}^{\infty}F_{\omega\vec{k}}^{(2)}dV_{ix}dV_{iy}=\sqrt{2\pi}\,V_{T}QM. 
\label{IntF_via_M}
\end{equation}
\end{linenomath*}
In order to simplify further calculations, we swap the order of integration
between $V_{ix}$ and $U_{x}$,
\begin{linenomath*}
\begin{align}
&  M=\int_{-\infty}^{\infty}dU_{x}\exp\left[  -~\frac{(U_{x}-U_{0})^{2}}
{2V_{Tx}^{2}}\right]  \int_{U_{x}}^{\infty}dV_{ix}\left[  \frac
{k_{x}(U_{x}-U_{0})}{V_{Tx}^{2}}+\frac{ik_{y}^{2}}{a_{i0}}\left(
U_{x}-V_{ix}\right)  \right] \nonumber\\
&  \times\exp\left\{  -\ \frac{k_{y}^{2}V_{Ty}^{2}}{2a_{i0}^{2}}
\left(U_{x}-V_{ix}\right)  ^{2}+\frac{i}{a_{i0}}\left[  \frac{k_{x}\left(  U_{x}
^{2}-V_{ix}^{2}\right)  }{2}-\omega_{i}\left(  U_{x}-V_{ix}\right)  \right]
\right\}  . 
\label{M_swapping}
\end{align}
\end{linenomath*}
Then, redistributing the terms inside both the pre-exponent and the exponents,
we rewrite equation~(\ref{M_swapping}) as
\begin{linenomath*}
\begin{align}
&  M=\int_{-\infty}^{\infty}dU_{x}\exp\left[  -~\frac{(U_{x}-U_{0})^{2}%
}{2V_{Tx}^{2}}-\frac{(\omega_{i}-k_{x}U_{x})^{2}}{2(k_{y}^{2}V_{\perp
}^{2}+ik_{x}a_{i0})}\right]  \int_{U_{x}}^{\infty}dV_{ix}\nonumber\\
&  \times\left[  \frac{k_{x}(U_{x}-U_{0})}{V_{Tx}^{2}}+\frac{k_{y}%
^{2}(\omega_{i}-k_{x}U_{x})}{k_{y}^{2}V_{Ty}^{2}+ik_{x}a_{i0}}-\frac
{ik_{y}^{2}}{a_{i0}}\left(  V_{ix}-\frac{k_{y}^{2}V_{Ty}^{2}U_{x}+iR
_{i}a_{i0}}{V_{Ty}^{2}k_{y}^{2}+ik_{x}a_{i0}}\right)  \right] \nonumber\\
&  \times\exp\left[  -\ \frac{k_{y}^{2}V_{Ty}^{2}+ik_{x}a_{i0}}{2a_{i0}%
^{2}}\left(  V_{ix}-\frac{k_{y}^{2}V_{Ty}^{2}U_{x}+i\omega_{i}a_{i0}}%
{k_{y}^{2}V_{Ty}^{2}+ik_{x}a_{i0}}\right)  ^{2}\right]  .
\label{M_integration}
\end{align}
\end{linenomath*}
This simplifies evaluating the internal integral over $V_{ix}$, so that we
obtain
\begin{linenomath*}
\begin{equation}
M=\int_{-\infty}^{\infty}\exp\left[  -~\frac{(U_{x}-U_{0})^{2}}%
{2V_{Tx}^{2}}\right]  K(U_{x})dU_{x}, 
\label{M_via_K}
\end{equation}
\end{linenomath*}
where
\begin{linenomath*}
\begin{align*}
K(U_{x})  &  \equiv\frac{\sqrt{\pi}\,a_{i0}}{\sqrt{2(k_{y}^{2}V_{Ty}%
^{2}+ik_{x}a_{i0})}}\left[  \frac{k_{x}(U_{x}-U_{0})}{V_{Tx}^{2}}%
+\frac{k_{y}^{2}(\omega_{i}-k_{x}U_{x})}{V_{Ty}^{2}k_{y}^{2}+ik_{x}a_{i0}%
}\right] \\
&  \times\ w\left[  \frac{\omega_{i}-k_{x}U_{x}}{\sqrt{2(k_{y}^{2}V_{Ty}%
^{2}+ik_{x}a_{i0})}}\right]  -\frac{ik_{y}^{2}a_{i0}}{k_{y}^{2}V_{Ty}%
^{2}+ik_{x}a_{i0}}.
\end{align*}
\end{linenomath*}
To calculate the remaining integral over $U_{x}$, we make the complex argument
of the $w$-function a new integration variable,
\begin{linenomath*}
\[
z=\frac{\omega_{i}-k_{x}U_{x}}{\sqrt{2(k_{y}^{2}V_{Ty}^{2}+ik_{x}a_{i0})}}.
\]
\end{linenomath*}
This transforms equation~(\ref{M_via_K}) to
\begin{linenomath*}
\begin{align}
M  &  =\frac{\sqrt{2\pi}a_{i0}}{k_{x}V_{Tx}^{2}}\int_{-\infty}%
^{\infty}\left[  \frac{k_{y}^{2}(V_{Tx}^{2}-V_{Ty}^{2}%
)-ik_{x}a_{i0}}{\sqrt{k_{y}^{2}V_{Ty}^{2}+ik_{x}a_{i0}}}\ z+\frac{R
_{i}-k_{x}U_{0}}{\sqrt{2}}\right] \nonumber\\
&  \times\exp\left\{  -~\frac{k_{y}^{2}V_{Ty}^{2}+ik_{x}a_{i0}}{k_{x}%
^{2}V_{Tx}^{2}}\left[  z-\frac{\omega_{i}-k_{x}U_{0}}{\sqrt{2(k_{y}%
^{2}V_{Ty}^{2}+ik_{x}a_{i0})}}\right]  ^{2}\right\}  w(z)dz\label{M_then}\\
&  -\frac{i\sqrt{2\pi}\,k_{y}^{2}a_{i0}V_{Tx}}{k_{y}^{2}V_{Ty}%
^{2}+ik_{x}a_{i0}},\nonumber
\end{align}
\end{linenomath*}
where, in a usual way, assuming the good analytic behavior of the integrand
and its integrability at infinities, we have reduced the new integration path
in the complex $z$-plane to its real axis.

Introducing temporary parameters
\begin{linenomath*}
\begin{subequations}
\label{AAABBB}%
\begin{align}
A  &  =\frac{\sqrt{k_{x}^{2}V_{Tx}^{2}+k_{y}^{2}V_{Ty}^{2}%
+ik_{x}a_{i0}}}{k_{x}V_{Tx}},\label{AAA}\\
B  &  =\frac{\omega_{i}-k_{x}U_{0}}{k_{x}V_{Tx}}\sqrt{\frac{k_{y}%
^{2}V_{Ty}^{2}+ik_{x}a_{i0}}{2(k_{x}^{2}V_{Tx}^{2}+k_{y}^{2}%
V_{Ty}^{2}+ik_{x}a_{i0})}}, \label{BBB}%
\end{align}
\end{subequations}
\end{linenomath*}
we have
\begin{linenomath*}
\[
\frac{k_{y}^{2}V_{Ty}^{2}+ik_{x}a_{i0}}{k_{x}^{2}V_{Tx}^{2}}%
=A^{2}-1,\qquad\frac{\omega_{i}-k_{x}U_{0}}{\sqrt{2(k_{y}^{2}V_{Ty}%
^{2}+ik_{x}a_{i0})}}=\frac{AB}{A^{2}-1}.
\]
\end{linenomath*}
so that we now can apply to (\ref{M_then}) equations (\ref{Final_integral_w()}) and (\ref{Int(z..)}).
After a straightforward algebra and by using equation~(\ref{IntF_via_M}), we
arrive at the sought-for result
\begin{linenomath*}
\begin{align}
&
{\displaystyle\iint\limits_{-\infty}^{\infty}}
F_{\omega\vec{k}}^{(2)}dV_{ix}dV_{iy}=-\ \frac{k^{2}V_{Tn}^{2}n_{0}%
\phi_{\omega k}}{\beta_{T}(k_{x}^{2}V_{Tx}^{2}+k_{y}%
^{2}V_{Ty}^{2}+i\vec{k}\cdot\vec{a}_{i0})}\nonumber\\
&  \times\left[  1+\frac{i\sqrt{\pi}(\omega_{i}-\vec{k}\cdot\vec{U}_{0})}%
{\sqrt{2(k_{x}^{2}V_{Tx}^{2}+k_{y}^{2}V_{Ty}^{2}%
+i\vec{k}\cdot\vec{a}_{i0})}}\ w\left(  \frac{\omega_{i}-\vec{k}\cdot\vec{U}%
_{0}}{\sqrt{2(k_{x}^{2}V_{Tx}^{2}+k_{y}^{2}V_{Ty}%
^{2}+i\vec{k}\cdot\vec{a}_{i0})}}\right)  \right]  , 
\label{Final_F^(2)}%
\end{align}
\end{linenomath*}
where, similarly to equation~(\ref{Int^(1)_Final}), we have expressed the
result in to the coordinate-invariant form, $k_{x}=k_{x}$,
$k_{y}=k_{y}$, and recalled that
\begin{linenomath*}
\begin{equation}
\vec{U}_{0}=\frac{\vec{a}_{i0}}{\nu_{in}} 
\label{U_0=a_0/nu_i}
\end{equation}
\end{linenomath*}

For $V_{Tx}^{2}=V_{Ty}^{2}=V_{Tn}^{2}$, $\vec{a}_{i0}=0$, $\vec{U}_{0}=0$,
equation~(\ref{Final_F^(2)}) reduces to
\begin{linenomath*}
\begin{equation}
{\displaystyle\iint\limits_{-\infty}^{\infty}}
F_{\omega\vec{k}}^{(2)}dV_{ix}dV_{iy}=-\ \frac{n_{0}\phi_{\omega k}}{\beta
_{T}}\left[  1+\sqrt{\frac{\pi}{2}}\ \frac{i\omega_{i}}{kV_{Tn}}\ w\left(
\frac{\omega_{i}}{\sqrt{2}kV_{Tn}}\right)  \right]  , 
\label{agrees}
\end{equation}
\end{linenomath*}
in full agreement with the corresponding no-field expression.

The non-standard case of heavily damped waves, $\gamma<-\nu_{in}$, i.e., of $\operatorname{Re}(R_i)<0$, 
for $F_{\omega\vec{k}}^{(2)}$ is fully analogous to that of $F_{\omega\vec{k}}^{(1)}$. 
Using the same arguments that led us to equation~(\ref{begin-}), instead 
of equation~(\ref{F^2_solution}) we obtain now the solution
\begin{linenomath*}
\begin{align*}
&  F_{\omega\vec{k}}^{(2)}=-Q\int_{V_{ix}}^{\infty}dU_{x}\left(  \frac
{k_{x}(U_{x}-U_{0})}{V_{Tx}^{2}}+\frac{k_{y}V_{iy}}{V_{Ty}^{2}%
}\right) \nonumber\\
&  \times\exp\left\{  -~\frac{(U_{x}-U_{0})^{2}}{2V_{Tx}^{2}}%
-\frac{V_{iy}^{2}}{2V_{Ty}^{2}}+\frac{i}{a_{i0}}\left[  \frac{k_{x}\left(
U_{x}^{2}-V_{ix}^{2}\right)  }{2}+(k_{y}V_{iy}-\omega_{i})\left(  U_{x}%
-V_{ix}\right)  \right]  \right\},
\end{align*}
\end{linenomath*}
which differs from equation~(\ref{F^2_solution}) by the integration limits and the sign in front of $Q$. 
As a result, making the same further steps that led us to equation~(\ref{Final_F^(2)}),
we obtain for the integral ${\iint\limits_{-\infty}^{\infty}}F_{\omega\vec{k}}^{(2)}dV_{ix}dV_{iy}$ 
the same expression, except that in the RHS of equation~(\ref{Final_F^(2)}) the $w$-function 
of its complex argument $z$ is replaced by $-w(-z)=w(z)-2\exp(-z^{2})$, 
in total analogy with equation~(\ref{Int^(1)_Final_-}) for $F_{\omega\vec{k}}^{(1)}$.

\section{Simplification of the General Solution for
Electrons\label{Simplification of the general solution for electrons}}

In this appendix, we simplify the exact general solution for $f_{\omega\vec{k}}^{e}(\xi)$ given by
equation~(\ref{w(x)_L_unique}) and preceding equations~(\ref{y(x)_via_w})--(\ref{mu==}). 
The goal of these simplifications is to make the solution more tractable for practical uses.
We will make a series of approximations based on the fact that the absolute value of the complex parameter $\mu$ is sufficiently large. 
In this appendix, we use a local variable $z$ which is not to be confused with such notation 
from section~\ref{No-field approximation for electrons}.

Firstly, we approximate the modified Bessel functions $\mathrm{I}_{\mu}(x/2)$
and $\mathrm{K}_{\mu}(x/2)$ by using the following exact integral
representations \cite{Bateman_Erdelyi:Higher3,Arfken:Mathematical},
\begin{linenomath*}
\begin{subequations}
\label{Bessel_I,K(x/2)_again_1}%
\begin{align}
\mathrm{I}_{\mu}\left(  \frac{x}{2}\right)   &  =\frac{1}{\sqrt{\pi}\,
\Gamma\left(  \mu+\frac{1}{2}\right)  }\left(  \frac{x}{4}\right)  ^{\mu}%
\int_{-1}^{1}\exp\left(  \frac{xt}{2}\right)  \left(  1-t^{2}\right)
^{\mu-\frac{1}{2}}dt,\label{Bessel_I(x/2)_again_1}\\
\mathrm{K}_{\mu}\left(  \frac{x}{2}\right)   &  =\frac{\sqrt{\pi}}%
{\Gamma\left(  \mu+\frac{1}{2}\right)  }\left(  \frac{x}{4}\right)  ^{\mu}%
\int_{1}^{\infty}\exp\left(  -\ \frac{xt}{2}\right)  \left(  t^{2}-1\right)
^{\mu-\frac{1}{2}}dt. 
\label{Bessel_K(x/2)_again_1}
\end{align}
\end{subequations}
\end{linenomath*}
These relations are valid for the real positive argument $x$ and arbitrary
complex order $\mu$ (whose real part is assumed positive). Analytically
continuing the real domain of integrations in
equations~(\ref{Bessel_I(x/2)_again_1}) and (\ref{Bessel_K(x/2)_again_1}) into
the complex plane of $t$, deforming the contours of integration in such a way
that they pass through the most appropriate saddle points with the direction
of passage aligned with the steepest descent, we will eventually arrive at the
following approximate expressions:
\begin{linenomath*}
\begin{subequations}
\label{Bessel_I,K_appoximate}
\begin{align}
&  \mathrm{I}_{\mu}\left(  \frac{x}{2}\right)  \approx\tilde{I}_{\mu}\left(
\frac{x}{2}\right) \nonumber\\
&  \equiv\frac{1}{\sqrt{\pi}\left(  4\eta^{2}+x^{2}\right)  ^{\frac{1}{4}}%
}\left(  \frac{\sqrt{4\eta^{2}+x^{2}}-2\eta}{x}\right)  ^{\eta+\frac{1}{2}%
}\exp\left(  \frac{\sqrt{4\eta^{2}+x^{2}}}{2}\right)
,\label{Bessel_I_appoximate}\\
&  \mathrm{K}_{\mu}\left(  \frac{x}{2}\right)  \approx\tilde{K}_{\mu}\left(
\frac{x}{2}\right) \nonumber\\
&  \equiv\frac{\sqrt{\pi}}{\left(  4\eta^{2}+x^{2}\right)  ^{\frac{1}{4}}%
}\left(  \frac{\sqrt{4\eta^{2}+x^{2}}+2\eta}{x}\right)  ^{\eta+\frac{1}{2}%
}\exp\left(  -\ \frac{\sqrt{4\eta^{2}+x^{2}}}{2}\right)  ,
\label{Bessel_K_appoximate}
\end{align}
\end{subequations}
\end{linenomath*}
where we temporarily introduced a new parameter
\begin{linenomath*}
\begin{equation}
\eta\equiv\mu-\frac{1}{2}. 
\label{eta_via_mu}
\end{equation}
\end{linenomath*}
Approximations given by equation~(\ref{Bessel_I,K_appoximate}) involve only
elementary functions.  While assuming $|\mu|\gg 1$, the difference between $\mu$ 
and $\eta$ may seem negligible, but in the process of derivations it is better to retain this 
difference and neglect it only at the very end. From equation~(\ref{Bessel_I,K(x/2)_again_1}), we can easily verify
that the approximations $\tilde{I}_{\mu}$ and $\tilde{K}_{\mu}$ are related as
\begin{linenomath*}
\begin{equation}
\mathrm{I}_{\mu}\left(  \frac{x}{2}\right)  \mathrm{K}_{\mu}\left(  \frac
{x}{2}\right)  \approx\tilde{I}_{\mu}\left(  \frac{x}{2}\right)  \tilde
{K}_{\mu}\left(  \frac{x}{2}\right)  =\frac{1}{\sqrt{4\eta^{2}+x^{2}}}.
\label{svoystvo}
\end{equation}
\end{linenomath*}
This approximate relation will dramatically simplify our results, as will
become clear soon.

As $|\eta| \rightarrow\infty$, the approximate functions
$\tilde{I}_{\mu}$ and $\tilde{K}_{\mu}$ asymptotically converge to the actual
modified Bessel functions. For the finite parameter $\eta$, the accuracy of
these approximations monotonically increases with both $|\eta|$ and $x$.
Direct numerical comparisons show that for sufficiently large $|\eta| $ these approximations are quite accurate. Below we
demonstrate analytically that the functions $\tilde{I}_{\mu}$ approximates
$\mathrm{I}_{\mu}$ within the 10\% accuracy for $|\eta|
\gtrsim5$ and 2\% for $|\eta| \gtrsim25$, while $\tilde
{K}_{\mu}$ approximates $\mathrm{K}_{\mu}$ within even a much better accuracy of
1\% for $|\eta| \gtrsim5$ and 0.1\% for $|\eta| \gtrsim25$.

To quantify the accuracy of these approximations in a systematic way, 
we consider the worst-accuracy limit of $x\ll1$. In this limit, equations~(\ref{Bessel_I_mu=Sum}) and
(\ref{Bessel_K_mu=Sum}) lead to
\begin{linenomath*}
\begin{subequations}
\label{Bessel_I,K,x<<1}
\begin{align}
\mathrm{I}_{\mu}\left(  \frac{x}{2}\right)   &  \approx\frac{1}{\Gamma(\mu
+1)}\left(  \frac{x}{2}\right)  ^{\mu}=\frac{1}{(\eta+1/2)  \Gamma(\eta+1/2)}\left(  \frac{x}{4}\right)  ^{\eta
+\frac{1}{2}}\label{Bessel_I,x<<1}\\
\mathrm{K}_{\mu}\left(\frac{x}{2}\right)  &  \approx\frac{\pi}{2\sin\left(  \mu
\pi\right)  \Gamma(1-\mu)}\left(  \frac{x}{2}\right)  ^{-\mu}=\frac{1}%
{2}\ \Gamma\left(  \eta+\frac{1}{2}\right)  \left(  \frac{x}{4}\right)
^{-\left(  \eta+\frac{1}{2}\right)  } \label{Bessel_K,x<<1}%
\end{align}
\end{subequations}
\end{linenomath*}
where we used $\mu=\eta+1/2$ and the well-known gamma-function relations
\begin{linenomath*}
\begin{equation}
\Gamma(z+1)=z\Gamma(z),\qquad\Gamma(z)\Gamma(1-z)=\frac{\pi}{\sin\mu\pi}.
\label{Gamma_sootno}
\end{equation}
\end{linenomath*}
In the same limit of $x\ll1$, the approximate functions $\tilde{I}_{\mu}$ and
$\tilde{K}_{\mu}$ are given by
\begin{linenomath*}
\begin{equation}
\tilde{I}_{\mu}\left(  \frac{x}{2}\right)  \approx\frac{1}{\sqrt{2\pi\eta}%
}\left(  \frac{x}{4\eta}\right)  ^{\eta+\frac{1}{2}}e^{\eta},\qquad\tilde
{K}_{\mu}\left(  \frac{x}{2}\right)  \approx\sqrt{\frac{\pi}{2\eta}}\left(
\frac{4\eta}{x}\right)  ^{\eta+\frac{1}{2}}e^{-\eta}, 
\label{I,K_approx_x<<1}
\end{equation}
\end{linenomath*}
so that for the ratios of the approximate functions to the actual ones we have
\begin{linenomath*}
\begin{equation}
\frac{\tilde{I}_{\mu}(x/2)  }{\mathrm{I}_{\mu}(x/2)}\approx\frac{1+1/(2\eta)}{\Lambda\left(
\eta\right)  },\qquad\frac{\tilde{K}_{\mu}(x/2)
}{\mathrm{K}_{\mu}(x/2)  }\approx\Lambda\left(
\eta\right)  , 
\label{prome}
\end{equation}
\end{linenomath*}
where
\begin{linenomath*}
\begin{equation}
\Lambda\left(  \eta\right)  \equiv\frac{\sqrt{2\pi}}{\Gamma\left(  \eta
+\frac{1}{2}\right)  }\left(  \frac{\eta}{e}\right)  ^{\eta}.
\label{Lambda(eta)}
\end{equation}
\end{linenomath*}
Using Stirling's formula with the first- and second-order corrections
\cite{Arfken:Mathematical},
\begin{linenomath*}
\begin{equation}
\Gamma(z+1)\approx\sqrt{2\pi z}\left(  \frac{z}{e}\right)  ^{z}\left(
1+\frac{1}{12z}+\frac{1}{288z^{2}}\right)  , 
\label{Stirling_accurate}
\end{equation}
\end{linenomath*}
where we assume $\left\vert z\right\vert \gtrsim1$, and expanding $\Lambda\left(
\eta\right)  $ in Taylor series with respect to $1/\eta$, we arrive at a
simple approximation
\begin{linenomath*}
\begin{equation}
\Lambda\left(  \eta\right)  \approx1+\frac{1}{24\eta}+\frac{5}{1152\eta^{2}}.
\label{Lambda(eta)_approximate}
\end{equation}
\end{linenomath*}
Then equations~(\ref{prome}) and (\ref{Lambda(eta)_approximate}) yield
\begin{linenomath*}
\begin{subequations}
\label{accuracy_I,K_x<<1}
\begin{align}
&  \frac{\tilde{I}_{\mu}(x/2)}{\mathrm{I}_{\mu}(x/2)}\approx1+\frac{11}{24\eta}-\frac{3}{128\eta^{2}}%
\approx1+\frac{0.46}{\eta}-\frac{0.023}{\eta^{2}},\label{accuracy_I_x<<1}\\
&  \frac{\tilde{K}_{\mu}(x/2)  }{\mathrm{K}_{\mu}(x/2)}\approx1+\frac{1}{24\eta}+\frac{5}{1152\eta^{2}}%
\approx1+\frac{0.0417}{\eta}+\frac{0.0043}{\eta^{2}}. \label{accuracy_K_x<<1}%
\end{align}
\end{subequations}
\end{linenomath*}
The inaccuracy of the approximations $\tilde{I}_{\mu}(x/2)$ 
and $\tilde{K}_{\mu}(x/2)$ is determined
by the corresponding highest-order corrections $\propto1/\eta$. Comparing them
with the second-order corrections $\propto1/\eta^{2}$, we see that these
expansions are valid for $|\eta| \gg0.1$, i.e., virtually for all values of interest. We see that the
function $\tilde{K}_{\mu}(x/2)  $ approximates the actual
function $\mathrm{K}_{\mu}(x/2)  $ within 10\% accuracy
for $\left\vert \eta\right\vert $ even as small as $|\eta| \sim0.5$. However, to provide the same 10\% accuracy for
$\tilde{I}_{\mu}(x/2)  $, the parameter $|\eta|$ should be at least an order of 
magnitude larger, $|\eta| \gtrsim5$. The physical estimates for the E-region
conditions ($|\eta| \approx|\mu|\gtrsim25$) made in Sec.~\ref{General case} 
put the parameter $\eta$ well within this bound. The inaccuracy of
the product given by equation~(\ref{svoystvo}) is determined by
\begin{linenomath*}
\begin{equation}
\frac{\tilde{I}_{\mu}(x/2)  \tilde{K}_{\mu}(x/2)}{\mathrm{I}_{\mu}(x/2)
\mathrm{K}_{\mu}(x/2)}\approx1+\frac{1}{2\eta}.
\label{product_inaccuracy}
\end{equation}
\end{linenomath*}
The accuracy of both approximations improves as $x$ increases, but
for all $x$ the accuracy of $\tilde{K}_{\mu}(x/2)$ always
remains significantly better than that of $\tilde{I}_{\mu}(x/2)$.

After substituting the original functions $\mathrm{I}_{\mu}$ and
$\mathrm{K}_{\mu}$ with their approximations in terms of elementary functions,
$\tilde{I}_{\mu}$ and $\tilde{K}_{\mu}$, the integrals in
equation~(\ref{w(x)_L_unique}) still cannot be evaluated exactly. This
requires further simplifications based on the specific behavior of these
functions within the integration domain. Of special importance is the fact that
the real part of the parameter $\eta=\mu-1/2$, according to
equations~(\ref{mu==}) and (\ref{kappa}) is positive and large. This leads to the fact
that for sufficiently large $|\eta|$, the integrands in
equation~(\ref{w(x)_L_unique}) mostly concentrate either near the upper
integration limit [for the first term in the RHS of
equation~(\ref{w(x)_L_unique})] or near the lower integration limit (for the
second term there).

This suggests the following procedure to further approximate the analytic
evaluation of the integrals in equation~(\ref{w(x)_L_unique}). The entire
$\delta$-function-like integrands can be locally approximated near the 
integration limit at $z=x$ by fictitious exponential functions that can be
integrated analytically. The fictitious exponents are determined by the
corresponding logarithmic derivatives of those integrands\ at $z=x$. Within
the main bulk of each function (in close proximities to the corresponding
integration limits), the integrands and the fictitious exponential functions
are reasonably close to each other. Outside those bulks they may differ
significantly, but the contributions of the corresponding integration domains to the
total integrals are relatively small. Then for each fictitious exponential
function, the other integration limit can be extended to the corresponding
infinity, so that the effect of that integration limit effectively 
disappears. As a result of the entire approximations, we might
expect that the integral of each fictitious exponential function will be
reasonable close to the corresponding exact integral in
equation~(\ref{w(x)_L_unique}). This procedure becomes asymptotically accurate
as $|\eta| \rightarrow\infty$. For the finite value of
$|\eta|$, the acceptable accuracy of the final results
may require somewhat larger values of $|\eta|$ than needed for approximating the
modified Bessel functions (see above) $|\eta| \gtrsim5$,
but mostly also well within the limits of the above physical estimates, $|\eta| \approx |\mu| \gtrsim 25$.

The described procedure of this local approximate integration leads to a
remarkable consequence: the result of each integration becomes proportional to
the integrand value at the upper (or lower) integration limit $x$ multiplied
by a simple elementary function of $x$. As a result, if the exact pre-integral
modified Bessel functions are also replaced by the corresponding approximate
functions $\tilde{I}_{\mu}\left(  z/2\right)  $ or $\tilde{K}_{\mu}\left(
z/2\right)  $, then in each integration term these two functions will merge
into a common factor $\tilde{I}_{\mu}\left(  z/2\right)  \tilde{K}_{\mu
}\left(  z/2\right)  $, for which we can apply the simple relation given by
equation~(\ref{svoystvo}). After making other minor approximations (see
below), the summation of the two terms yields the function $L\left(  x\right)
$ divided by a simple polynomial. This will allow us to obtain the final
result in terms of the error functions (or, equivalently, the $Z$-functions).

Bearing in mind equations~(\ref{Bessel_I,K_appoximate}) and (\ref{L(x)}), we
see that each integrand in the RHS of equation~(\ref{w(x)_L_unique}) is
proportional to a fraction in the $\eta$-th power, an exponential function,
the pre-exponential multiplier $(4\eta^{2}+z^{2})^{-\frac{1}{4}}$, a factor $\sqrt{z}$ preceding the
modified Bessel functions in equation~(\ref{w(x)_L_unique}), as well as
$z^{5/4}$ and powers of $z$ stemming from $L(z)$. Within the narrow bulk of
the integrands, the $\eta$-th powers and exponential functions play the
major role, while the other, much slower varying, functions can be pulled out
of the integrals by approximately replacing their integration variable $z$
with the corresponding integration limit $x$ (this will lead to an additional small
loss of accuracy).

Temporarily denoting the essential functions in the corresponding integrals by
$f(z)$ and $g(z)$, according to the procedure
described above, we obtain
\begin{linenomath*}
\begin{equation}
\int_{0}^{x}f\left(  z\right)  dz\approx\frac{f\left(  x\right)  }{\frac
{d}{dx}\ln f\left(  x\right)  },\qquad\int_{x}^{\infty}g\left(  z\right)
dz\approx\frac{g\left(  x\right)  }{-\frac{d}{dx}\ln g\left(  x\right)  }.
\label{f,g}
\end{equation}
\end{linenomath*}
To calculate the logarithmic derivatives, we temporarily disregard in
$f_{n}(x)  $ and $g_{n}(x)  $ all multipliers that
exert no effect on the logarithmic derivatives and leave only the essential
functions:
\begin{linenomath*}
\begin{subequations}
\label{f,g_kappa}
\begin{align}
f\left(  x\right)   &  =\left(  \frac{\sqrt{4\eta^{2}+x^{2}}-2\eta}{x}\right)
^{\eta+\frac{1}{2}}\exp\left(  \frac{\sqrt{4\eta^{2}+x^{2}}}{2}-\left(
\alpha+\lambda\right)  x\right)  ,\label{f_kappa}\\
g\left(  x\right)   &  =\left(  \frac{\sqrt{4\eta^{2}+x^{2}}+2\eta}{x}\right)
^{\eta+\frac{1}{2}}\exp\left(  -\ \frac{\sqrt{4\eta^{2}+x^{2}}}{2}-\left(
\alpha+\lambda\right)  x\right)  . \label{g_kappa}%
\end{align}
\end{subequations}
\end{linenomath*}
Then the corresponding logarithmic derivatives become
\begin{linenomath*}
\begin{subequations}
\label{(d/dx)f,g_kappa}
\begin{align}
\frac{d}{dx}\ln f\left(  x\right)   &  =\frac{4\eta^{2}+2\eta+x^{2}-2\left(
\alpha+\lambda\right)  x\sqrt{4\eta^{2}+x^{2}}}{2x\sqrt{4\eta^{2}+x^{2}}%
},\label{(d/dx)f_kappa}\\
-\frac{d}{dx}\ln g\left(  x\right)   &  =\frac{4\eta^{2}+2\eta+x^{2}+2\left(
\alpha+\lambda\right)  x\sqrt{4\eta^{2}+x^{2}}}{2x\sqrt{4\eta^{2}+x^{2}}}.
\label{(d/dx)g_kappa}
\end{align}
\end{subequations}
\end{linenomath*}

Our final approximation is neglecting in the numerators of
equations~(\ref{(d/dx)f_kappa}) and (\ref{(d/dx)g_kappa}) the term 
$2\eta$ compared to the largest constant term $4\eta^{2}$. While leading to an
additional relatively small loss of accuracy, this neglect will significantly simplify
both equation~(\ref{(d/dx)f,g_kappa}) and all expressions that will follow,
\begin{linenomath*}
\begin{subequations}
\label{(d/dx)f,g_kappa_simplified}
\begin{align}
\frac{d}{dx}\ln f\left(  x\right)   &  \approx\frac{\sqrt{4\eta^{2}+x^{2}}-
2\left(  \alpha+\lambda\right)  x}{2x},\label{(d/dx)f_kappa_simplified}\\
-\,\frac{d}{dx}\ln g\left(  x\right)   &  \approx\frac{\sqrt{4\eta^{2}+x^{2}}+
2\left(  \alpha+\lambda\right)  x}{2x}. 
\label{(d/dx)g_kappa_simplified}
\end{align}
\end{subequations}
\end{linenomath*}
These expressions lead to the following approximate integral evaluations:
\begin{linenomath*}
\begin{align*}
\int_{0}^{x}\sqrt{z}\,\mathrm{I}_{\mu}\left(  \frac{z}{2}\right)  L(z)dz  &
\approx\frac{2x^{\frac{3}{2}}\tilde{I}_{\mu}(x/2)
L(x)}{\sqrt{4\eta^{2}+x^{2}}-2\left(  \alpha+\lambda\right)  x},\\
\int_{x}^{\infty}\sqrt{z}\,\mathrm{K}_{\mu}\left(  \frac{z}{2}\right)  L(z)dz
&  \approx\frac{2x^{\frac{3}{2}}\tilde{K}_{\mu}(x/2)
L(x)}{\sqrt{4\eta^{2}+x^{2}}+2\left(  \alpha+\lambda\right)  x}.
\end{align*}
\end{linenomath*}

Now we apply these expressions to the full solution given by
equation~(\ref{w(x)_L_unique}), where we replace the exact pre-integral
functions $\mathrm{I}_{\mu}\left(  x/2\right)  $ and $\mathrm{K}_{\mu}(x/2)
$ by their approximations $\tilde{I}_{\mu}(x/2) $
and $\tilde{K}_{\mu}\left(  x/2\right)  $. As a result, in each of the two
terms, the two different modified Bessel functions -- one in the integrand and
the other in the pre-integral factor -- combine together to the the common
factor $\tilde{I}_{\mu}\left(  x/2\right)  \tilde{K}_{\mu}\left(  x/2\right)
$. After applying equation~(\ref{svoystvo}), we arrive at the following
approximate solution for $\tilde{w}(x)$:
\begin{linenomath*}
\begin{equation}
\tilde{w}(x)\approx-\ \frac{x^{2}L(x)}{\mu^{2}+[1/4-(
\alpha+\lambda)  ^{2}]  x^{2}}=\frac{(N_{0}\xi^{2}+N_{1}
\xi+N_{2})  e^{-\alpha x}x^{-\gamma}e^{-\xi}\phi_{\omega\vec{k}}}
{\mu^{2}+[1/4-(\alpha+\lambda)  ^{2}]  x^{2}},
\label{w(x)_approximate}
\end{equation}
\end{linenomath*}
where we have finally neglected the difference between $\eta$ and $\mu$. Then the
original function $f_{\omega\vec{k}}^{e}(\xi)$ becomes
\begin{linenomath*}
\[
f_{\omega\vec{k}}^{e}(\xi)=e^{\alpha x}x^{\gamma}\tilde{w}(x)
\approx\frac{N_{0}\xi^{2}+N_{1}\xi+N_{2}}{\mu^{2}+[1/4-(
\alpha+\lambda)  ^{2}]  x^{2}}\ e^{-\xi}\phi_{\omega\vec{k}}.
\]
\end{linenomath*}
The validity of the approaches used above, including the approximate integrations, 
can be additionally verified by applying them to the exact integral relations 
given by equation~(\ref{Bessel_I,K_via_K,I}).

Neglecting in the expression for $\mu$ the term $9/16$ compared to the large
parameter $\mathcal{P}$, see equation~(\ref{mu==}), and using the expressions
of $\lambda=$ $\xi/x$ and $\alpha$ given by equation~(\ref{gamma,alpha,kappa}), we obtain
\begin{linenomath*}
\begin{align*}
\mu^{2}  &  \approx\mathcal{P}=\frac{\Omega_{e}^{2}V_{Te}^{2}}{4\nu_{0}
a_{e0}^{2}}\left(3i\Delta_{\omega\vec{k}}+\frac{k_{x}^{2}V_{Te}^{2}
}{\nu_{0}}\right)  ,\\
\alpha+\lambda &  =\frac{1}{2\sin\varphi}\left(i\cos\varphi+\frac{a_{e0}%
}{k_{y}V_{Te}^{2}}\right)  ,\qquad x^{2}=\frac{4k_{y}^{2}V_{Te}%
^{4}\sin^{2}\varphi}{a_{e0}^{2}}\ \xi^{2},
\end{align*}
\end{linenomath*}
so that the denominator in equation~(\ref{w(x)_approximate}) becomes
\begin{linenomath*}
\begin{align*}
&  \mu^{2}+\left[  \frac{1}{4}-\left(  \alpha+\lambda\right)  ^{2}\right]
x^{2}\\
&  \approx\frac{\Omega_{e}^{2}V_{Te}^{2}}{4\nu_{0}a_{e0}^{2}}\left(
3i\Delta_{\omega\vec{k}}+\frac{k_{x}^{2}V_{Te}^{2}}{\nu_{0}}\right)
+\frac{k_{y}^{2}V_{Te}^{4}\xi^{2}}{a_{e0}^{2}}\left(  1-\frac{2ia_{e0}%
\cos\varphi}{k_{y}V_{Te}^{2}}-\frac{a_{e0}^{2}}{k_{y}^{2}V_{Te}^{4}%
}\right)
\end{align*}
\end{linenomath*}
Recalling the definitions of the parameters $N_{0,1,2}$
given by equation~(\ref{pPCxiABD_1}), we obtain
\begin{linenomath*}
\begin{align}
&  f_{\omega\vec{k}}^{e}(\xi)\approx\frac{n_{0}e^{-\xi}}{\left(  2\pi\right)
^{3/2}V_{Te}^{3}}\nonumber\\
&  \times\frac{4\left(  k_{y}^{2}V_{Te}^{2}-2ik_{y}a_{e0}\cos
\varphi\right)  \xi^{2}+20ik_{y}\xi a_{e0}\cos\varphi+k_{x}%
^{2}V_{Te}^{2}(\Omega_{e}^{2}/\nu_{0}^{2})}{4\left(  k_{y}^{2}V_{Te}%
^{2}-2ik_{y}a_{0}\cos\varphi-a_{0}^{2}/V_{Te}^{2}\right)  \xi^{2}+\left(
3i\Delta_{\omega\vec{k}}\nu_{0}+k_{x}^{2}V_{Te}^{2}\right)
(\Omega_{e}^{2}/\nu_{0}^{2})}\ \phi_{\omega\vec{k}}.
\label{Final_f(xi)}
\end{align}
\end{linenomath*}
In the $a_{e0}\rightarrow0$ limit, this expression reduces to
equation~(\ref{classical_1}) which was obtained by neglecting in
equation~(\ref{full_equation_xi_new_1}) all derivatives of $f_{\omega\vec{k}}^{e}(\xi)$ with respect to $\xi$.

\acknowledgments
At various stages of this research, it was supported by NSF Grants 1755350, 1007789, 0442075, 2301644, 9812679,  0332354,
2350366, and 2531480; NASA Grants 80NSSC21K1322, 80NSSC19K0080, and 80NSSC25K7753.

%
%


%
%
%
%
%

\bibliography{kinetic_TFBI}

\end{document}